# New Approaches to Soliton Quantization and Existence for Particle Physics

Paul J. Werbos
Room 675, National Science Foundation[*]
Arlington, Va. 22230
pwerbos@nsf.gov
2/25/98

**Abstract**

This paper provides mathematical details related to another new paper which suggests: (1) new approaches to the analysis of soliton stability; (2) families of Lagrangian field theories where solitons might possibly exist even without topological charge; (3) alternative approaches to quantizing solitons, with testable nuclear implications. This paper evaluates the possibility of strong energy-minimizing states in four families of systems, two promising and two not promising. In these examples, it presents new methods for second-order stability analysis, and describes the phenomenon of persistent multifurcation. Section 6 presents three alternative formalisms for quantizing solitons (topological or nontopological), all of which have major implications for the foundations of quantum theory: (1) the standard formalism, based on functional integration, reinterpreted as an imaginary Markhov Random Field (iMRF) across time and space, with parallels to fuzzy logic; (2) two radically conservative formalisms, consistent with the core of Einstein's vision, based on a true MRF model. Bell's Theorem, bosonization, time-symmetry and macroscopic asymmetry are discussed, along with a variety of testable alternative possibilities and heresies, such as nonDoppler redshift.

## 1. Introduction

Many theoretical physicists have tried to explain the existence and properties of elementary particles by representing them as stable localized patterns in continuous, classical field theory [1]. They call these particles "solitons" (with apologies to the mathematicians, who use the term in a different way). The procedures for quantizing these solitons -- for building quantum field theories (QFTs) which represent particles as solitons -- are straightforward, in principle, within the modern formulation of QFT[1-3], which has grown out of Feynman's path integral approach and Schwinger's source theory approach.

This paper is the technical companion to a new paper [4] which re-examines both the existence and quantization of such solitons. That paper had four major objectives: (1) it provides more rigorous concepts of stability, making contact with nonlinear system dynamics, for use in studying the existence of solitons; (2) it suggests possible families of Lagrangian field theories in which stable solitons might possibly exist even without topological charge, based on a careful re-evaluation of the established arguments against this possibility; (3) it provides an heretical new approach to the quantization of solitons, consistent with the philosophy of realism, with testable nuclear implications; (4) it provides a detailed review of mathematical and physical prerequisites to this work.

This paper will not repeat the reviews and overviews of that companion paper. Instead, it will proceed directly to provide certain mathematical details which were omitted from [4] for reasons of length. Sections 2-5 will evaluate the possibility of strong energy-minimizing equilibria in a series of 4 families of examples:
 (1) Section 2 will address the possibility of stable oscillation in a nonrelativistic 1+1-D

---

[*] The views herein are those of the author, not those of NSF. However, since this was written on government time, it is in the public domain subject to proper citation and retention of this caveat.
Many thanks are due to Dr. Ludmila Dolmatova, without whose help this would not have been written.

Lagrangian field theory previously proposed in [5];

(2) Section 3 will address the possibility of stable states in a relativistic 1+1-D theory;

(3) Section 4 will address the possibility of "energy" minimization in a 3-D spatial system, where dynamics are not considered;

(4) Section 5 will address the possibility of stable states in dynamical systems -- like general relativity -- where the Lagrangian depends in part on the second derivatives of the field.

In all four cases, I will focus on "Strong Second Order Stability" (SSOS) -- the search for equilibria which minimize energy in a <u>strong</u> sense, such that its Hessian is positive definite in all directions except for those directions which represent translation, rotation or a gauge shift. I will restrict attention to field theories where the energy density is positive definite in the neighborhood of the vacuum state, the state where $\varphi = \partial_t \varphi = 0$ and where the energy density is zero. (For alternative forms of stability, see [4].) Sections 2 and 5 will present negative results, suggesting that SSOS is impossible in those cases; however, the original strategy discussed in section 2 may yet be useful, when carried over to more promising systems like those of section 3 and 4. See [4] for ideas about how to extend the work of sections 3 and 4, <u>if successful</u>, so as to construct nontopological solitons in 3+1-D relativistic Lagrangian theories.

Section 2 will also describe an approach to second-order stability analysis complementary to that of Gelfand and Fomin[6], as discussed in [4]. Section 3 will contain comments about the tremendous variety of relativistic systems.

Finally, section 6 will present three alternative formalisms for quantizing solitons (topological or nontopological), all of which have major implications for the foundations of quantum theory: (1) the standard formalism, based on functional integration, reinterpreted as an imaginary Markhov Random Field (iMRF) across time and space, with parallels to fuzzy logic; (2) two radically conservative formalisms, consistent with the core of Einstein's vision, based on a true MRF model. Bell's Theorem, bosonization, time-symmetry and macroscopic asymmetry are discussed, along with a variety of testable alternative possibilities and heresies, such as nonDoppler redshift. These formalisms do not violate Bell's Theorem, because local time-symmetric MRFs cannot be represented as local Markhov processes (like the class of theories ruled out by Bell's Theorem experiments).

## **2. General Lessons from Analysis of a Nonrelativistic 1+1-D System**

The following section will use a simple nonrelativistic example in order to illustrate key issues and techniques relevant to nontopological solitons in general. Section 2.1 will work out the dynamics of the system, and develop some notation. Section 2.2 will work out the first-order conditions for the "soliton" states which would minimize the energy H of the system. Section 2.3 will briefly summarize the kind of reasoning which helped me select the parameters of this system. Section 2.4 will evaluate the significance of a general problem -- "persistent multifurcation" -- which may well appply to <u>all</u> nontopological solitons; this problem <u>appears</u> to imply an ill-posed dynamical system, but -- on further analysis -- the problem <u>is not</u> so severe as it appears. Section 2.5 gives my original plan to construct a numerical example of a soliton in this system. Section 2.6 describes two numerical methods for second-order stability analysis, one of them completely new, which can be applied in general to 1+1-D or radially defined solitons; the analytical form of the first method can be used to rule out the possibility of SSOS stability in a large class of systems (including equation 2). Finally, section 2.7 will describe how an alternative mechanism of stability -- based on minimizing another conserved quantity H' instead of H -- can be evaluated as well in this system; however, the results of the evaluation are negative.

The system to be studied here was originally proposed in [5]. In [5], I asked whether it would be possible to construct stable oscillatory "solitons" in <u>some</u> kind of Lagrangian system. I proposed that we study the following Lagrangian (with inconsequential sign changes) for a scalar field:

$$\mathcal{L} = |\dot{\varphi}|^2 + b|\varphi|^2|\dot{\varphi}|^2 + c|\dot{\varphi}|^2 + d|\dot{\varphi}|^2|\varphi_x|^2 - g(|\varphi|^2) - h|\varphi_x|^2 \qquad (1)$$

Even though this system does not actually permit such solitons, it is a useful context for developing concepts which may be useful to systems which do.

## 2.1. Basic Dynamics and Notation for the System

The Lagrangian in equation 1 may be written more compactly as:

$$\mathcal{L} = s^2 + b\rho s + cs + dsz - g(\rho) - hz \qquad (2)$$

where I define, for convenience:

$$\begin{aligned}
\varphi &= \varphi_1 + i\varphi_2 = re^{i\theta} \\
\rho &= |\varphi|^2 = \varphi\overline{\varphi} = \varphi_1^2 + \varphi_2^2 = r^2 \\
s &= |\dot{\varphi}|^2 = \dot{\varphi}(\partial_t\overline{\varphi}) = \dot{\varphi}_1^2 + \dot{\varphi}_2^2 = \dot{r}^2 + r^2\dot{\theta}^2 \\
z &= |\varphi_x|^2 = \varphi_x\overline{\varphi}_x = \varphi_{1,x}^2 + \varphi_{2,x}^2 = r_x^2 + r^2\theta_x^2
\end{aligned} \qquad (3)$$

For a complex scalar field, the usual conjugate momentum with respect to time may be derived:

$$\Pi_t^{[2]} = \frac{\delta\mathcal{L}}{\delta(\partial_t\overline{\varphi})} = 4s\dot{\varphi} + 2b\rho\dot{\varphi} + 2(c + dz)\dot{\varphi} \qquad (4)$$

The superscript expression "[2]" is simply a reminder that this is the usual conjugate momentum for 1+1=2 dimensional space-time. Inserting this into the usual Hamiltonian expression, we get:

$$\begin{aligned}
\mathcal{H} &= (\Pi_t^{[2]}\dot{\overline{\varphi}}) - \mathcal{L} = (4s^2 + 2b\rho s + 2cs + 2dzs) - (s^2 + b\rho s + cs + dsz - g(\rho) - hz) \\
&= 3s^2 + b\rho s + cs + dsz + g(\rho) + hz
\end{aligned} \qquad (5)$$

Notice that this expression for the energy density can easily be cross-checked by the laborious procedure of working out and using the conjugate momenta for the real variables $\varphi_1$ and $\varphi_2$ or r and $\theta$.

In general, the Lagrange-Euler equation for this kind of system can be expressed as:

$$\partial_t \Pi_t^{[2]} + \partial_x \Pi_x^{[2]} = \frac{\delta\mathcal{L}}{\delta\overline{\varphi}} \qquad (6)$$

Substituting equations 2 and 4 into this equation, we get:

$$(4s + 2b\rho + 2c + 2dz)\ddot{\varphi} + (4\dot{s} + 2b\dot{\rho} + 2d\dot{z})\dot{\varphi} + \partial_x \Pi_x^{[2]} = 2bs\varphi - g'(\rho)(2\varphi) \qquad (7)$$

With a bit of algebra, we can easily derive that:

$$4\dot{s}\dot{\varphi} = 8s\ddot{\varphi} \qquad (8)$$

Substituting equation 8 back into equation 12, we may deduce that:

$$(12s + 2b\rho + 2c + 2dz)\ddot{\varphi} = X \qquad , \qquad (9)$$

where X is some function of the state of the system ($\varphi$ and $\partial_t\varphi$ across space).

The major problematic aspect of this system is that this equation may become singular if the coefficient expression on the left becomes zero. In fact, that is true <u>precisely</u> at any state of nonstatic equilibrium, as we will see. This is an example of the general problem called "persistent multifurcation" in [5].

## 2.2. First-Order Conditions for Energy-Minimizing States

An energy-minimizing state of this system would consist of a state of the two independent functions, $\varphi$ and $\partial_t\varphi$, across space, at some t, which minimizes the integral over space of the energy density (equation 5). The Lagrange-Euler equations for that minimization problem are:

$$\partial_x\left(\frac{\delta\mathcal{H}}{\delta\overline{\varphi}_x}\right) = \frac{\delta\mathcal{H}}{\delta\overline{\varphi}} \tag{10}$$

and

$$\partial_x\left(\frac{\delta\mathcal{H}}{\delta(\dot{\overline{\varphi}})_x}\right) = \frac{\delta\mathcal{H}}{\delta\dot{\overline{\varphi}}} \tag{11}$$

Substituting equation 5 into equations 10 and 11, we get:

$$\partial_x(2ds\varphi_x + 2h\varphi_x) = 2bs\varphi + g'(\rho)(2\varphi) \tag{12}$$

where g' is just the derivative of g with respect to its argument, and (with just a little more algebra):

$$\partial_x(0) = (12s + 2b\rho + 2c + 2dz)\dot{\varphi} \tag{13}$$

For a <u>nonstatic</u> solution, a solution such that $\partial_t\varphi$ 0, equation 13 essentially requires:

$$12s + 2b\rho + 2c + 2dz = 0, \tag{14}$$

leading to the persistent multifurcation problem mentioned in section 2.1.
(For a complete rigorous analysis, we can only conclude that the real line is partitioned into closed intervals, where alternately $\partial_t\varphi=0$ and equation 14 is satisfied, and where there is at least one interval where equation 14 is satisfied and $\varphi$ is nonzero, for a nonstatic soliton. I believe that the logic below still goes through, based on an analysis of conditions within any partition, but in a more tedious fashion.)
Furthermore, Lagrange/Hamilton theory applied to this same minimization problem tells us that the following quantity (which is <u>not</u> the usual energy density) must be constant over all x:

$$\begin{aligned}\boldsymbol{H}^{[1]} &= \Pi^{[1]}_\varphi \overline{\varphi}_x + \Pi^{[1]}_{\dot{\varphi}} \dot{\overline{\varphi}}_x - \mathcal{H} = \left(\frac{\delta\mathcal{H}}{\delta\overline{\varphi}_x}\right)\overline{\varphi}_x + \left(\frac{\delta\mathcal{H}}{\delta\dot{\overline{\varphi}}_x}\right)\dot{\overline{\varphi}}_x - \mathcal{H} \\ &= (2ds\varphi_x + 2h\varphi_x)\overline{\varphi}_x + 0 - \mathcal{H} \\ &= 2dsz + 2hz - (3s^2 + b\rho s + cs + dsz + g(\rho) + hz) \\ &= dsz + hz - 3s^2 - b\rho s - cs - g(\rho)\end{aligned} \tag{15}$$

The superscript "[1]" is used here to remind us that these are quantities related to a minimization across <u>one</u> dimension (space), which should not be confused with the usual Lagrangian dynamics across <u>two</u>-dimensional space-time.

An alternative, equivalent way to analyze this minimization problem is to consider the energy density (equation 5) as a function of the four real variables r and θ (defined in equation 2) and $\partial_t r$ and $\partial_t \theta$. The Lagrange-Euler equations for θ and r here are:

$$\partial_x \Pi_\theta^{[1]} = \frac{\delta \mathcal{H}}{\delta \theta} \tag{16}$$

and:

$$\partial_x \Pi_r^{[1]} = \frac{\delta \mathcal{H}}{\delta r} \tag{17}$$

which by the usual substitutions reduce to:

$$\partial_x((ds+h)(2r^2 \theta_x)) = 0 \tag{18}$$

and:

$$\partial_x((ds+h)2r_x) = 2bsr + 2rg'(\rho) \tag{19}$$

Equation 18 tells us that the following quantity is "conserved" (i.e., constant over x):

$$I = (ds+h)(2r^2 \theta_x) \tag{20}$$

Equations 16, 18 and 20 are essentially an easy way to derive the conserved charge of Noether's Theorem [8] in this case.

Not surprisingly, the Lagrange-Euler equations for $\partial_t r$ and $\partial_t \theta$ again reduce to equation 14, for a nonstatic soliton.

## 2.3. Simplifications of the System

In the initial study of this system, my goal was to construct a numerical example of localized stable oscillation. Therefore, I was careful not to exclude serious possibilities for combinations of b,c,d,g and h, without careful arguments that these possibilities could not be made to work. This led to a lot of involved logic, whose details [9] are no longer relevant. The methods of Gelfand [6] can be used to rule out SSOS "solitons" in all versions of this system -- including even the possibility of static solitons, or solitons which possess an alternate form of stability called "H' stability" in [4]. Therefore, this section will only provide a brief discussion of the arguments which I originally used to simplify the system. Some of these arguments would have been easier if I had simply assumed a stronger version of the localization assumption; however, some theorems in physics assume localization so strict that even the electron (clothed in Coulomb force) would not qualify. I did not want to miss the possibility of an electron-like soliton!

First, I multiplied equation 14 by s/2, and subtracted it from equation 5, so as to derive a new expression for $\mathcal{H}$ (valid only for the oscillating soliton state). Next, I exploited the localization requirements (i.e. ρ and $\mathcal{H} \to 0$ as $x \to \pm\infty$), the assumption of zero vacuum energy (implying g(0)=0), and the assumption of positive definiteness near the vacuum, for the two cases h=0 and h 0, in order to deduce c=0 and deduce $s \to 0$ and $z \to 0$ as $x \to \pm\infty$. Note that the case h=d=0 could be disregarded, because it reduces the system to ODE, which cannot yield solitons.

Second, without loss of generality, I assumed that ρ reaches a maximum at x=0, such that ρ(0)>0 and $\rho_x(0)=0$.

Third, again considering the limit as $x \to \pm\infty$, I deduced that I=0 in equation 20, and that $\theta_x=0$ at all points x where it is meaningful (i.e. where ρ 0). Because the stability requirements would have to be met both by the present state of the system and its future states, I could also deduce that $\partial_t \theta_x = 0$. This then led to the requirement that:

$$\varphi(x) = f(x)e^{ia_1}$$
$$\dot{\varphi}(x) = \psi(x)e^{ia_2} \qquad , \qquad (21)$$

where $a_1$ and $a_2$ are real constants, and where f and $\psi$ are also real. Substituting this back into the time-differentiated version of equation 12, I found only two possibilities: (1) if $a_2=a_1\pm n\pi$, the soliton would be purely real and static, in effect -- not what I was looking for; (2) if not, the equation can be decomposed into two equations which each must be satisfied, one involving only f and the other only $\psi$. The second case allows us to deduce that:

$$\psi(x) = \nu f(x) \quad , \qquad (22)$$

where $\nu$ is a real constant. From this we may deduce that $\varphi$ must take the simple form:

$$\varphi(x,t) = r(x)e^{i\nu t} \qquad (23)$$

Substituting this back into equation 14 at the point x=0, and recalling that z=0 at x=0, we may deduce that:

$$\nu^2 = -\frac{b}{6} \qquad (24)$$

Substituting this back into equation 5, and integrating over space (x), we arrive at a relatively simply expression for total energy $H^{[2]}$ as a function of $\nu$, which then leads to the following simple but necessary condition for energy minimization:

$$\frac{\partial H^{[2]}}{\partial(\nu^2)} = 6\nu^2 \int_{-\infty}^{\infty} f^4 dx + b \int_{-\infty}^{\infty} f^4 dx + d \int_{-\infty}^{\infty} f^2 f_x^2 dx = 0 \qquad (25)$$

Inserting equation 24 into equation 25, we have a choice between assuming that $f^2 f_x^2=0$ for all x (which does not produce a soliton!) or else d=0.

Finally, by considering dimensional analysis (i.e. possibilities for rescaling $\varphi$ and x and t), we may choose $\nu^2=1$ and h=1 without loss of generality. (Again, a choice of zero for either of these quantities would not allow solitons.) Equation 24 then tells us that b=-6.

In the end, the only choice left to make, in searching for oscillating solitons, is the choice of g. Equation 2 is now reduced to:

$$\mathcal{L}^{[2]} = s^2 - 6\rho s - g(\rho) - z \qquad (26)$$

## 2.4. Persistent Multifurcation and Well-Posedness

At first glance, it is extremely disturbing that the Lagrangian proposed here yields a dynamical law -- equation 9 -- which becomes ill-posed <u>precisely</u> at the proposed state of "stable equilibrium." Furthermore, as discussed in [5], this sort of behavior may be expected in <u>any</u> nonstatic soliton, generated by any nontopological theory. One begins to wonder if nature will only allow three types of particles -- particles based on point singularities (ala Lorenz, Infeld, or traditional QFTs), particles based on tying wormhole-like knots into space, or particles based on this kind of singular multifurcation. (Some of the later work of Einstein and Infeld suggested geometric models with similar paradoxes.) All of this is enough to make one wonder if there might ultimately be some hope for ideas of representing elementary particles as unit records [10] in some kind of intelligent computer system[11,12]; however, that would be a difficult approach to implement in a useful way, and the situation is not so bad, yet.

On careful examination [9], I would argue that systems displaying this kind of persistent multifurcation should be fully admissible as possible models of physics, even in a hard-core Realist approach.

To begin with, it should be recalled that the assumptions of Lagrangian field theory include more information than just the dynamical equation (equation 9). The Lagrangian assumptions also lead to the conservation of energy. In regions away from the soliton state of apparent multifurcation (where equation 14 is satisfied), the dynamics are well-posed. In regions near the soliton, energy conservation by itself is enough to ensure that the system is not permitted to jump to far away states, which would possess higher energy. More precisely, the stability arguments in [4] would tell us that we would have a stable state here, despite the odd-looking dynamical equation. The main effect of the singularity near the stable state is that it theoretically permits the solitons to "shed" or "emit" or "dissipate" deviations from the stable state at an infinite speed, if these deviations are small enough. The dynamics of the stable state itself are totally determined by the requirement of energy conservation, since the minimum energy can only be maintained at that state (modulo a phase shift). (Translations are presumably not allowed, because adding momentum would add energy, but I have not studied that issue in detail.)

Nevertheless, this situation still seems somewhat paradoxical. Until and unless nontopological solitons have actually been constructed numerically [4], one might wonder if such properties might be extended even further to arrive at an argument for impossibility by reduction to absurdity! However, there are several reasons to hope that this would not be possible.

First, it is possible to generate constructive examples of some of these paradoxical phenomena even in simple ODE models. For example, in [9], I considered the Lagrangian of equation 1, in the special case where d=h=0 (which reduces the system to ODE) and where $g(\rho)$ is quadratic in $\rho$. In the special case of that special case where $\varphi$ is real, the conditions for H-minimization can be integrated to yield the restriction:

$$g(\varphi) = \frac{b^2}{12}\varphi^4 + \frac{bc}{6}\varphi^2 + k \quad , \quad (27)$$

where we may assume k=0 without loss of generality, since k has no effect on the dynamics in the ODE case. From the requirement for energy conservation, we may deduce that the dynamics in that case are exactly the same as those of a simple harmonic oscillator with a Lagrangian L of $(1/2)(\partial_t\varphi)^2-(1/12)b\varphi^2$ ! In other words, we end up with a complicated nonlinear representation of a simple, familiar linear system! The "energy" of the nonlinear system happens to equal $H^2+cH$, where "H" is the energy according to the simpler representation. Unlike the simpler representation, however, the complex representation does allow for an H-minimizing orbit. There is no emission mechanism in energy-conserving ODE comparable to the emission mechanism in PDE[4] which makes it possible to achieve strong stability in energy-minimizing states. The dynamics are also well-defined for the complex version of the ODE example [9], which is closer in flavor to the full system.

Second, I have found a way to address the following puzzle: why is the information about energy conservation not present already, implicitly, in the Lagrange-Euler equations? The classical derivation showing that $\partial_t H=0$ for ODE [13] is based solely on a few definitions and on the Lagrange-Euler equations applied at time t. A careful but tedious double-check shows that this same logic applies here, in detail, as one would expect. But this reasoning must use more than just the fact that $\partial_t H(t)=0$ at time t; it exploits energy conservation across all time. The situation here would be very paradoxical if we assumed that all possible combinations of $\varphi$ and $\partial_t\varphi$ were allowed at time t; however, the variational assumption does restrict the possible combinations of $\varphi$ and $\partial_t\varphi$ in this case, when $\varphi$ is at the multifurcation point. In principle, this is like the usual restriction that we are not allowed to consider all possible combinations of $\varphi$, $\partial_t\varphi$ and $\partial_t^2\varphi$; the only unusual aspect is that the Lagrangian assumption imposes different restrictions in different situations -- such that the definition of the "state" of the system may be first-order or second-order, depending on what the state itself is.

## 2.5 Original Proposal For Constructing Numerical Example

Based on the simplifications in section 2.3, I proposed a strategy [9] for constructing possible numerical examples of solitons in this system. The hope was to follow the heuristic search strategy described in this section, and then test the resulting examples for second-order stability. Dr. Robert Kozma, in correspondence, has stated that he was able to construct many numerical examples which fulfilled the first part of this program, but that no examples were able to pass the test of SSOS[14]. At approximately the same time, I was able to explain why the second-order test cannot be passed here[4], based on concepts from Gelfand and Fomin[6]. Nevertheless, the general approach here might also be useful as a kind of warmup exercise, to prepare for the construction of numerical examples in more complex systems (like those of sections 3 and 4) where there is still some hope of passing the SSOS tests.

After the simplifications of section 2.3, the Lagrange--Euler equations for the minimization problem reduce to the simple equation:

$$\frac{d^2 f}{dx^2} = -6f^3 + g'(f^2)f \quad , \tag{28}$$

where "g'" is simply the derivative of g with respect to its argument. Without loss of generality, we can also assume that f(x) reaches a maximum (not necessarily unique!) at x=0. This tells us that $f_x(0)=0$, giving us one of the boundary conditions we need to set up a numerical example. If we subtract s/2 times equation 14 from equation 15, and recall that $H^{[1]}$ must equal zero (because it must have the same value at all points x, including where x→±∞), we may deduce the requirement that:

$$H^{[1]} = 3\rho^2(0) - g(\rho(0)) = 0 \quad , \tag{29}$$

where we again define $\rho=f^2$. After we have selected the function g, this can be used to give us the other boundary condition we need at x=0. <u>Starting from these two boundary conditions, we can use ordinary (Runge-Kutta) integration of equation 36 in order to construct the entire function f(x) numerically.</u>

To prove that an energy minimizing state exists in this class of systems, we would only need to select a suitable function g, perform the numerical integration, and then use my method or Gelfand's to verify the second-order stability conditions. Unfortunately, as discussed [4,section 4.3.3], Gelfand's test shows directly that we cannot pass that test. Still, it is interesting to ask: how could we have found a suitable function g, if there did exist a function g sufficient to do the job?

To begin answering these questions, one might first attempt a quick numerical proof of existence, based on a heuristic effort to find a suitable function g. To simplify this search, we may define:

$$g_0(\rho) = g(\rho) - 3\rho^2 \quad , \tag{30}$$

and then look for a suitable function $g_0$. From the various stability conditions[9], we may deduce that $g(0)=g_0(0)=0$ and $\partial_x(g_0(0))>0$. Equation 29 also tells us that $g_0(\rho(0))=0$. Thus to choose $g_0$ we must choose a curve which rises up to the right from the origin, and then comes back down again to zero; the value of ρ where it crosses the x axis again defines the initial value ρ(0), to be used in Runge-Kutta integration. The key challenge would be to find such a function which also lets us meet the second-order conditions. (I can imagine performing such calculations on a spreadsheet, and experimenting with the parameters used to define g -- perhaps even using backpropagated derivatives [15] of the results to guide the adjustment of parameters.) Roughly speaking, the key trick may be to keep the <u>average</u> value of $\partial_x^2 g_0$ positive, as a weighted average, despite the negative slope of $g_0$ as it descends back to the x axis. Similar search procedures may be useful in other, more promising examples.

Given the wide qualitative differences in behavior to be expected between the inner, nonlinear zone of such a soliton, and the outer asymptotic, more linear zone, it would probably be desirable to use an adaptive step-size algorithm here [16]. This would also have the advantage of providing some indication of error bounds.

## 2.6. Second Order Stability Analysis for 1-D or Radial Systems

### 2.6.1. Introduction

This section will discuss two basic, generalized numerical techniques which can be used to decide whether a proposed energy-minimizing state actually meets the second-order conditions for a local minimum. Both techniques are based on considering possible perturbations, $\underline{\xi}$, to the proposed state, $\underline{\psi}$. One technique, based on Gelfand and Fomin[6], basically tries to find out whether there exist perturbations $\underline{\xi}$ other than the null perturbation ($\underline{\xi}(x)=0$ for all x) which meet the Lagrange-Euler conditions for a minimum. The other technique, based on my own explorations [9], basically uses dynamic programming to find the perturbations which minimize energy. Both techniques are exact and sufficient and computationally efficient, in theory; however, both require special analysis to handle the special features of soliton physics (e.g. translational invariance), and the two methods are best used together, because of numerical approximation issues which have never been quantified or bounded. Both techniques require further mathematical analysis. Because the work of Gelfand and Fomin is extremely well-known, there probably exist other extensions in the literature which I have yet to search for; in addition, while writing [4], I located a reference to [17], which is probably also well-worth studying further.

Dr. Robert Kozma has reported [14] that he sucessfully applied my method to the analysis of the system in equation 2. He developed an implementation in MatLab, and applied it to many concrete examples. The method did successfully pinpoint instability in all cases; the theoretical reason for this instability is discussed in section 2.6.2.

After this introduction, sections 2.6.2 and 2.6.3 will describe how the two methods are used when evaluating a proposed minimum of the integral of energy over a finite interval [a,b] in one-dimensional space. Section 2.6.2 will also include a few important words about translation, and the reasons [4] why Gelfand's criteria cannot be satisfied for the system in equation 1. Section 2.6.4 will then discuss the application to soliton stability analysis.

In my earlier work here, I fully hoped to avoid the perturbation approach. I hoped to find energy-minimizing states by approximating energy itself as a function of φ on a fnite-element grid. I hoped to approximate the energy-minimizing function φ(x) directly, by minimizing energy exactly on the grid, and verifying that the second-order conditions are met. Unfortunately, I found that the minimization process would simply "find" the high-frequency modes (near the natural frequency of the grid) where the finite-element approximation breaks down. However, in this work, I approximated derivatives in the usual naive way, by considering differences between φ(j+1) and φ(j); perhaps more sophisticated finite-element approximations [16] might have made it possible to take this approach.

Following the notation of [4], we may represent the energy function to be minimized as:

$$\boldsymbol{H} = \int_{-\infty}^{\infty} \mathcal{H}(\underline{\psi}, \underline{\psi}_x) d\boldsymbol{x} \quad , \tag{31}$$

where "$\underline{\psi}$" represents the state of the system. For a complete description of systems like equation 1, $\underline{\psi}$ would be a two-component vector consisting of φ and $\partial_t$φ. However, when we can prove analytically that the perturbations to $\partial_t$φ must be zero, in order to minimize energy, then it may be good enough to consider the problem of minimization with respect to φ alone (i.e. to choose $\underline{\psi}$ to simply be φ).

Both in [4] and in [1], $\mathcal{H}(\underline{\psi}+\underline{\xi}, \underline{\psi}_x+\underline{\xi}_x)$ is expanded in a Taylor series. The second order term in this expansion, $\tfrac{1}{2}\delta^2\mathcal{H}$, is essentially a quadratic form or "Hessian matrix" in the perturbations ($\underline{\xi}, \underline{\xi}_x$). Both Gelfand and I focus on the problem of how to minimize $\delta^2H$, in effect, as a function of the perturbations, when the base state $\underline{\psi}$ is held fixed. In practical calculations, it is usually simpler to use the term "$\underline{\xi}$" to refer to perturbations of $\underline{\phi}$ rather than $\underline{\psi}$.

Note that Gelfand defines "$\delta^2\mathcal{H}$", in effect, as <u>one half</u> of the Hessian (i.e. as the second term in the Taylor series approximation to H), while I define it as the Hessian itself. The latter definition permits a consistent treatment of "δ" itself as an operator, and is more convenient in calculations. For example, Makhankov et al cite Gelfand's definition in their theoretical introduction [1, p.50], but use the latter definition in their detailed calculations (e.g. [1, Appendix D].) This factor of ½ has no effect, of course, on the kery issue of positivity.

In general, for most families of systems, one can derive considerable insight by simply studying the algebraic expression for $\delta^2\mathcal{H}$, and applying analytical methods to it [1,4]. For the system in equation 1, I used equation 14 to simplify the expression for $\delta^2\mathcal{H}$, and then obtained [9]:

$$\delta^2\mathcal{H} = (ds+h)|\xi_x|^2 + (bs+g'(\rho))\xi^2 + 2g''(\rho)(\xi_1\varphi_1 + \xi_2\varphi_2)^2$$
$$+ 4b(\xi_1\varphi_1 + \xi_2\varphi_2)(\dot{\xi}_1\dot{\varphi}_1 + \dot{\xi}_2\dot{\varphi}_2) + 4d(\xi_{1,x}\varphi_{1,x} + \xi_{2,x}\varphi_{2,x})(\dot{\xi}_1\dot{\varphi}_1 + \dot{\xi}_2\dot{\varphi}_2) \quad (32)$$
$$+ 12(\dot{\xi}_1\dot{\varphi}_1 + \dot{\xi}_2\dot{\varphi}_2)^2$$

(Note that it is valid to apply equation 14 to the proposed soliton state, $\varphi$, even though the perturbed state itself is not bound by equation 14. Also note that I have not rechecked this old calculation -- though the results here are not sensitive to the exact details.)
From this equation, it is easy to deduce that we require ds+h≥0, but beyond that, the implications are far from obvious. In general, when a system is a promising place to look for solitons, one expects to find a mix of positive terms (giving some hope for the positivity of $\delta^2$H!) and of negative terms ("holding the particle together," and permitting $\delta^2$H=0 as required by relativity for translational modes $\underline{\xi}=\underline{\varphi}_x$). Decisive numerical techniques can then become very useful, to find out whether the positive outweighs the negative in particular numerical examples.

### 2.6.2. Method of Gelfand and Fomin

I will now try to define two such techniques, to address the general problem as represented in equation 31, where $\underline{\psi}$ is considered to be a vector made up of n real components and where $\underline{\xi}$ represents a perturbation of $\underline{\psi}$. This does not imply a loss of generality, since the components of a complex number can be treated as separate compoennts of $\underline{\psi}$.

Let us begin by representing the second variation as:

$$\tfrac{1}{2}\delta^2\mathcal{H} = \tfrac{1}{2}\underline{\xi}^T\mathcal{H}_{11}\underline{\xi} + \underline{\xi}^T\mathcal{H}_{12}\underline{\xi}_x + \tfrac{1}{2}\underline{\xi}_x^T\mathcal{H}_{22}\underline{\xi}_x \quad (33)$$

Gelfand and Fomin prove that their tests are valid only on the assumption that $\mathcal{H}_{22}$ is positive definite at all points x, and that $\mathcal{H}_{12}$ is a symmetric matrix [6,ch.5]. They also limit their discussion to the problem of minimizing the integral of $\mathcal{H}$ over a finite interval [a,b]. They define the n-by-n matrices [6, section 29.1]:

$$P(x) = \tfrac{1}{2}\mathcal{H}_{22}$$
$$Q(x) = \tfrac{1}{2}(\mathcal{H}_{11} - \partial_x\mathcal{H}_{12}) \quad (34)$$

which reduces the minimization problem to the problem of minimizing:

$$J_0 = \int_a^b (\underline{\xi}_x^T P \underline{\xi}_x + \underline{\xi}^T Q \underline{\xi}) dx \quad (35)$$

which leads to the Lagrange-Euler equation:

$$P(x)\underline{\xi}_{xx} = Q(x)\underline{\xi} \quad (36)$$

In essence, they show that $J_0$ will be positive definite if and only if equation 36 does not possess any solution $\underline{\xi}(x)$ across x with the following three properties:

$$\underline{\xi}(a) = 0$$

$$\underline{\xi}(a') = 0, \quad \text{for some } a < a' \leq b \tag{37}$$

$$\underline{\xi}(x) \neq 0 \quad \text{for some } a < x < b$$

To test for the existence of such solutions, they propose that we first solve equation 36 n different times, with different boundary conditions, over the interval [a,b]. More precisely, for each j from 1 to n, solve equation 36 with the boundary conditions:

$$\underline{\xi}(a) = 0$$

$$\underline{\xi}_x(a) = e_j \tag{38}$$

where $e_j$ is just the usual basis vector of elementary vector algebra. If the n different versions of $\underline{\xi}(a')$ are linearly dependent on each other, for any point a' in the interval [a,b], then we know that $J_0$ is not positive definite. (We know this, because it is easy to construct a solution which fits equation 37, by using whatever linear dependence exists between the solutions at the point a'.) We can test for this linear dependence by evaluating the determinant of the matrix formed by concatenating the n different solutions, at every point x in the interval [a,b].

This procedure can be translated into a numerical algorithm simply by approximating these n solution trajectories on a finite grid, and checking the determinant at each grid point to see if it equals zero. Gelfand and Fomin[6] says nothing about this numerical possibility, however. They leave us with questions about how close the approach to zero must be, how tight the grid must be, and so on. Presumably one would want to program special subroutines to double-check places where the determinant seems to approach zero, perhaps by using a finer grid in those places or performing other tests.

Gelfand's discussion of "Jacobi's necessary condition" (plus the results of section 2.6.3) suggest that the same procedure might work even when $\mathcal{H}_{12}$ is not symmetric, if we used the Lagrange-Euler equation for the original expression in equation 33.

Gelfand and Fomin also pay careful attention to the case where $J_0$ might be positive semidefinite (as we require here!). As $a \to -\infty$ and $b \to +\infty$, for the 1+1-D soliton application, their analysis should yield a single perturbation mode $\underline{\xi}$ which comes closer and closer to satisfying $\underline{\xi}(b)=0$, and which corresponds to the translational mode $\underline{\xi}=\underline{\varphi}_x$. This should be easy to account for in any implementation of this method. In cases where the translation mode itself leads to $\underline{\xi}=0$ at a finite x (as in the example of this section, for x=0!), then we know that the system is unstable.

If the interval from a to b is divided up into T intervals, in the finite-element approximation, then the calculations required by this method would cost on the order of $kn^3$ operations per grid point, or $kn^3T$ in all.

## 2.6.3. An Alternative Approach

This section will describe the vector generalization of a new method which I originally proposed as a way of evaluating the stability of the system in equation 2 [9].

In essence, the idea is to use dynamic programming to find the path for $\underline{\xi}$ which minimizes the integral of $\delta^2 \mathcal{H}$ over some interval [a,b]. In its general form, developed by Bellman, dynamic programming is a general method for minimization over time (or over any other dimension) when there is random disturbance added at every point in time. Dynamic programming is the foundation of modern approaches to building brain-like intelligent systems[11]. The noise-free continuous-time special case of the Bellman equation is essentially identical to older formulations by Hamilton and Jacobi [6, chapters 4 and 6] and Pontryagin. In the situation here, it is more convenient to use the noise-free discrete-time case of dynamic programming.

In the method of Gelfand and Fomin, it is natural to think of the algorithm as moving "forward" from the start of the interval, "a," up to the "end" of the interval, "b." (Actually, one can run either method in either direction.) In my method, it is most natural to think of us as moving <u>backwards</u> from "b" back to "a." If we map the interval [a,b] into a series of grid points, labelled by the integer t = 0,1,...,T, we may proceed as follows. First define the grid interval or cell size as:

$$\Delta = \frac{b-a}{T} \tag{39}$$

Then approximate equation 33 (dx) by:

$$U(t,\underline{\xi}(t),\underline{\xi}(t+1)) = \underline{\xi}^T(t)A(t)\underline{\xi}(t) + \underline{\xi}^T(t)B(t)(\underline{\xi}(t+1) - \underline{\xi}(t))$$
$$+(\underline{\xi}^T(t+1) - \underline{\xi}^T(t))C(t)(\underline{\xi}(t+1) - \underline{\xi}(t)) \tag{40}$$

where:

$$A(t) = \tfrac{\Delta}{2} H_{11}(x(t))$$
$$B(t) = H_{12}(x(t))$$
$$C(t) = \tfrac{1}{2\Delta} H_{22}(x(t)) \tag{41}$$
$$x(t) = a + (b-a)\tfrac{t}{T}$$

For each t, from 0 to T-1, define:

$$J(t,\underline{\xi}) = \underset{\underline{\xi}(t+1),...,\underline{\xi}(T-1)}{\text{minimum}} \sum_{j=t}^{T-1} U(j,\underline{\xi}(j),\underline{\xi}(j+1)) \quad, \tag{42}$$

subject to the boundary conditions $\underline{\xi}(t)=\underline{\xi}$ and $\underline{\xi}(T)=0$. This kind of "J" function -- sometimes called the "cost to go" function, and sometimes called the "strategic utility function" -- is the key function which we use the Bellman equation to solve, in dynamic programming. In this example, the Bellman equation takes the form:

$$J(t-1,\underline{\xi}(t-1)) = \underset{\underline{\xi}(t)}{\min} \left( U(t-1,\underline{\xi}(t),\underline{\xi}(t-1)) + J(t,\underline{\xi}(t)) \right) \tag{43}$$

We can solve this equation by using the ansatz:

$$J(t,\underline{\xi}) = \underline{\xi}^T Q(t) \underline{\xi} \tag{44}$$

Substituting this and equation 40 into equation 43, we can solve the minimization problem by use of elementary calculus, which yields:

$$(C(t-1) + Q(t))\underline{\xi}(t) = (C(t-1) - \tfrac{1}{2}B^T(t-1))\underline{\xi}(t-1), \tag{45}$$

which (again in equation 43) leads to the recurrence relation:

$$Q(t-1) = A(t-1) - B(t-1) + C(t-1)$$
$$- (C(t-1) - \tfrac{1}{2}B(t-1)(C(t-1)+Q(t))^{-1}(C(t-1) - \tfrac{1}{2}B^T(t-1)) \tag{46}$$

When we are minimizing with respect to the boundary condition ξ(b)=0, we use this recurrence relation in a <u>backwards</u> sweep, from t=T-1 to t=0, starting from:

$$\mathbf{Q(T) = 0} \tag{47}$$

The recurrence relation in equation 46 may seem somewhat strange at first; however, it does cross-check with the corresponding equations (58-60 of [9]) for the system of equation 2, which do demonstrate the kind of behavior one might expect intuitively[9]. (In [9], however, I described the possibility of a forwards sweep -- the exact same calculations, but reversing the roles of "a" and "b".) The minimization which leads to equation 45 is valid if and only if the matrix C+Q is positive definite; however, the matrix $\mathcal{H}_{22}$ is normally positive and, when Δ is very small, the matrix C will be positive definite and large. Thus C will dominate Q, and ensure that C+Q is positive definite, <u>unless</u> there is a condition which I have called a "divergent point [9]."

In the continuous-time version of the Bellman equation, with constant coefficients, it is easy to see that the true value of the matrix Q can diverge at some points to contain an infinite negative eigenvalue. This may be called a "divergent point." Intuitively, if a' is a divergent point, this implies that there exists a function ξ(x) which satisfies ξ(a')=0 and ξ(b)=0, but has a <u>negative</u> integral of $\delta^2 \mathcal{H}$ in the interval [a',b]. Thus there is no finite limit to how negative $\delta^2 \mathcal{H}$ can be, with finite boundary conditions; we can always multiply the negative solution by any scalar as large as we like, and still meet the boundary conditions (with small adjustments). Of course, this also implies that a' corresponds exactly to what Gelfand and Fomin call a "conjugate point" -- a point a' where a zero solution first exists. (Strictly speaking, one might think of the difference between the divergence point and the conjugate point as infinitesimal rather than zero, but that is a matter for future formal mathematical elaborations.)

Since divergences of Q to negative infinity are supposed to occure at exactly the same points where Gelfand's determinant is expected to equal zero, it makes sense to use both tests together, and see if indeed both phenomena do appear to occur together in numerical work. As with Gelfand's test, instability is equivalent to the proposition that one encounters such a point by the time one's sweep has reached the opposite end ("a") of the interval [a,b]. Also, as with Gelfand's test, one can do some extra checking at points which appear to fail the test; for example, by making the grid finer in that region (by using smaller Δ there), one can verify the divergence more stringently.

Note that the computational cost of this method, like that of Gelfand and Fomin, is on the order of $kn^3T$.

## 2.6.4. Application to 1-D (Line-Based or Radial) Solitons

The previous two sections mainly addressed the problem of determining positivity for the integral of energy over a finite interval [a,b]. But with solitons, these kinds of one-dimensional calculations are useful mainly in two contexts: (1) evaluating positivity for total energy over the interval (-∞,+∞), for 1+1-D Lagrangian systems; (2) evaluating positivity for total energy over the interval [0,∞), for 3-D (or 3+1-D) systems reduced to one dimension by analytical means (as in [4]).

For the line, one way to use these methods is simply to let a→-∞ and b→+∞, as discussed in section 2.6.2 (which can be applied to my method as well). Naturally, this will work best if we actually choose a and b to be very large; for example, if the original soliton construction was based on a variable step-size method, the largest value of "x" represented on the grid should be fairly large. Certainly one would want a and b to be far into the region of "positive potential" (or positive Gelfand-Q).

In practice, there is usually a better way to proceed, with my method. First, pick the origin (x=0) to be a natural "center" of symmetry of the soliton (as in the construction in section 2.5). The maximum grid point x should be large enough that a linear approximation to the Lagrange-Euler equations (and positive quadratic energy) should be a decent approximation for x'>x. Thus one can actually try TWO calculations: (1) one with the boundary condition Q(T)=0 (as in equation 47); (2) another with Q(T) estimated based on an analytical calculation of $Q_+(x)$ -- the value of Q(x) which would represent the sum of U from b to +∞, based on the linear/quadratic approximation. The second would be the "real" estimate, and the former would only be a kind of cross-check. One would then work back only from x to 0, not all the way to -x! After that, one would perform the same calculation in the opposite direction, working up from -x to 0. One would then SUM the two resulting Q matrices, and check the sum for positive definiteness. In this way, one can avoid the need to deal with divergent situations in the test. (However, if a divergence is encountered, and verified by the Gelfand-Fomin method, the conclusion would still be straightforward.)

In order to reduce calculations, one usually does not have to repeat the numerical analysis for negative x as for positive x. In some cases, the two are absolutely symmetric, i.e. $\varphi(-x)=\varphi(x)$. In those cases, the Gelfand method immediately rules out stability (as discussed in 2.6.2). But for promising cases, like the one in section 3, we may have $\varphi(-x)=M\varphi(x)$, where M is a simple, known matrix; in that case, we can use this relation to "recalculate" Q based on negative x, simply by performing a couple of matrix operations on the Q which emerges from the positive calculation.

For the case of radial systems, the Gelfand-Fomin equations and my recurrence relation both remain valid. Neither method assumes an "autonomous" Lagrangian or utility function. Nevertheless, I have not thought carefully about the details of the best way to handle the boundary conditions in that case. In some cases, in order to solve the Bellman equation, it may be necessary to replace equation 44 by an ansatz like:

$$J(t,\underline{\xi}) = (\underline{\xi} - \underline{\xi}'(t))^T Q(t)(\underline{\xi} - \underline{\xi}'(t)) + c(t) \quad , \tag{48}$$

where recurrence relations are needed for the <u>combination</u> of Q(t), $\underline{\xi}$'(t) and c(t).

## 2.7. H' Stability and Uniqueness

Energy minimization (the minimization of H) is not the only mechanism which can lead to strong stability, discussed in [4]. Another possible mechanism is the minimization of a different conserved quantity, H'(λ), for some real number λ, which may be defined as:

$$\boldsymbol{H'}(\lambda) = \int \mathcal{H}'(\lambda,\underline{x})d^D\underline{x} = \int (\mathcal{H}(\underline{x}) + \lambda q(\underline{x}))d^D\underline{x} \quad , \tag{49}$$

where "q" is any local conserved quantity (such as charge) at point x. The stability results given in [4] will still apply to any proposed localized soliton state $\underline{\varphi}(\underline{x})$ which meets the three basic requirements: (1) $\underline{\varphi}(\underline{x})$ solves the Lagrange-Euler equations for minimizing H' over D-dimensional space; (2) H' is positive definite near the vacuum state; (3) H' is positive semidefinite in a strong way ("SSOS") near $\underline{\varphi}$. For systems which contain more than one "charge," the generalization of equation 49 is obvious.

According to Noether's Theorem, if a Lagrangian $\mathcal{L}$ is symmetric with respect to a group of transformations G(α), where α is a real number, then the corresponding conserved charge is:

$$q = \sum_j \frac{\delta \mathcal{L}}{\delta \dot{\varphi}_j} \cdot \frac{\partial \varphi_j}{\partial \alpha} \tag{50}$$

In the example of equation 2, $\mathcal{L}$ is invariant with respect to a simple global gauge change:

$$\varphi' = G(\alpha)\varphi = e^{i\alpha}\varphi \tag{51}$$

If we represent φ for the moment as a two-component real vector, where the first component is r and the second component is θ, then equation 50 leads very quickly to the result that:

$$q = \frac{\delta \mathcal{L}}{\delta \dot{\varphi}} = (2s + b\rho + c + dz)(2r^2 \dot{\theta}) \qquad (52)$$

This results in:

$$\mathcal{H}' = \mathcal{H} + \lambda q = 3s^2 + b\rho s + cs + dsz + g(\rho) + hz \\ + \lambda(2s + b\rho + c + dz)(2r^2 \dot{\theta}) \qquad (53)$$

When we recalculate the Lagrange-Euler equations to minimize H', with respect to $\partial_t r$ and $\partial_t \theta$, we get:

$$0 = (6s + b\rho + c + dz + 4\lambda r^2 \dot{\theta})(2\dot{r}) \qquad (54)$$

and:

$$0 = (6s + b\rho + c + dz + 4\lambda r^2 \dot{\theta})(2r^2 \dot{\theta}) + 2r^2 \lambda (2s + b\rho + c + dz) \qquad (55)$$

If λ=0, this reduces to the previous analysis, as expected. However, if λ 0 and if the coefficient of ($\partial_t r$) in equation 54 is zero, then the right-hand side of equation 55 is both positive definite and required to equal zero! Therefore, we cannot find a nonzero soliton by that route. Therefore, for a stable soliton with λ 0, equation 54 tells us that we still require $\partial_t r$=0. The derivation of equation 18 is also again valid, and we appear to be driven again to the solution form of equation 23. This once again reduces our minimization problem to a scalar, real minimization problem, where the Gelfand methods (section 2.6.2 and [4]) rule out a stable minimum.

Strictly speaking, it should be noted that the three conditions mentioned above are still not quite enough to ensure stability. If there exists another value of λ, λ', nearby to λ, which also satisfies all three conditions, but for a different set of possible states, then there is a uniqueness problem, related to metastability. This is discussed further in [4]. This problem is not relevant to the example of equation 2, since the three conditions are not met in any case.

## 3. Relativistic Systems, Promising and Otherwise

This section will discuss the most important class of systems considered in this paper or in [4] in the search for nontopological solitons: relativistic Lagrangian systems such that $\mathcal{L}$ is a function of a vector field φ and of the derivatives $\partial_t \varphi$. However, this is not the most important section for most people to read. One reason for this paradox is that section 5 of [4] already gives a strategy for constructing numerical examples of solitons in such systems. It places primary emphasis on a system which I call "static real $\mathcal{L}$ ($f_0, f_2, f_4$)," but it discusses other possibilities as well.

This section will mainly contain auxiliary information, starting with relatively important material and decreasing from there. Section 3.1 will make general observations about the problem of constructing nontopological solitons in relativistic theories; even though these are heuristic observations, they may be of some use, in case there should be difficulties in the $\mathcal{L}$ ($f_0, f_2, f_4$) case, even after exploiting the kinds of techniques discussed in section 2 and in [4]. Section 3.2 will give the details of the derivation of $\delta_d^2 \mathcal{H}$ for $\mathcal{L}$ ($f_0, f_2, f_4$); these details are straightforward in principle, but considerable art is needed to avoid being swamped by the complexity of the algebra. Sections 3.4 and 3.3 will describe the basis of negative results in two systems which turn out to be less promising than initially hoped -- $\mathcal{L}$ ($f_0, f_2$) with H' stability and oscillation, and static real $\mathcal{L}$ ($f_0, f_1, f_2, f_3$). In the end, oscillatory solitons may be more relevant to particle physics than are static solitons; however, the static case may be an easier place to start, as described in [4].

## 3.1. How Promising Are They, Really?

Conventional wisdom holds that stable nontopological solitons cannot exist in relativistic systems. This belief has two main sources: (1) theoretical results; and (2) practical experience.

In [4], I review these theoretical results in some detail. My final conclusion is that these results are not decisive at the present time. It is possible to construct families of systems where the known mechanisms of instability do not apply. On the other hand, no one has actually constructed a numerical example of such a stable nontopological soliton, either, in a relativistic system (for the strong concepts of stability discussed in [4]). Therefore the issue is undecided. Both in this paper and in [4], I emphasize the possibilities for constructing numerical examples. In the end, it is quite conceivable that the real product of such research would be a better understanding of the mechanisms of instability, followed by rigorous grand impossibility theorems. Nevertheless, the new systems suggested for exploration in [4] would overcome the known theoretical obstacles to constructing a numerical example.

The negative practical experience can be explained in part by the natural human tendency to explore similar families of systems over and over again. The basic difficulties here seem to involve two kinds of problems -- problems associated with three spatial dimensions (as in the classic Hobart-Derrick Theorem [1,4]), and problems associated with relativistic field theory as such (which can exist even in 1+1-D). Section 4 will address the issue of three spatial dimensions, while this section will address relativity as such.

For relativistic theories based on a single scalar field $\varphi$, there is not much flexibility. A relativistic field theory generally requires a Lagrangian of the form $\mathcal{L}(f_0,f_2)$, where $f_0$ is $|\varphi|^2$ and $f_2$ is $|\partial_\mu\varphi|^2$. It has not been proven that solitons are impossible under such conditions, but simple spherically symmetric solitons <u>are</u> impossible, and there are special difficulties even in the general case [4]. Therefore, solitons based on covariant vector fields seem more promising at present. Still, in the 1+1-D case, there might be some value in a startup exercise, proving that a strict concept of stability can be met in <u>some</u> Lagrangian system, based on two real scalars A and $\varphi$ which are not constrained by relativity.

A covariant vector field may be represented as $\varphi_\mu$, where $=0,1,...D$. For convenience, I will often refer to the time-like component of the field, $\varphi_0$, as "A." I will refer to the space-like component of the field (made up of $\varphi_1$ through $\varphi_D$) as the D-dimensional vector $\underline{\varphi}$.

To conform with special relativity, the Lagrangian $\mathcal{L}$ must be a real relativistically invariant scalar. In order to construct an invariant scalar out of a covariant vector $\varphi_\mu$ and its derivatives, our only hope is to <u>combine</u> vectors and tensors by a form of tensor multiplication, following rules which are extremely well-known in physics[18]. After we do this, any function of such invariant scalars will also be an invariant scalar. Thus for any relativistic theory, we may write:

$$\mathcal{L} = \mathcal{L}(f_0,...,f_k,...) \quad , \tag{56}$$

where each scalar $f_k$ represents an elementary, irreducible invariant scalar, generated by the usual rules. The systems which I have studied were based on the following five invariant scalars:

$$\begin{aligned}
f_0 &= \varphi^\mu\varphi_\mu = A^2 - |\underline{\varphi}|^2 \\
f_1 &= \partial^\mu\varphi_\mu = \dot{A} - \mathbf{div}\,\underline{\varphi} \\
f_2 &= (\partial^\mu\varphi^\nu)(\partial_\mu\varphi_\nu) = \dot{A}^2 - |\nabla A|^2 - |\underline{\dot{\varphi}}|^2 + |\nabla\underline{\varphi}|^2 \\
f_3 &= (\partial^\mu\varphi^\nu)(\partial_\nu\varphi_\mu) = \dot{A}^2 + |\nabla\underline{\varphi}|^2 - 2(\underline{\dot{\varphi}}\cdot\nabla A) \\
f_4 &= (\partial_\mu f_0)(\partial^\mu f_0) = (A\dot{A} - (\underline{\varphi}\cdot\underline{\dot{\varphi}}))^2 - |A\nabla A - \underline{\varphi}^T\nabla\underline{\varphi}|^2 = w_0^2 - |\underline{w}|^2 \quad ,
\end{aligned} \tag{57}$$

where for convenience I define:

$$w_\mu = A \partial_\mu A - (\underline{\varphi} \cdot \partial_\mu \underline{\varphi}) \tag{58}$$

In each of these equations, my first definition uses the standard Einstein summation convention, which requires the use of the simple Minkowski metric tensors $g^{\mu\nu}$ and $g_{\mu\nu}$ to raise and lower indices. Appendix D of MRS[1] uses similar-looking notation, but does not use the Minkowski tensors, because the analysis is done in three-dimensional space, not space-time. In a very strict sense, one might complain that I am actually representing a contravariant field vector in covariant notation, but the class of theories being examined is equivalent in any case. In section 3.2, I will also use the Einstein convention, but only for Greek letter indices representing 0,...,D.

In the past, when physicists tried to construct examples of relativistic vector field theories, it was natural for them to use the invariants $f_0$, $f_2$ and $f_3$, because of their historical importance. For example, the nonlinear Klein-Gordon equations can be expressed in terms of $f_0$ and $f_2$. But section 3.4 will explain why no Lagrangian made up from these fundamental invariants can yield stable solitons! There are certain symmetry properties present which make it impossible to meet the SSOS criterion. However, when I tried to find a more general impossibility theorem here, I found that Lorenz invariance as such has very little power to restrict the possibilities for $\delta^2 H$. After I understood the symmetry problem with $\delta^2 H$ based on $\mathcal{L}(f_0, f_1, f_2, f_3)$, I then chose to introduce $f_4$, in order to break the troublesome symmetries. In fact, these obvious problems did go away in the $\mathcal{L}(f_0, f_2, f_4)$ theories. But special relativity, more generally, permits a much larger set of invariants to draw on than $f_0...f_4$!

As a general matter, the number of invariant scalars which could be constructed by the usual combination rules appears to be infinite, even if we limit ourselves to irreducible terms using $\varphi_\mu$ and its first derivatives, and postpone consideration of the additional invariants which could be constructed by using the antisymmetric Levi-Civita tensor. A complete list could be constructed by considering $f_0$, $f_1$, and all terms of the form:

$$(\partial_{\alpha 1} \varphi_{\beta 1})(\partial_{\alpha 2} \varphi_{\beta 2})...(\partial_{\alpha m} \varphi_{\beta m}) g^{\cdot\cdot} g^{\cdot\cdot} ... g^{\cdot\cdot} \quad , \tag{59}$$

where m>1, where the indices of each matrix "g" (here shown as dots) are chosen to connect one of the indices $\alpha k$ or $\beta k$ to one of the indices $\alpha(k+1)$ or $\beta(k+1)$. In addition, we could multiply by g one more time in order to contract the two remaining free indices (one from the first term and one from the last), or we could contract them using $\varphi \cdot \varphi$. Clearly there is a huge variety of potential theories here. (It would be interesting to try to characterize them in terms of the matrix $\partial_\mu \varphi_\nu$.)

If, in the end, we cannot construct stable solitons in the 3+1-D $\mathcal{L}(f_0, f_2, f_4)$ system, I proposed in [4] that we might consider possible stable states which do not fit the "radial vector ansatz (RVA)" solution. But before doing that, we might try to add additional invariant scalars which allow us to preserve the RVA, but essentially "craft" $\delta^2 H$ so that we can independently adjust the coefficients of the three basic eigenfunction components to be discussed in section 4.

## 3.2. Explanation of $\delta_d^2 \mathcal{H}$ and Application to $\mathcal{L}(f_0, f_2, f_4)$

The discussion of the $\mathcal{L}(f_0, f_2, f_4)$ system in [4] depends on a derivation of $\delta_d^2 \mathcal{H}$ for that system. This section will first explain what $\delta_d^2 \mathcal{H}$ is, in the general case, and then derive what it is for this system.

### 3.2.1. Definitions and Calculations for General $\mathcal{L}(\varphi_\mu, \partial_\nu \varphi_\mu)$

For any Lagrangian $\mathcal{L}$ which is a function of $\varphi_\mu$ and $\partial_\nu \varphi_\mu$, the standard expression for the Hamiltonian density $\mathcal{H}$ can be written as:

$$\mathcal{H} = \frac{\delta \mathcal{L}}{\delta \dot{\varphi}_\alpha} \dot{\varphi}_\alpha - \mathcal{L} \tag{60}$$

In general, for any functional $f(\underline{\varphi}, \partial_t \underline{\varphi}, \nabla \underline{\varphi})$, we may define the perturbation operator $\delta[\underline{\xi}]$ by:

$$\delta f(\underline{\varphi}, \underline{\dot{\varphi}}, \nabla \underline{\varphi}) = \frac{\delta f}{\delta \varphi_\alpha} \xi_\alpha + \frac{\delta f}{\delta \dot{\varphi}_\alpha} \dot{\xi}_\alpha + \sum_i \frac{\delta f}{\delta(\partial_i \varphi_\alpha)}(\partial_i \xi_\alpha) \tag{61}$$

Before we apply this operator to equation (60), note that:

$$\frac{\delta \mathcal{H}}{\delta \dot{\varphi}_\beta} = \frac{\delta^2 \mathcal{L}}{\delta \dot{\varphi}_\beta \delta \dot{\varphi}_\alpha} \dot{\varphi}_\alpha \tag{62}$$

When we calculate the Hessian of $\mathcal{H}$, the derivatives of equation 62 with respect to $\varphi_\alpha$ or $\partial_i \varphi_\alpha$ will always equal zero, in the static case (where $\partial_t \underline{\varphi}=0$), because of the term $(\partial_t \varphi_\alpha)$.
Thus when we perform a straightforward second differentiation of equation 60
(i.e. when we calculate $\delta(\delta \mathcal{H})$), we arrive at:

$$\begin{aligned}\delta^2 \mathcal{H} &= \frac{\delta^2 \mathcal{L}}{\delta \dot{\varphi}_\alpha \delta \dot{\varphi}_\beta} \dot{\xi}_\alpha \dot{\xi}_\beta - \sum_{i,j} \frac{\delta^2 \mathcal{L}}{\delta(\partial_i \varphi_\alpha)\delta(\partial_j \varphi_\beta)}(\partial_i \xi_\alpha)(\partial_j \xi_\beta) \\ &\quad - \sum_i \frac{2\delta^2 \mathcal{L}}{\delta(\partial_i \varphi_\alpha)\delta \varphi_\beta}(\partial_i \xi_\alpha)\xi_\beta - \frac{\delta^2 \mathcal{L}}{\delta \varphi_\alpha \delta \varphi_\beta}\xi_\alpha \xi_\beta\end{aligned} \tag{63}$$

The Generalized Legendre Condition (GLC[4]) tells us that the portion of this expression which depends only on the <u>derivatives</u> of $\xi_\mu$ must be nonnegative <u>at all points x</u>, in order for stability to be possible. From inspection of equation 63, we can see that this portion is:

$$\delta_d^2 \mathcal{H} = \frac{\delta^2 \mathcal{L}}{\delta \dot{\varphi}_\alpha \delta \dot{\varphi}_\beta} \dot{\xi}_\alpha \dot{\xi}_\beta - \sum_{i,j} \frac{\delta^2 \mathcal{L}}{\delta(\partial_i \varphi_\alpha)\delta(\partial_j \varphi_\beta)}(\partial_i \xi_\alpha)(\partial_j \xi_\beta) \tag{64}$$

Equation 64 can also be obtained by defining and applying the partial perturbation operator $\delta_d[\underline{\xi}]$:

$$\delta_d f(\underline{\varphi}, \underline{\dot{\varphi}}, \nabla \underline{\varphi}) = \frac{\delta f}{\delta \dot{\varphi}_\alpha} \dot{\xi}_\alpha + \sum_i \frac{\delta f}{\delta(\partial_i \varphi_\alpha)}(\partial_i \xi_\alpha) \tag{65}$$

In order to apply this to Lagrangians of the form shown in equation 56, we must define:

$$\mathcal{L}_k = \frac{\partial \mathcal{L}}{\partial f_k} \quad ; \quad \mathcal{L}_{kl} = \frac{\partial^2 \mathcal{L}}{\partial f_k \partial f_l} \quad ; \quad \text{etc.} \tag{66}$$

### 3.2.2. Derivation of $\delta_d^2 \mathcal{H}$ for $\mathcal{L}(f_0, f_2, f_4)$

The derivation of $\delta_d^2 \mathcal{H}$ for this system is straightforward but tedious. In order to simplify the appearance of the equations, I will have to define a number of auxiliary quantities to represent common subexpressions.

To begin with, we may use the chain rule and the definitions of equation 66 to deduce:

$$\frac{\delta}{\delta\dot{\varphi}_\alpha} \mathcal{L}(f_0, f_2, f_4) = \mathcal{L}_0 \frac{\delta f_0}{\delta\dot{\varphi}_\alpha} + \mathcal{L}_2 \frac{\delta f_2}{\delta\dot{\varphi}_\alpha} + \mathcal{L}_4 \frac{\delta f_4}{\delta\dot{\varphi}_\alpha} \qquad (67)$$

It is easiest to treat this as two separate cases:

$$\frac{\delta}{\delta\dot{A}} \mathcal{L}(f_0, f_2, f_4) = \mathcal{L}_0 \frac{\delta f_0}{\delta\dot{A}} + \mathcal{L}_2 \frac{\delta f_2}{\delta\dot{A}} + \mathcal{L}_4 \frac{\delta f_4}{\delta\dot{A}}$$

$$\frac{\delta}{\delta\dot{\varphi}_i} \mathcal{L}(f_0, f_2, f_4) = \mathcal{L}_0 \frac{\delta f_0}{\delta\dot{\varphi}_i} + \mathcal{L}_2 \frac{\delta f_2}{\delta\dot{\varphi}_i} + \mathcal{L}_4 \frac{\delta f_4}{\delta\dot{\varphi}_i} \qquad (68)$$

From equations 57, we may derive:

$$\frac{\delta f_0}{\delta\dot{A}} = \frac{\delta f_0}{\delta\dot{\varphi}_i} = 0$$

$$\frac{\delta f_2}{\delta\dot{A}} = 2\dot{A}$$

$$\frac{\delta f_4}{\delta\dot{A}} = 2w_0 A \qquad (69)$$

$$\frac{\delta f_2}{\delta\dot{\varphi}_i} = -2\dot{\varphi}_i$$

$$\frac{\delta f_4}{\delta\dot{\varphi}_i} = -2w_0 \varphi_i$$

Substituting equations 69 into equations 68, we get:

$$\Pi_A^{[t]} = \frac{\delta}{\delta\dot{A}} \mathcal{L}(f_0, f_2, f_4) = 2\mathcal{L}_2 \dot{A} + 2\mathcal{L}_4 w_0 A$$

$$\Pi_{\varphi_i}^{[t]} = \frac{\delta}{\delta\dot{\varphi}_i} \mathcal{L}(f_0, f_2, f_4) = -2\mathcal{L}_2 \dot{\varphi}_i - 2\mathcal{L}_4 w_0 \varphi_i \qquad (70)$$

It is possible to calculate the complete set of second derivatives of $\mathcal{L}$ and then substitute them into equation 64. However, a more direct approach can also be used. First, we can substitute equations 70 back into the definition of $\mathcal{H}$, which yields:

$$\begin{aligned}\mathcal{H} &= (2\mathcal{L}_2 \dot{A} + 2\mathcal{L}_4 w_0 A)\dot{A} + \sum_i (-2\mathcal{L}_2 \dot{\varphi}_i - 2\mathcal{L}_4 w_0 \varphi_i)\dot{\varphi}_i - \mathcal{L} \\ &= 2\mathcal{L}_2(\dot{A}^2 - |\underline{\dot{\varphi}}|^2) + 2\mathcal{L}_4 w_0^2 - \mathcal{L} = 2\mathcal{L}_2 s_0 + 2\mathcal{L}_4 w_0^2 - \mathcal{L}\end{aligned} \qquad (71)$$

where for convenience I now define:

$$f_2 = s_0 - \sum_k s_k$$
$$s_0 = \dot{A}^2 - |\dot{\underline{\varphi}}|^2 \quad (72)$$
$$s_k = (\partial_k A)^2 - |\partial_k \underline{\varphi}|^2$$

Next I would want to work out $\delta_d \mathcal{H}$, the <u>first</u> variation of $\mathcal{H}$ as given in equation 71 but containing only those terms which are <u>linear</u> in the <u>derivatives</u> of $\xi$. However, before doing so, I must first work out $\delta_d()$ for the various terms which appear in equation 71:

$$\delta_d s_0 = 2\dot{A}\dot{\xi}_0 - 2(\dot{\underline{\varphi}} \cdot \dot{\underline{\xi}})$$
$$\delta_d w_0 = A\dot{\xi}_0 - (\underline{\varphi} \cdot \dot{\underline{\xi}})$$
$$\delta_d f_2 = \delta_d s_0 - \sum_k \delta_d s_k = 2\dot{A}\dot{\xi}_0 - 2(\dot{\underline{\varphi}} \cdot \dot{\underline{\xi}}) - 2\sum_k ((\partial_k A)(\partial_k \xi_0) - (\partial_k \underline{\varphi} \cdot \partial_k \underline{\xi})) \quad (73)$$
$$\delta_d f_4 = 2w_0(\delta_d w_0) - 2\sum_i w_i(\delta_d w_i)$$
$$= 2w_0(A\dot{\xi}_0 - (\underline{\varphi} \cdot \dot{\underline{\xi}})) - 2\sum_i w_i(A(\partial_i \xi_0) - \sum_k \varphi_k(\partial_i \xi_k))$$

We can simplify the appearance of these expressions by defining:

$$\alpha_0 = \dot{A}\dot{\xi}_0 - (\dot{\underline{\varphi}} \cdot \dot{\underline{\xi}})$$
$$\alpha_k = (\partial_k A)(\partial_k \xi_0) - \sum_j (\partial_k \varphi_j)(\partial_k \xi_j) \quad (74)$$

which, when substituted into equation 70, results in:

$$\delta_d s_0 = 2\alpha_0$$
$$\delta_d f_2 = 2\alpha_0 - 2\sum_k \alpha_k \quad (75)$$

Now we may calculate the derivative-based portion of the first variation of $\mathcal{H}$, for $\mathcal{H}$ as expressed in equation 71:

$$\delta_d \mathcal{H} = (2\mathcal{L}_{22}s_0 + 2\mathcal{L}_{42}w_0^2 - \mathcal{L}_2)\delta_d f_2 + (2\mathcal{L}_{24}s_0 + 2\mathcal{L}_{44}w_0^2 - \mathcal{L}_4)\delta_d f_4$$
$$+ 2\mathcal{L}_2(\delta_d s_0) + 4\mathcal{L}_4 w_0(\delta_d w_0) \quad (76)$$

Substituting the results and definitions from equations 73 to 75 into equations 76, and performing some algebra, we get:

$$\tfrac{1}{2}\delta_d \mathcal{H} = (2\mathcal{L}_{22}s_0 + 2\mathcal{L}_{42}w_0^2 + \mathcal{L}_2)\alpha_0 - \sum_k \alpha_k(2\mathcal{L}_{22}s_0 + 2\mathcal{L}_{42}w_0^2 - \mathcal{L}_2)$$
$$+ w_0(A\dot{\xi}_0 - (\underline{\varphi}\cdot\underline{\dot{\xi}}))(2\mathcal{L}_{24}s_0 + 2\mathcal{L}_{44}w_0^2 - \mathcal{L}_4) \tag{77}$$
$$- \sum_i w_i\left(A(\partial_i\xi_0) - \sum_k \varphi_k(\partial_i\xi_k)\right)(2\mathcal{L}_{24}s_0 + 2\mathcal{L}_{44}w_0^2 - \mathcal{L}_4)$$

Performing the same sort of expansion once again, we derive:

$$\tfrac{1}{2}\delta_d^2 \mathcal{H} = \alpha_0 \delta_d\left(2\mathcal{L}_{22}s_0 + 2\mathcal{L}_{42}w_0^2 + \mathcal{L}_2\right) - \sum_k \alpha_k \delta_d\left(2\mathcal{L}_{22}s_0 + 2\mathcal{L}_{42}w_0^2 - \mathcal{L}_2\right)$$
$$+ (2\mathcal{L}_{22}s_0 + 2\mathcal{L}_{42}w_0^2 + \mathcal{L}_2)(\delta_d \alpha_0) - \sum_k (\delta_d \alpha_k)(2\mathcal{L}_{22}s_0 + 2\mathcal{L}_{42}w_0^2 - \mathcal{L}_2)$$
$$+ w_0(A\dot{\xi}_0 - (\underline{\varphi}\cdot\underline{\dot{\xi}}))(\delta_d(2\mathcal{L}_{24}s_0 + 2\mathcal{L}_{44}w_0^2 - \mathcal{L}_4))$$
$$+ (2\mathcal{L}_{24}s_0 + 2\mathcal{L}_{44}w_0^2 - \mathcal{L}_4)(\delta_d(w_0(A\dot{\xi}_0 - (\underline{\varphi}\cdot\underline{\dot{\xi}})))) \tag{78}$$
$$- \sum_i w_i\left(A(\partial_i\xi_0) - \sum_k \varphi_k(\partial_i\xi_k)\right)(\delta_d(2\mathcal{L}_{24}s_0 + 2\mathcal{L}_{44}w_0^2 - \mathcal{L}_4))$$
$$- (2\mathcal{L}_{24}s_0 + 2\mathcal{L}_{44}w_0^2 - \mathcal{L}_4)\sum_i \delta_d\left(w_i\left(A(\partial_i\xi_0) - \sum_k \varphi_k(\partial_i\xi_k)\right)\right)$$

Now that we have written down the full second-order expansion, we may exploit our assumption that we are studying a static soliton solution, such that $\partial_t A = \partial_t \varphi_k = 0$. (If we had tried to exploit this assumption when working with the first variation, we would not have gone on to deduce the correct second variation, because we would have failed to account for the effect of perturbations, which need not satisfy $\partial_t \xi_\mu = 0$!) At this point, we can exploit the fact that $w_0 = s_0 = \delta_d s_0 = \alpha_0 = 0$ in a static solution. This immediately reduces equation 78 to:

$$\tfrac{1}{2}\delta_d^2 \mathcal{H} = \left(\sum_k \alpha_k\right)\delta_d \mathcal{L}_2 + \mathcal{L}_2(\delta_d \alpha_0) + \sum_k (\delta_d \alpha_k)\mathcal{L}_2 + \mathcal{L}_4 \delta_d\left(w_0(A\dot{\xi}_0 - (\underline{\varphi}\cdot\underline{\dot{\xi}}))\right)$$
$$- \sum_i w_i\left(A(\partial_i\xi_0) - \sum_k \varphi_k(\partial_i\xi_k)\right)\delta_d\left(2\mathcal{L}_{44}w_0^2 + \mathcal{L}_{24}s_0 - \mathcal{L}_4\right) \tag{79}$$
$$+ \mathcal{L}_4 \sum_i \delta_d\left(w_i\left(A(\partial_i\xi_0) - \sum_k \varphi_k(\partial_i\xi_k)\right)\right)$$

Furthermore, from $w_0 = s_0 = \delta_d s_0 = \alpha_0 = 0$, we may deduce that:

$$\delta_d\left(w_0(A\dot{\xi}_0 - (\underline{\varphi}\cdot\underline{\dot{\xi}}))\right) = (\delta_d w_0)(A\dot{\xi}_0 - (\underline{\varphi}\cdot\underline{\dot{\xi}})) = (A\dot{\xi}_0 - (\underline{\varphi}\cdot\underline{\dot{\xi}}))^2$$
$$\delta_d\left(2\mathcal{L}_{44}w_0^2 + \mathcal{L}_{24}s_0 - \mathcal{L}_4\right) = -\delta_d \mathcal{L}_4 \tag{80}$$

Also note that:

$$\delta_d\left(w_i\left(A(\partial_i\xi_0) - \sum_k \varphi_k(\partial_i\xi_k)\right)\right) = (\delta_d w_i)\left(A(\partial_i\xi_0) - \sum_k \varphi_k(\partial_i\xi_k)\right) \quad , \tag{81}$$

because the $\delta_d$ variation of the rightmost term is zero, because its variation contains no terms which are bilinear in the <u>derivatives</u> of $\xi_\mu$.

Likewise, by direct use of our definitions above, we may derive the variations:

$$\begin{aligned}
\delta_d\alpha_0 &= \delta_d\left(\dot{A}\dot{\xi}_0 - (\dot{\underline{\varphi}}\cdot\dot{\underline{\xi}})\right) = \dot{\xi}_0^2 - \left|\dot{\underline{\xi}}\right|^2 \\
\delta_d\alpha_i &= \delta_d\left((\partial_i A)(\partial_i\xi_0) - \sum_k (\partial_i\varphi_k)(\partial_i\xi_k)\right) = (\partial_i\xi_0)^2 - \sum_k (\partial_i\xi_k)^2 \\
\delta_d w_i &= \delta_d\left(A(\partial_i A) - \sum_k \varphi_k(\partial_i\varphi_k)\right) = A(\partial_i\xi_0) - \sum_k \varphi_k(\partial_i\xi_k)
\end{aligned} \tag{82}$$

Substituting all this back into equation 79, we get:

$$\begin{aligned}
\tfrac{1}{2}\delta_d^2 \mathcal{H} &= \left(\sum_k \alpha_k\right)\delta_d \mathcal{L}_2 + \mathcal{L}_2(\dot{\xi}_0^2 - |\dot{\underline{\xi}}|^2) + \mathcal{L}_2\sum_i\left((\partial_i\xi_0)^2 - \sum_k(\partial_i\xi_k)^2\right) \\
&\quad + \mathcal{L}_4(A\dot{\xi}_0 - (\underline{\varphi}\cdot\dot{\underline{\xi}}))^2 + (\delta_d\mathcal{L}_4)\sum_i\left(A(\partial_i\xi_0) - \sum_k \varphi_k(\partial_i\xi_k)\right) \\
&\quad + \mathcal{L}_4\sum_i\left(A(\partial_i\xi_0) - \sum_k \varphi_k(\partial_i\xi_k)\right)^2 \\
&= \mathcal{L}_2\left(\dot{\xi}_0^2 - |\dot{\underline{\xi}}|^2 + \sum_i\left((\partial_i\xi_0)^2 - \sum_k(\partial_i\xi_k)^2\right)\right) \\
&\quad + \mathcal{L}_4\left(A\dot{\xi}_0 - (\underline{\varphi}\cdot\dot{\underline{\xi}}))^2 + \sum_i\left(A(\partial_i\xi_0) - \sum_k \varphi_k(\partial_i\xi_k)\right)^2\right) \\
&\quad + (\delta_d\mathcal{L}_4)\sum_i\left(A(\partial_i\xi_0) - \sum_k \varphi_k(\partial_i\xi_k)\right) + \left(\sum_k \alpha_k\right)\delta_d\mathcal{L}_2
\end{aligned} \tag{83}$$

Finally, following the same kind of procedures we used to deduce equation 76 (but simpler), we may work out the two unexpanded terms in equation 83:

$$\begin{aligned}
\delta_d\mathcal{L}_2 &= \mathcal{L}_{22}(\delta_d f_2) + \mathcal{L}_{24}(\delta_d f_4) \\
&= 2\mathcal{L}_{22}\left(\alpha_0 - \sum_k \alpha_k\right) + 2\mathcal{L}_{24}\left(w_0(A\dot{\xi}_0 - (\underline{\varphi}\cdot\dot{\underline{\xi}})) - \sum_i w_i(A(\partial_i\xi_0) - \sum_k \varphi_k(\partial_i\xi_k))\right) \\
\delta_d\mathcal{L}_4 &= \mathcal{L}_{42}(\delta_d f_2) + \mathcal{L}_{44}(\delta_d f_4) \\
&= 2\mathcal{L}_{42}\left(\alpha_0 - \sum_k \alpha_k\right) + 2\mathcal{L}_{44}\left(w_0(A\dot{\xi}_0 - (\underline{\varphi}\cdot\dot{\underline{\xi}})) - \sum_i w_i(A(\partial_i\xi_0) - \sum_k \varphi_k(\partial_i\xi_k))\right)
\end{aligned} \tag{84}$$

Before inserting equation 84 into equation 83, we may again exploit the fact that $w_0=\alpha_0=0$ in static equilibrium states, which reduces equation 84 to:

$$\delta_d \mathcal{L}_2 = -2\mathcal{L}_{22}\sum_k \alpha_k - 2\mathcal{L}_{24}\sum_i w_i\left(A(\partial_i \xi_0) - \sum_k \varphi_k(\partial_i \xi_k)\right)$$
$$\delta_d \mathcal{L}_4 = -2\mathcal{L}_{42}\sum_k \alpha_k - 2\mathcal{L}_{44}\sum_i w_i\left(A(\partial_i \xi_0) - \sum_k \varphi_k(\partial_i \xi_k)\right) \quad (85)$$

Finally, when we substitute this back into equation 83, we may deduce:

$$\tfrac{1}{2}\delta_d^2 \mathcal{H} = \mathcal{L}_2(\dot{\xi}_0^2 - \underline{\dot{\xi}}^2 + \sum_i((\partial_i \xi_0)^2 - \sum_k(\partial_i \xi_k)^2))$$
$$+ \mathcal{L}_4((A\dot{\xi}_0 - \underline{\varphi}\cdot\underline{\dot{\xi}})^2 + \sum_i(A(\partial_i \xi_0) - \sum_k \varphi_k(\partial_i \xi_k))^2) \quad (86)$$
$$- 2\mathcal{L}_{44} C^2 - 2\mathcal{L}_{22} D^2 - 4\mathcal{L}_{24} CD,$$

where for convenience I define:

$$C = \sum_i w_i\left(A(\partial_i \xi_0) - \sum_k \varphi_k(\partial_i \xi_k)\right)$$
$$D = \sum_i\left((\partial_i A)(\partial_i \xi_0) - \sum_k(\partial_i \varphi_k)(\partial_i \xi_k)\right) = \sum_i \alpha_i = \sum_k \alpha_k \quad (87)$$

For a discussion of the implications of equation 86, see [4]. In order to perform second-order stability analysis of the numerical examples proposed in [4], one would first apply the analytical methods to be discussed in section 4 below, followed by the one-dimensional methods of section 2, as discussed in [4].

### 3.3. Impossibility Results for Static Real $\mathcal{L}$ ($f_0$, $f_1$, $f_2$, $f_3$)

For a real vector field $\varphi_\mu$ with a Lagrangian of the form $\mathcal{L}(f_0, f_1, f_2, f_3)$, over a space of dimension D>1, there cannot exist a static equilibrium state which possesses SSOS (strong secnd-order stability) as discussed in [4]. The argument is based solely on the analysis of $\delta_d^2 \mathcal{H}$ for such a field theory. Section 3.3.1 will derive an expression for $\delta_d^2 \mathcal{H}$ for this family of field theories. Section 3.3.2 will analyze its implications, and show why SSOS is impossible here.

### 3.3. Derivation of $\delta_d^2 \mathcal{H}$ for Static Real $\mathcal{L}$ ($f_0$, $f_1$, $f_2$, $f_3$)

The direct derivation of $\delta_d^2 \mathcal{H}$ in this case is far more complex than the derivation in section 3.2.2. Therefore, in order to minimize the algebra, we need to use two additional tricks. First, we may use equation 64 to deduce:

$$\delta_d^2 \mathcal{H} = \delta_t^2 \mathcal{L} - \delta_x^2 \mathcal{L} \quad , \quad (88)$$

where I define two new general operators:

$$\delta_t f(\underline{\varphi}, \underline{\dot\varphi}, \nabla\underline{\varphi}) = \sum_\alpha \frac{\delta f}{\delta \dot\varphi_\alpha} \dot\xi_\alpha$$
$$\delta_x f(\underline{\varphi}, \underline{\dot\varphi}, \nabla\underline{\varphi}) = \sum_{i,\alpha} \frac{\delta f}{\delta(\partial_i \varphi_\alpha)} (\partial_i \xi_\alpha)$$
(89)

where I do not use the Einstein convention for summation over $\alpha$, in part because I want to emphasize the fact that these definitions can be applied to any collection of fields, not just relativistic covariant vectors.

Before beginning, I will first go back to the definitions in equations 57, and work out:

$$\delta_d f_1 = \delta_d\left(\dot A - \sum_i \partial_i \varphi_i\right) = \dot\xi_0 - \sum_i \partial_i \xi_i = \dot\xi_0 - \mathbf{div}\ \underline\xi$$

$$\delta_d f_3 = \delta_d\left(\dot A^2 + |\nabla\underline\varphi|^2 - 2(\underline{\dot\varphi}\cdot\nabla A)\right)$$
$$= 2\dot\xi_0 \dot A + 2\sum_{i,j}(\partial_i \varphi_j)(\partial_i \xi_j) - 2(\underline{\dot\varphi}\cdot\nabla\xi_0) - 2(\underline{\dot\xi}\cdot\nabla A)$$
(90)
$$= 2\dot\xi_0 \dot A + 2S - 2(\underline{\dot\varphi}\cdot\nabla\xi_0) - 2(\underline{\dot\xi}\cdot\nabla A)\ ,$$

using the new auxiliary definition:

$$S = \sum_{i,j}(\partial_i \varphi_j)(\partial_i \xi_j)$$
(91)

Again, recall that $\delta_t$ singles out <u>that portion</u> of $\delta_d$ which involves time-derivative <u>of $\underline\xi$</u>.

In order to work out $\delta_d^2 \mathcal{H}$, as in equation 88, I will begin by working out $\delta_t^2 \mathcal{L}$. First I will apply the operator $\delta_t$ directly to $\mathcal{L}$:

$$\delta_t \mathcal{L} = \mathcal{L}_0(\delta_t f_0) + \mathcal{L}_1(\delta_t f_1) + \mathcal{L}_2(\delta_t f_2) + \mathcal{L}_3(\delta_t f_3)$$
$$= \mathcal{L}_0(0) + \mathcal{L}_1(\dot\xi_0) + \mathcal{L}_2(2\alpha_0) + \mathcal{L}_3\left(2\dot\xi_0 \dot A - 2(\underline{\dot\xi}\cdot\nabla A)\right)$$
(92)
$$= \dot\xi_0 \mathcal{L}_1 + 2\alpha_0 \mathcal{L}_2 + 2\mathcal{L}_3\left(\dot\xi_0 \dot A - (\underline{\dot\xi}\cdot\nabla A)\right)$$

Next, as I apply the operator $\delta_t$ to equation 92, I will simply not write down those terms which are obviously zero in the static case, where $\partial_t A = \partial_t \varphi = s_0 = \alpha_0 = \delta_d s_0 = 0$. (This would be like jumping directly from equation 77 to equation 79 in section 3.2.2, without bothering to write down equation 78.) This yields:

$$\delta_t^2 \mathcal{L} = \dot\xi_0(\delta_t \mathcal{L}_1) + 2\mathcal{L}_2(\delta_t \alpha_0) + \mathcal{L}_3(2\dot\xi_0^2) - 2(\underline{\dot\xi}\cdot\nabla A)(\delta_k \mathcal{L}_3)$$
(93)

By analogy to equation 92, we may deduce

$$\delta_t \mathcal{L}_1 = \mathcal{L}_{11}\dot{\xi}_0 + 2\alpha_0 \mathcal{L}_{12} + 2\mathcal{L}_{13}(\dot{\xi}_0 \dot{A} - (\dot{\underline{\xi}} \cdot \nabla A))$$
$$\delta_t \mathcal{L}_2 = \mathcal{L}_{21}\dot{\xi}_0 + 2\alpha_0 \mathcal{L}_{22} + 2\mathcal{L}_{23}(\dot{\xi}_0 \dot{A} - (\dot{\underline{\xi}} \cdot \nabla A)) \quad (94)$$
$$\delta_t \mathcal{L}_3 = \mathcal{L}_{31}\dot{\xi}_0 + 2\alpha_0 \mathcal{L}_{32} + 2\mathcal{L}_{33}(\dot{\xi}_0 \dot{A} - (\dot{\underline{\xi}} \cdot \nabla A))$$

Substituting equations 94 and 82 back into equation 93, and deleting the terms which are obviously zero in the static case, we get:

$$\begin{aligned}\delta_t^2 \mathcal{L} &= \dot{\xi}_0\Big(\mathcal{L}_{11}\dot{\xi}_0 + 2\mathcal{L}_{13}(-(\dot{\underline{\xi}} \cdot \nabla A))\Big) + 2\mathcal{L}_2(\dot{\xi}_0^2 - |\dot{\underline{\xi}}|^2) \\ &\quad + 2\mathcal{L}_3 \dot{\xi}_0^2 - 2(\dot{\underline{\xi}} \cdot \nabla A)\Big(\mathcal{L}_{31}\dot{\xi}_0 + \mathcal{L}_{33}(-2(\dot{\underline{\xi}} \cdot \nabla A))\Big) \\ &= \dot{\xi}_0^2(\mathcal{L}_{11} + 2\mathcal{L}_2 + 2\mathcal{L}_3) - 4\mathcal{L}_{13}\dot{\xi}_0(\dot{\underline{\xi}} \cdot \nabla A) - 2\mathcal{L}_2|\dot{\underline{\xi}}|^2 + 4\mathcal{L}_{33}(\dot{\underline{\xi}} \cdot \nabla A)^2 \end{aligned} \quad (95)$$

Next I will work out $\delta_x^2 \mathcal{L}$. First I will apply the operator $\delta_x$ directly to $\mathcal{L}$:

$$\begin{aligned}\delta_x \mathcal{L} &= \mathcal{L}_0(\delta_x f_0) + \mathcal{L}_1(\delta_x f_1) + \mathcal{L}_2(\delta_x f_2) + \mathcal{L}_3(\delta_x f_3) \\ &= \mathcal{L}_0(0) + \mathcal{L}_1(-\mathbf{div}\ \underline{\xi}) + \mathcal{L}_2(-2\sum_k \alpha_k) + \mathcal{L}_3\Big(2S - 2(\dot{\underline{\varphi}} \cdot \nabla \xi_0)\Big) \\ &= -\mathcal{L}_1(\mathbf{div}\ \underline{\xi}) - 2\mathcal{L}_2 \sum_k \alpha_k + 2\mathcal{L}_3\Big(S - (\dot{\underline{\varphi}} \cdot \nabla \xi_0)\Big)\end{aligned} \quad (96)$$

Once again, I will apply the operator a second time and exploit the fact that $\partial_t \varphi = 0$ in the sttaic case:

$$\delta_x^2 \mathcal{L} = -(\mathbf{div}\ \underline{\xi})(\delta_x \mathcal{L}_1) - 2\sum_k \alpha_k (\delta_x \mathcal{L}_2) - 2\mathcal{L}_2 \sum_k (\delta_x \alpha_k) + 2(\delta_x \mathcal{L}_3) + 2\mathcal{L}_3(\delta_x s) \quad (97)$$

(Note that $\delta_x$ applied to $(\partial_t \varphi)$ yields zero, because terms like $\partial_i \varphi_\alpha$ do not appear in $\partial_t \varphi$.)
By analogy to equation 96, we may deduce:

$$\delta_x \mathcal{L}_1 = -\mathcal{L}_{11}(\mathbf{div}\ \underline{\xi}) - 2\mathcal{L}_{12}\sum_k \alpha_k + 2\mathcal{L}_{13}(S - (\dot{\underline{\varphi}} \cdot \nabla \xi_0))$$
$$\delta_x \mathcal{L}_2 = -\mathcal{L}_{21}(\mathbf{div}\ \underline{\xi}) - 2\mathcal{L}_{22}\sum_k \alpha_k + 2\mathcal{L}_{23}(S - (\dot{\underline{\varphi}} \cdot \nabla \xi_0)) \quad (98)$$
$$\delta_x \mathcal{L}_3 = -\mathcal{L}_{31}(\mathbf{div}\ \underline{\xi}) - 2\mathcal{L}_{32}\sum_k \alpha_k + 2\mathcal{L}_{33}(S - (\dot{\underline{\varphi}} \cdot \nabla \xi_0))$$

Also, from equation 82, we can easily see that:

$$\delta_x \alpha_i = (\partial_i \xi_0)^2 - \sum_k (\partial_i \xi_k)^2 = |\nabla \xi_0|^2 - \sum_j (\partial_i \xi_j)^2 \quad (99)$$

From the definition of S in equation 91, we easily deduce:

$$\delta_x S = \sum_{i,j} (\partial_i \xi_j)^2 \quad (100)$$

Substituting equations 98 through 100 into equation 97, and again exploiting the fact that $\partial_t \varphi = 0$ here, we get:

$$\begin{aligned}
\delta_x^2 \mathcal{L} = &-(\mathbf{div}\ \underline{\xi})(-\mathcal{L}_{11}(\mathbf{div}\ \underline{\xi}) - 2\mathcal{L}_{12}D + 2\mathcal{L}_{13}S) \\
&- 2D(-\mathcal{L}_{21}(\mathbf{div}\ \underline{\xi}) - 2\mathcal{L}_{22}D + 2\mathcal{L}_{23}S) \\
&+ 2S(-\mathcal{L}_{31}(\mathbf{div}\ \underline{\xi}) - 2\mathcal{L}_{32}D + 2\mathcal{L}_{33}S) \\
&- 2\mathcal{L}_2 \left( |\nabla \xi_0|^2 - \sum_{i,j} (\partial_i \xi_j)^2 \right) + 2\mathcal{L}_3 \sum_{i,j} (\partial_i \xi_j)^2 \quad,
\end{aligned} \qquad (101)$$

using the auxiliary quantity "D" defined in equation 87. Collecting terms, we end up with:

$$\begin{aligned}
\delta_x^2 \mathcal{L} = &\mathcal{L}_{11}(\mathbf{div}\ \underline{\xi})^2 + 4(\mathcal{L}_{12}D - \mathcal{L}_{13}S)(\mathbf{div}\ \underline{\xi}) \\
&+ 4\mathcal{L}_{22}D^2 + 8\mathcal{L}_{23}DS + 4\mathcal{L}_{33}S^2 \\
&- 2\mathcal{L}_2 |\nabla \xi_0|^2 + 2(\mathcal{L}_2 + \mathcal{L}_3)\sum_{i,j}(\partial_i \xi_j)^2
\end{aligned} \qquad (102)$$

From the definitions in equations 87 and 91, we may also deduce:

$$D = (\nabla A \cdot \nabla \xi_0) - S \qquad (103)$$

Finally, by substituting equations 95, 102 and 103 into equation 88, we obtain:

$$\begin{aligned}
\delta_d^2 \mathcal{H} = &\dot{\xi}_0^2(\mathcal{L}_{11} + 2\mathcal{L}_2 + 2\mathcal{L}_3) - 4\mathcal{L}_{13}\dot{\xi}_0(\underline{\dot{\xi}} \cdot \nabla A) - 2\mathcal{L}_2 |\underline{\dot{\xi}}|^2 + 4\mathcal{L}_{33}(\underline{\dot{\xi}} \cdot \nabla A)^2 \\
&- \mathcal{L}_{11}(\mathbf{div}\ \underline{\xi})^2 - 4\mathcal{L}_{12}(\nabla A \cdot \nabla \xi_0)(\mathbf{div}\ \underline{\xi}) + 4(\mathcal{L}_{12} + \mathcal{L}_{13})S(\mathbf{div}\ \underline{\xi}) \\
&- 4\mathcal{L}_{22}(\nabla A \cdot \nabla \xi_0)^2 - (4\mathcal{L}_{22} - 8\mathcal{L}_{23} + 4\mathcal{L}_{33})S^2 - S(\nabla A \cdot \nabla \xi_0)(8\mathcal{L}_{23} - 8\mathcal{L}_{22}) \\
&+ 2\mathcal{L}_2 |\nabla \xi_0|^2 - 2(\mathcal{L}_2 + \mathcal{L}_3)\sum_{i,j}(\partial_i \xi_j)^2
\end{aligned} \qquad (104)$$

### 3.3.2. Why SSOS Is Impossible In This System

In order to satisfy the Generalized Legendre Condition (GLC) [4], it is necessary that $\delta_d^2 \mathcal{H}$ must be nonnegative for <u>any</u> perturbation derivatives at <u>any</u> point $\mathbf{x}$ in space; in effect, at each point $\mathbf{x}$, $\delta_d^2 \mathcal{H}$ is a bilinear form in $\{\partial_\mu \xi_\nu\}$, a vector of length 16 (for 3+1-D systems), whose components are all assumed to be independent. We require that this 16-by-16 bilinear form must be nonnegative.

To begin with, consider what this requires for perturbations where $\partial \underline{\xi}$ is nonzero, but all the other components of $\{\partial_\mu \xi_\nu\}$ are zero. All of the terms in equation 104 go to zero for perturbations of this type, except for the third major term in the right; therefore, the GLC requires that this term be nonnegative for these perturbations, which implies:

$$\mathcal{L}_2 \leq 0 \quad \text{at all points } \mathbf{x} \qquad (105)$$

On the other hand, consider what GLC requires for perturbations where all of the derivatives $\partial_\mu \xi_\nu = 0$ except for $\nabla \xi_0$. When D>1, we can always pick $\nabla \xi_0$ to be nonzero <u>and</u> orthogonal to $\nabla A$. In this case, all of the terms in equation 104 again become zero, except for the next-to-last term, which tells us that:

$$\mathcal{L}_2 \geq 0 \text{ at all points } \underline{\mathbf{x}} \qquad (106)$$

This is the major destabilizing symmetry which I mentioned earlier in this section! Of course, it tells us immediately that $\mathcal{L}_2=0$. This by itself is enough to rule out the possibility of Strong Second-Order Stability, since $\delta_d^2\mathcal{H}$ (and $\delta^2\mathcal{H}$) becomes totally indefinite with respect to perturbations $\partial_t\underline{\xi}$ at all points $\underline{\mathbf{x}}$! H itself is highly indefinite, since we can choose any particular function at all across space for the perturbation of $\partial_t\underline{\varphi}$, and still leave $\delta^2H=0$, so long as we do not perturb $\varphi_\mu$ or $\partial_t A$ at the same time.

Note that this argument does not rule out "stability" in the broadest sense. Even with $\delta^2H$ indefinite, one could imagine possible mechanisms for stability based on $\delta^4H$ and so on. However, further analysis shows even more constraints of equation 104. In any case, the claim about SSOS stability has been proven here.

## 3.4. Lack of Nonstatic H' Stability in Complex 1+1-D $\mathcal{L}$ ($f_0$, $f_2$)

This section will evaluate the possibility of stable states -- static or dynamic -- in the complex $\mathcal{L}$ ($f_0$, $f_2$) system, in 1+1-D. (The 1+1-D case is chosen for the sake of simplicity.) The conclusion is that SSOS stability is impossible, either in minimizing H or in minimizing H' (a concept discussed in section 2.7).

## 3.4.1 Derivation of $\delta_d^2\mathcal{H}$ and of First-Order Minimization Conditions

First I must establish some definitions:

$$\begin{aligned}
A &= r_A e^{i\theta_A} \\
\varphi &= r_\varphi e^{i\theta_\varphi} \\
f_0 &= |A|^2 - |\varphi|^2 = r_A^2 - r_\varphi^2 \\
f_2 &= |\dot{A}|^2 - |\nabla A|^2 - |\dot{\varphi}|^2 + |\nabla \varphi|^2 = s_0 - s_1 \\
s_0 &= |\dot{A}|^2 - |\dot{\varphi}|^2 = \dot{r}_A^2 + r_A^2\dot{\theta}_A^2 - \dot{r}_\varphi^2 - r_\varphi^2\dot{\theta}_\varphi^2
\end{aligned} \qquad (107)$$

Next I must work out the usual conjugate momenta:

$$\begin{aligned}
\Pi_{r_A}^{[t]} &= \frac{\delta\mathcal{L}}{\delta\dot{r}_A} = \mathcal{L}_2 \cdot \frac{\delta f_2}{\delta\dot{r}_A} = \mathcal{L}_2(2\dot{r}_A) \\
\Pi_{r_\varphi}^{[t]} &= \mathcal{L}_2(-2\dot{r}_\varphi) \\
\Pi_{\theta_A}^{[t]} &= \frac{\delta\mathcal{L}}{\delta\dot{\theta}_A} = \mathcal{L}_2(2r_A^2\dot{\theta}_A) \\
\Pi_{\theta_\varphi}^{[t]} &= \mathcal{L}_2(-2r_\varphi^2\dot{\theta}_\varphi)
\end{aligned} \qquad (108)$$

Substituting this into equation 60, we derive:

$$\begin{aligned}
\mathcal{H} &= (2\mathcal{L}_2\dot{r}_A)\dot{r}_A + (-2\mathcal{L}_2\dot{r}_\varphi)\dot{r}_\varphi + (2\mathcal{L}_2 r_A^2\dot{\theta}_A)\dot{\theta}_A + (-2\mathcal{L}_2 r_\varphi^2\dot{\theta}_\varphi)\dot{\theta}_\varphi - \mathcal{L} \\
&= 2\mathcal{L}_2(\dot{r}_A^2 + r_A^2\dot{\theta}_A^2 - \dot{r}_\varphi^2 - r_\varphi^2\dot{\theta}_\varphi^2) - \mathcal{L} = 2\mathcal{L}_2 s_0 - \mathcal{L}
\end{aligned} \qquad (109)$$

Then, in order to consider the possibility of H' stability, we must work out the conserved charges of this system. The Lagrangian $\mathcal{L}$ ($f_0$, $f_2$) is invariant with respect to two global gauge transformations -- $G_1(\alpha)(A,\varphi)=(e^{i\alpha}A,\varphi)$ and $G_2(\alpha)(A,\varphi)=(A,e^{i\alpha}\varphi)$. Following equation 50, we may derive the two conserved charges associated with these two symmetries:

$$q_A = \frac{\delta \mathcal{L}}{\delta \dot{\theta}_A} = \mathcal{L}_2(2r_A^2\dot{\theta}_A)$$

$$q_\varphi = \frac{\delta \mathcal{L}}{\delta \dot{\theta}_\varphi} = \mathcal{L}_2(2r_\varphi^2\dot{\theta}_\varphi)$$

(110)

Taking the same approach as in equation 49, but slightly extended (and with a slight rescaling of the $\lambda$ factors), we may define:

$$\mathcal{H}' = \mathcal{H} + \lambda_A \mathcal{L}_2 r_A^2 \dot{\theta}_A + \lambda_\varphi \mathcal{L}_2 r_\varphi^2 \dot{\theta}_\varphi$$
$$= \mathcal{L}_2(2s_0 + \lambda_A r_A^2 \dot{\theta}_A + \lambda_\varphi r_\varphi^2 \dot{\theta}_\varphi) - \mathcal{L}$$
$$= Q\mathcal{L}_2 - \mathcal{L} \quad,$$

(111)

where I define yet another auxiliary quantity:

$$Q = 2s_0 + \lambda_A r_A^2 \dot{\theta}_A + \lambda_\varphi r_\varphi^2 \dot{\theta}_\varphi$$

(112)

Note that H-stability is simply the special case of H' stability in which we choose $\lambda_A=\lambda_\varphi=0$. We will see below that there are <u>no</u> states which meet the test of SSOS for <u>any</u> choice of $\lambda_A$ and $\lambda_\varphi$.

In order to fulfill the Generalized Legendre Condition (GLC), it is necessary (but not sufficient!) that $\delta_t^2 \mathcal{H}'$ be nonnegative at every point in space. We may begin by operating on equation 111:

$$\delta_t \mathcal{H}' = \mathcal{L}_2(\delta_t Q) + (Q\mathcal{L}_{20} - \mathcal{L}_0)(\delta_t f_0) + (Q\mathcal{L}_{22} - \mathcal{L}_2)(\delta_t f_2)$$

(113)

It is easy to see that:

$$\delta_t f_0 = 0$$
$$\delta_t f_2 = \delta_t s_0 = 2(\dot{r}_A \dot{\eta}_A + r_A^2 \dot{\theta}_A \dot{\xi}_A - \dot{r}_\varphi \dot{\eta}_\varphi - r_\varphi^2 \dot{\theta}_\varphi \dot{\xi}_\varphi) = 2\alpha_0$$
$$\delta_t Q = \delta_t \left(2s_0 + \lambda_A r_A^2 \dot{\theta}_A + \lambda_\varphi r_\varphi^2 \dot{\theta}_\varphi\right)$$
$$2(2\alpha_0) + \lambda_A r_A^2 \dot{\xi}_A + \lambda_\varphi r_\varphi^2 \dot{\xi}_\varphi = P \quad,$$

(114)

where I use "$\eta$" to indicate a perturbation of r, and "$\xi$" to indicate a perturbation of $\theta$, and where I am introducing two new definitions (just for section 3.4) for the auxiliary quantities $\alpha_0$ and P. In this notation, equation 113 reduces to:

$$\delta_t \mathcal{H}' = P\mathcal{L}_2 + 2\alpha_0(Q\mathcal{L}_{22} - \mathcal{L}_2)$$

(115)

Applying the $\delta_t$ operator again to equation 115, we get:

$$\delta_t^2 \mathcal{H}' = \left(P\mathcal{L}_{22} + 2\alpha_0(Q\mathcal{L}_{222} - \mathcal{L}_{22})\right)(\delta_t f_2)$$
$$+ \mathcal{L}_2(\delta_t P) + 2(Q\mathcal{L}_{22} - \mathcal{L}_2)(\delta_t \alpha_0) + 2\alpha_0 \mathcal{L}_{22}(\delta_t Q) \tag{116}$$

Performing the same substitutions as with equation 115 (but <u>not</u> assuming a static solution in this section!), and accounting for the definitions of P and $\alpha_0$, we get:

$$\delta_t^2 \mathcal{H}' = \left(P\mathcal{L}_{22} + 2\alpha_0(Q\mathcal{L}_{222} - \mathcal{L}_{22})\right)(2\alpha_0)$$
$$+ 4\mathcal{L}_2(\delta_t \alpha_0) + 2(Q\mathcal{L}_{22} - \mathcal{L}_2)(\beta_0) + 2\alpha_0 \mathcal{L}_{22} P \quad, \tag{117}$$

where I define a new auxiliary quantity:

$$\beta_0 = \delta_t \alpha_0 = \dot{\eta}_A^2 + r_A^2 \dot{\xi}_A^2 - \dot{\eta}_\varphi^2 + r_\varphi^2 \dot{\xi}_\varphi^2 \tag{118}$$

Collecting terms, and substituting in the definition of P, and performing some minor algebra, we get:

$$\delta_t^2 \mathcal{H}' = 4\alpha_0^2(Q\mathcal{L}_{222} + 3\mathcal{L}_{22}) + 4\alpha_0(\lambda_A r_A^2 \dot{\xi}_A + \lambda_\varphi r_\varphi^2 \dot{\xi}_\varphi)\mathcal{L}_{22} + 2(Q\mathcal{L}_{22} + \mathcal{L}_2)\beta_0 \tag{119}$$

Finally, the analysis below will also require that we consider the first-order conditions (Lagrange-Euler equations) for a minimum of H'. For the minimization of H', integrated across space, the Lagrange Euler-equations for the variable $\partial_t r_A$ are:

$$\partial_x \frac{\delta \mathcal{H}'}{\delta(\partial_x \dot{r}_A)} = \mathbf{0} = \frac{\delta \mathcal{H}'}{\delta \dot{r}_A} = (Q\mathcal{L}_{22} - \mathcal{L}_2)\frac{\delta f_2}{\delta \dot{r}_A} + \mathcal{L}_2 \frac{\delta Q}{\delta \dot{r}_A} = 2(Q\mathcal{L}_{22} + \mathcal{L}_2)\dot{r}_A \tag{120}$$

Likewise, for $\partial_t r_\varphi$, we may deduce the requirement:

$$(Q\mathcal{L}_{22} + \mathcal{L}_2)\dot{r}_\varphi = \mathbf{0} \tag{121}$$

### 3.4.2. Why SSOS Stability Is Impossible Here

Again, the GLC requires that equation 119 be nonnegative at all points in space.

First, consider the possibility that $\partial_t r_A = \partial_t r_\varphi = 0$ at some point. At such a point, consider perturbations whose derivatives are all zero except for $\partial_t \eta_A$ and $\partial_t \eta_\varphi$. From the definition of $\alpha_0$ in equation 114, $\alpha_0 = 0$ for such perturbations at such points. Thus the only term which can be nonzero for such perturbations in equation 119 is the last term. But, from the definition of $\beta_0$, we can easily find perturbations of this type which set $\beta_0$ to a positive number or to a negative number. Thus the only way that equation 119 can be nonnegative with respect to such perturbations is if:

$$Q\mathcal{L}_{22} + \mathcal{L}_2 = \mathbf{0} \tag{122}$$

In summary, if $\partial_t r_A = \partial_t r_\varphi = 0$ at any point, then the GLC requires that equation 122 must be obeyed. But on the other hand, suppose that $\partial_t r_A$ 0 or $\partial_t r_\varphi$ 0. Then the first order conditions for a minimum of H', as given in equations 120 and 121, <u>also</u> require that equation 122 must be obeyed. In summary, equation 122 must be obeyed in any case, for there to be any possibility of H' minimization. This reduces equation 119 to:

$$\delta_t^2 \mathcal{H}' = 4\alpha_0^2(Q\mathcal{L}_{222} + 3\mathcal{L}_{22}) + 4\alpha_0(\lambda_A r_A^2 \dot{\xi}_A + \lambda_\varphi r_\varphi^2 \dot{\xi}_\varphi)\mathcal{L}_{22} \qquad (123)$$

From the definition of $\alpha_0$, it is easy to see that we can always find nonzero combinations of $\partial_t \eta_A$ and $\partial_t \eta_\varphi$ which set $\alpha_0$ to zero, and set $\delta_t^2 \mathcal{H}'$ to zero as well. Thus at each point in space, we have at least one degree of freedom in defining a perturbation for which $\delta_t^2 \mathcal{H}' = 0$; in fact, since we are considering perturbations for which $\xi_A = \xi_\varphi = \eta_A = \eta_\varphi = \partial_t \xi_A = \partial_t \xi_\varphi = 0$, we can therefore "connect" this degree of freedom across points in space, such that we have an infinite family of perturbation functions for which $\delta^2 H' = 0$. There are many other constraints which can be derived for this system, but this by itself is sufficient to demonstrate that SSOS is impossible here.

The reader may note that in [4] I recommended the use of the "interaction picture" in order to locate stable oscillatory solitons. Indeed, the calculations with the interaction picture would have been much simpler here. Calculations with the interaction picture provide a simpler path to demonstrating that stable oscillatory solutions exist, when they exist. However, the approach in this section is more powerful in ruling out stable solutions, because it rules out any kind of attractor set which minimizes energy, with the SSOS property.

## 4. A 3-D Example Addressing Rybakov's Conjecture

In [4], I proposed a multi-step strategy for constructing numerical examples of static stable states in 3+1-D relativistic Lagrangian field theories. The first two steps could be carried out independently, in parallel. In one step, we would try to construct stable static examples in a 1+1-D relativistic field theory, drawing from the family $L(f_0, f_2, f_4)$ discussed in section 3. (Actually, a 2-scalar nonrelativistic example in 1+1-D might be worth attempting as a preliminary to this step.) In a parallel step, we would try to construct a local 3-dimensional "energy" functional, V, which is not derived as the Hamiltonian of a Lagrangian field theory. We would try to disprove, by example, Rybakov's conjecture that no local energy functional $V(\underline{\psi}, \nabla \underline{\psi})$ in dimensions $D \geq 3$ can ever have a true local minimum, except on the basis of topological charge[4].

To construct such an example, I proposed that we look for functionals $V(\underline{\psi}, \nabla\underline{\psi})$ of the form $V(f_0, f_2)$, where $\underline{\psi}$ is a real three-dimensional vector field, and where we use new definitions (just for this section, section 4):

$$f_0 = \sum_{i=1}^{3} \psi_i^2$$

$$f_1 = \sum_{ij} (\partial_i \psi_j)^2 = \sum_{i=1}^{3} |\nabla \psi_j|^2 \qquad (124)$$

It is straightforward to deduce:

$$\tfrac{1}{2}\delta^2 V = 2V_{00}(\underline{\psi}\cdot\underline{\xi})^2 + 4V_{01}(\underline{\psi}\cdot\underline{\xi})S + 2V_{11}S^2 + V_0|\underline{\xi}|^2 + V_1 \sum_{j=1}^{3}|\nabla \xi_j|^2 \quad , \qquad (125)$$

where for convenience I define:

$$S \equiv \sum_{ij}(\partial_i \psi_j)(\partial_i \xi_j) \qquad (126)$$

The challenge here is to find a functional V and a state of the function $\underline{\psi}(\underline{x})$ such that $\underline{\psi}(\underline{x})$ provides a local minimum of V*, the integral of V over three-dimensional space. I proposed that we look for such a state by considering functions $\underline{\psi}(\underline{x})$ of the form $\underline{\psi}(\underline{x}) = \underline{x}u(r)$, and then solving the Lagrange-Euler equations in one dimension (r).

The chief difficulty in this approach will be to determine whether the second order conditions for stability are met, for specific numerical solutions based on this form of $\underline{\psi}$. To overcome this difficulty, I proposed that we conduct a complete eigenanalysis of $\delta^2 V^*$. More precisely, I claim that $\delta^2 V^*$ has a complete eigenfunction decomposition based on eigenfunctions of the forms:

$$\underline{\xi}(\underline{x}) = G(r)\underline{x}Y_l^m(\underline{x}) + E(r)\nabla Y_l^m(\underline{x})$$

$$\text{or } = H(r)\left(\underline{x} \times \nabla Y_l^m(\underline{x})\right) \ , \tag{127}$$

where $Y_l^m$ is a spherical harmonic[19].

Section 4.1 will verify this claim. Section 4.2 will show how $\delta^2 V^*$ can be expressed as a functional in $G(r)$ and $E(r)$, to be integrated over one dimension (r), for each value of l. Given such a one-dimensional representation, the methods of section 2.6 can be applied directly. Section 4.3 will fill in one final mathematical detail, associated with Rybakov's conjecture: it will give the derivation for the dipole moments, discussed in [4], which provide at least one loophole in Rybakov's arguments.

## 4.1. Eigenvector Analysis of $\delta^2 V^*$

### 4.1.1. Goals of This Section

The goal of section 4.1 is to verify that equation 127 gives us a complete set of eigenvectors -- or, more precisely, eigenfunctions -- $\underline{\xi}$ for the bilinear form $\delta^2 V^*(\underline{\xi},\underline{\xi})$, which is initially defined as:

$$\delta^2 V^*(\underline{\xi},\underline{\xi}) = \int \delta^2 V(\underline{\xi}(\underline{x}), \nabla \underline{\xi}(\underline{x})) \, d^3\underline{x} \ , \tag{128}$$

where $\delta^2 V$ is defined as in equation 125. For the moment, I suppress the dependence on $\underline{\psi}$, because $\underline{\psi}$ is treated as fixed in this perturbation analysis.

The first step is to observe that $\delta^2 V^*(\underline{\xi},\underline{\xi})$ can always be expressed equivalently as:

$$\delta^2 V^* = \int \underline{\xi}^T(\underline{x}) \underline{\eta}(\underline{x}) \, d^3\underline{x} \tag{129}$$

where:

$$\underline{\eta} = M\underline{\xi} \ , \tag{130}$$

for some linear operator M, obtained by applying integration by parts to equation 125. Notice that I am using "$\underline{\xi}$" without an argument "$\underline{x}$" as a way of referring to the entire function $\underline{\xi}$ across all space; in other words, it refers to something like a vector in Hilbert space. Thus in physicists' notation:

$$\delta^2 V^*(\underline{\xi},\underline{\xi}) = \langle \underline{\xi} | M\underline{\xi} \rangle \tag{131}$$

From equation 131, $\delta^2 V^*$ will be partially positive definite ("SSOS") if and only if the operator M is. (Recall [4] that "SSOS" means that $\delta^2 V^*$ is positive definite except in directions corresponding to a rotation, translation or gauge transformation; in those directions, $\delta^2 V^*$ is zero.) Furthermore, $\delta^2 V^*$ will be SSOS if and only if the eigenvalues of M along the eigenvector directions have positive $\delta^2 V^*$ (except in directions where $\delta^2 V^*$ must be zero). Thus to prove that $\delta^2 V^*$ is SSOS, for a particular choice of functional V, it would be enough to prove: (1) that functions $\underline{\xi}$ of the form of equation 127 provide a complete set of eigenvectors (eigenfunctions) of M; and (2) that $\delta^2 V^*$ is positive definite (or rather SSOS) with respect to that particular class of functions. This section will prove the first part, in general, for all choices of V, for $\underline{\psi}$ of the form $\underline{\psi}=\underline{x}u(r)$. Section 4.2 will describe how to prove or disprove the second part, numerically, for particular choices of V. The test of section 4.2 will be "decisive" for all systems where $\underline{\psi}=\underline{x}u(r)$; in other words, it will be passed if and only if the system is SSOS, when $\underline{\psi}$ is of this form.

## 4.1.2. Derivation of M and of Its Eigenvector Equation

It is possible to derive M and its eigenvector equation by brute force, by applying integration by parts to equation 128 directly, using the formal strategies of first-year calculus. However, there exists an equivalent approach which is more straightforward and easier to apply. We can derive the eigenvector equation directly, as the system of Lagrange-Euler equations for the minimization of:

$$L = \underline{\xi}^T Q(\lambda) \underline{\xi} = \tfrac{1}{2} \underline{\xi}^T V^* \underline{\xi} - \lambda |\underline{\xi}|^2 = \int V'(\underline{x})\, d^3\underline{x}\quad , \tag{132}$$

where

$$V'(\underline{x}) = \tfrac{1}{2} \delta^2 V(\underline{\xi}(\underline{x}), \nabla \underline{\xi}(\underline{x})) - \lambda |\underline{\xi}(\underline{x})|^2 \tag{133}$$

In other words, the system of Lagrange-Euler equations for minimizing the integral of V' are the eigenvector equation we seek, for eigenvalue $\lambda$. Intuitively, this follows directly from the fact that the usual Lagrange-Euler equations are simply the result of an integration by parts of functionals such as equation 133 [6].

More formally, we know that the bilinear form $\tfrac{1}{2}V^*$ is real and symmetric. Therefore, its eigenvectors are orthogonal and its eigenvalues are real, and the same must be true of $Q(\lambda)$ for all real $\lambda$. (More precisely, we know that the form possesses at least one such eigenvector decomposition.) Any eigenvector of $\tfrac{1}{2}V^*$ of eigenvalue $\lambda$ must be an eigenvector of $Q(\lambda)$ with eigenvalue zero. But for such an eigenvector, it is well known that the first variation of $\underline{\xi}^T Q(\lambda) \underline{\xi}$ must equal zero. (This follows from the fact that a small perturbation $\underline{\delta}$ of $\underline{\xi}$ can be written as the sum of two parts -- a part for which $\underline{\delta}$ is orthogonal to $\underline{\xi}$, where $\underline{\delta}^T Q(\lambda) \underline{\xi}=0$, and a part in the direction of $\underline{\xi}$, for which $Q(\lambda)\underline{\delta}=0$.) Since all eigenvectors of $Q(\lambda)$ must obey the first-order conditions for minimizing equation 132, for the corresponding eigenvalue $\lambda$, they must also obey the Lagrange-Euler equations for that minimization problem.

In summary, we could derive the eigenvector equation for $\delta^2 V^*$ in the general case simply by substituting equation 125 into equation 133, and working out the corresponding Lagrange-Euler equations. However, our goal in this section is not to address the general case, for all possible functions $\underline{\psi}$; rather, we are focusing on the special case of functions $\underline{\psi}$ of the form $\underline{x}u(r)$.

In order to work out the eigenvector equation for this special case, the obvious procedure is simply to substitute $\underline{\psi}=\underline{x}u(r)$ into equation 125, and then substitute the result into equation 133, and then work out the Lagrange-Euler equations. However, it is possible to make equation 125 more tractable by decomposing the function $\underline{\xi}(\underline{x})$ into two components, which I would call the <u>radial</u> ($\underline{\xi}^R(\underline{x})$) and <u>antiradial</u> ($\underline{\xi}^A(\underline{x})$) components, defined as follows:

$$\underline{\xi}^R(\underline{x}) = \frac{(\underline{x}\cdot\underline{\xi})}{r^2}\underline{x} = \underline{x}\,g(\underline{x})$$
$$\underline{\xi}^A(\underline{x}) = \underline{\xi}(\underline{x}) - \underline{\xi}^R(\underline{x}) \tag{134}$$

Note that equations 134 are simply a Gramm-Schmidt decomposition of $\underline{\xi}$ into two subspaces, one being the line which connects $\underline{x}$ to the origin, and the other being the plane orthogonal to that line. Also note that equations 134 define a new auxiliary quantity, $g(\underline{x})$, which characterizes the <u>radial</u> part of the perturbation $\underline{\xi}$. Notice that this decomposition of a vector field into radial and antiradial components can be applied to <u>any</u> 3-D vector field here.

Our first major task is to rewrite equation 125 for the case $\underline{\psi}=\underline{x}u(r)$, <u>with</u> $\underline{\xi}(\underline{x})$ expressed in terms of $g(\underline{x})$ and $\underline{\xi}^A(\underline{x})$. This requires many auxiliary calculations, which I will present in order, starting from definitions. Note that subscripts attached to the scalars V, g and u will indicate differentiation. Also note that $(\underline{x},\underline{v})=0$ for any antiradial vector $\underline{v}$, a fact which I exploit over and over again below.

$$\psi_j = x_j u(r) \qquad j=1,2,3 \tag{135}$$

$$\xi_j = x_j g(\underline{x}) + \xi_j^A(\underline{x}) \qquad j=1,2,3 \tag{136}$$

$$\partial_i \psi_j = \delta_{ij} u + \frac{x_i x_j}{r} u_r \tag{137}$$

$$(\underline{\psi} \cdot \underline{\xi}) = \sum_{i=1}^{3} \psi_i \xi_i = \sum_{i=1}^{3} \left( x_j^2 u(r) g(\underline{x}) + x_j u(r) \xi_j^A \right)$$
$$= u(r) g(\underline{x}) \sum_{i=1}^{3} x_i^2 + u(r) \sum_{i=1}^{3} x_i \xi_i^A = r^2 u(r) g(\underline{x}) \tag{138}$$

$$\left| \underline{\xi} \right|^2 = \left| \underline{x} g(\underline{x}) \right|^2 + \left| \underline{\xi}^A \right|^2 = r^2 g^2(\underline{x}) + \left| \underline{\xi}^A \right|^2 \tag{139}$$

$$S = \sum_{i,j} (\partial_i \psi_j)(\partial_i \xi_j) = \sum_{i,j} \left( \delta_{ij} u (\partial_i \xi_j) + \frac{x_i x_j}{r} u_r (\partial_i \xi_j) \right)$$
$$= u \sum_i (\partial_i \xi_i) + \frac{u_r}{r} \sum_{i,j} x_i x_j (\partial_i \xi_j) \tag{140}$$

$$u \sum_i \partial_i \xi_i = u \sum_i \partial_i \left( x_i g(\underline{x}) + \xi_i^A \right) = u \sum_i \left( g(\underline{x}) + x_i g_i + \partial_i \xi_i^A \right)$$
$$= 3ug + urg_r + u \sum_i \partial_i \xi_i^A \tag{141}$$

$$\sum_{i,j} x_i x_j (\partial_i \xi_j) = \sum_{i,j} x_i x_j (\partial_i (x_j g(\underline{x}) + \xi_j^A))$$
$$= \sum_{i,j} x_i x_j \left( \delta_{ij} g + x_j g_i + \partial_i \xi_j^A \right) = \sum_i \left( x_i^2 g + x_i r^2 g_i + x_i \sum_j x_j \partial_i \xi_j^A \right) \tag{142}$$
$$= r^2 g + r^3 g_r + \sum_i x_i \sum_j x_j \partial_i \xi_j^A$$

$$0 = \partial_i (0) = \partial_i \left( \sum_j x_j \xi_j^A \right) = \sum_j \partial_i (x_j \xi_j^A)$$
$$= \sum_j \left( \delta_{ij} \xi_j^A + x_j \partial_i \xi_j^A \right) = \xi_i^A + \sum_j x_j \partial_i \xi_j^A \tag{143}$$

Substituting equation 143 into equation 142:

$$\sum_{i,j} x_i x_j (\partial_i \xi_j) = r^2 g + r^3 g_r + \sum_i x_i (-\xi_i^A) = r^2 g + r^3 g_r \tag{144}$$

Substituting equations 144 and 141 into 140:

$$S = (3u + ru_r) g + (ur + r^2 u_r) g_r + u \sum_i \partial_i \xi_i^A \tag{145}$$

For the last unresolved term in equation 125, we may begin as in equation 142, to get:

$$\sum_{i,j}(\partial_i\xi_j)^2 = \sum_{i,j}(\delta_{ij}g + x_j g_i + \partial_i\xi_j^A)^2$$
$$= \sum_{i,j}(\delta_{ij}g^2 + 2\delta_{ij}x_j gg_i + 2\delta_{ij}g\partial_i\xi_j^A + x_j^2 g_i^2 + 2x_j g_i\partial_i\xi_j^A + (\partial_i\xi_j^A)^2) \quad (146)$$
$$= 3g^2 + \sum_i\left(2x_i gg_i + 2g\partial_i\xi_i^A + r^2 g_i^2 + 2g_i\sum_j x_j\partial_i\xi_j^A + \sum_j(\partial_i\xi_j^A)^2\right)$$

Exploiting equation 143 and performing further algebra, we get:

$$\sum_{i,j}(\partial_i\xi_j)^2 = 3g^2 + 2rgg_r + 2g\,\mathrm{div}\,\underline{\xi}^A + r^2\sum_i g_i^2 - 2\sum_i g_i\xi_i^A + \sum_{i,j}(\partial_i\xi_j^A)^2 \quad (147)$$

Finally, we may substitute the results above into equations 125 and 133, to derive:

$$V' = 2V_{00}r^4 u^2 g^2 + 4V_{01}r^2 ugS + 2V_{11}S^2 + (V_0 - \lambda)(r^2 g^2 + |\underline{\xi}^A|^2) + V_1\sum_{i,j}(\partial_i\xi_j)^2, \quad (148)$$

where S and the sum in the last term now represent the auxiliary quantities given in equations 145 and 147.

As discussed above, the eigenvector equation for $\delta^2 V^*$ is just the set of Lagrange-Euler equations for equation 148. (These are somewhat redundant, because $\xi^A$ is constrained, but this does not affect the validity of the usual procedures here.) More precisely, we want the Lagrange-Euler equations for minimizing the integral of V' over space with respect to the two functions, $\xi^A$ and g, which we now use to characterize the perturbations.

To derive the Lagrange-Euler equations for the variables $\xi_j^A$, we begin by deriving the usual conjugate momenta:

$$\Pi^i_{\xi_j^A} = \frac{\delta V'}{\delta(\partial_i\xi_j^A)} = (4V_{01}r^2 ug + 4V_{11}S)\frac{\delta S}{\delta(\partial_i\xi_j^A)} + V_1\frac{\delta}{\delta(\partial_i\xi_j^A)}\left(\sum_{i,j}(\partial_i\xi_j)^2\right) \quad (149)$$

Differentiating equations 145 and 147 with respect to $\partial_i\xi_j^A$, and substituting into equation 149, we get:

$$\Pi^i_{\xi_j^A} = (4V_{01}r^2 ug + 4V_{11}S)u\delta_{ij} + 2V_1(g\delta_{ij} + \partial_i\xi_j^A) \quad (150)$$

This leads to a Lagrange-Euler equation of:

$$\sum_i \partial_i\left(\delta_{ij}(4V_{01}r^2 u^2 g + 4V_{11}Su + 2V_1 g) + 2V_1\partial_i\xi_j^A\right)$$
$$= \frac{\delta V'}{\delta\xi_j^A} = 2(V_0 - \lambda)\xi_j^A + V_1(-2g_j) \quad (151)$$

This reduces to:

$$\partial_j\left(2V_{01}r^2 u^2 g + 2V_{11}uS + V_1 g\right) + \sum_i \partial_i(V_1\partial_i\xi_j^A) = (V_0 - \lambda)\xi_j^A - V_1 g_j \quad (152)$$

The second major term on the left may be further expanded:

$$\sum_i \partial_i(V_1 \partial_i \xi_j^A) = \sum_i V_1 \partial_i^2 \xi_j^A + \sum_i (\partial_i V_1)(\partial_i \xi_j^A)$$

$$= V_1 \Delta \xi_j^A + \sum_i (\frac{x_i}{r}(\partial_r V_1))(\partial_i \xi_j^A) = V_1 \Delta \xi_j^A + \frac{\partial_r V_1}{r}\sum_i x_i \partial_i \xi_j^A \quad (153)$$

$$= V_1 \Delta \xi_j^A + \frac{\partial_r V_1}{r}(r \partial_r \xi_j^A) = V_1 \Delta \xi_j^A + (\partial_r V_1)(\partial_r \xi_j^A)$$

Using equations 153 and 145, we may reduce equation 152 to the primary eigenvector equation:

$$\nabla\left(f_1(r)g + f_2(r)g_r + 2V_{11}u^2 \text{ div } \underline{\xi}^A\right) + V_1 \Delta \underline{\xi}^A + (\partial_r V_1)(\partial_r \underline{\xi}^A) = (V_0 - \lambda)\underline{\xi}^A - V_1 \nabla g \quad (154)$$

where I define:

$$\begin{aligned} f_1(r) &= 2V_{01}r^2u^2 + V_1 + 2V_{11}u(3u + ru_r) \\ f_2(r) &= 2V_{11}u(ur + r^2 u_r) \end{aligned} \quad (155)$$

Note the importance of the fact that these are all functions of r (i.e. all spherically symmetric) because of the assumption that $\underline{\psi}(\underline{x})$ takes the form $\underline{x}u(r)$ in this particular exercise.

Next, to derive the Lagrange-Euler equation for g, we again begin by deriving the conjugate term:

$$\Pi_g^i = \frac{\delta V'}{\delta(\partial_i g)} = (4V_{01}r^2 ug + 4V_{11}S)\frac{\delta S}{\delta(\partial_i g)} + V_1 \frac{\delta}{\delta(\partial_i g)}\left(\sum_{i,j}(\partial_i \xi_2)^2\right) \quad (156)$$

which expands to:

$$\begin{aligned} \tfrac{1}{2}\Pi_g^i &= (2V_{01}r^2 ug + 2V_{11}S)(\frac{x_i}{r}(ur + r^2 u_r)) + \tfrac{1}{2}V_1(\frac{x_i}{r}(2rg) + 2r^2 g_i - 2\xi_i^A) \\ &= x_i p_1(r)g + x_i p_2(r)g_r + x_i p_3(r) \text{ div } \underline{\xi}^A + V_1 r^2 g_i - V_1 \xi_i^A \end{aligned} \quad (157)$$

where I have used the definition in equation 145 again and defined:

$$\begin{aligned} p_1(r) &= V_1 + (u + ru_r)(2V_{01}r^2 u + 2V_{11}(3u + ru_r)) \\ p_2(r) &= (u + ru_r)(2V_{11}(ur + r^2 u_r)) \\ p_3(r) &= (u + ru_r)(2V_{11}u) \end{aligned} \quad (158)$$

With these definitions, the Lagrange-Euler equation for g may be written as:

$$\tfrac{1}{2}\frac{\delta V'}{\delta g} = \sum_i \partial_i \left( x_i p_1(r)g + x_i p_2(r)g_r + x_i p_3(r) \text{ div } \underline{\xi}^A + V_1 r^2 g_i - V_1 \xi_i^A \right)$$

$$= p_1(r)g + p_2(r)g_r + p_3(r) \text{ div } \underline{\xi}^A$$

$$+ \sum_i x_i \left( \frac{x_i}{r} \partial_r \left( p_1(r)g + p_2(r)g_r + p_3(r) \text{ div } \underline{\xi}^A \right) \right)$$

$$+ \sum_i \frac{x_i}{r} \left( \partial_r (V_1 r^2) g_i - \partial_r (V_1) \xi_i^A \right) + V_1 r^2 \Delta g - V_1 \text{ div } \underline{\xi}^A$$

$$= p_1(r)g + p_2(r)g_r + (p_3(r) - V_1) \text{ div } \underline{\xi}^A + V_1 r^2 \Delta g$$

$$+ r \partial_r \left( p_1(r)g + p_2(r)g_r + p_3(r) \text{ div } \underline{\xi}^A \right) \tag{159}$$

$$+ \frac{\partial_r (V_1 r^2)}{r} \sum_i x_i g_i - \frac{\partial_r V_1}{r} \sum_i x_i \xi_i^A$$

Collecting terms, this reduces to:

$$g\left(p_1(r) + r(\partial_r p_1)\right) + g_r\left(p_2(r) + rp_1(r) + r(\partial_r p_2) + \partial_r(V_1 r^2)\right) + g_{rr}(rp_2(r))$$
$$+ V_1 r^2 \Delta g + \text{ div } \underline{\xi}^A \left(p_3(r) - V_1 + r(\partial_r p_3)\right) + rp_3(r) \partial_r (\text{div } \underline{\xi}^A) = \tfrac{1}{2}\frac{\delta V'}{\delta g} \tag{160}$$

The right-hand side of this may also be expanded:

$$\tfrac{1}{2}\frac{\delta V'}{\delta g} = \tfrac{1}{2}\begin{pmatrix} 4V_{00}r^4 u^2 g + 4V_{01}r^2 uS + 2(V_0 - \lambda)r^2 g \\ + (4V_{01}r^2 ug + 4V_{11}D)\frac{\delta S}{\delta g} + V_1 \frac{\delta}{\delta g}\sum_{i,j}(\partial_i \xi_j)^2 \end{pmatrix}$$

$$= 2V_{00}r^4 u^2 g + 2V_{01}r^2 uS + (V_0 - \lambda)r^2 g \tag{161}$$
$$+ (2V_{01}r^2 ug + 2V_{11}S)(3u + ru_r) + \tfrac{1}{2}V_1(6g + 2rg_r + 2 \text{ div } \underline{\xi}^A)$$

Collecting terms, we get:

$$\tfrac{1}{2}\frac{\delta V'}{\delta g} = g\left(2V_{00}r^4 u^2 + (V_0 - \lambda)r^2 + 2V_{01}r^2 u(3u + ru_r) + 3V_1\right)$$
$$+ S\left(2V_{01}r^2 u + 2V_{11}(3u + ru_r)\right) + rV_1 g_r + V_1 \text{div } \underline{\xi}^A \tag{162}$$

Substituting in from the definition of S from equation 145, this expands to:

$$\tfrac{1}{2}\frac{\delta V'}{\delta g} = s_1(r)g + s_2(r)g_r + s_3(r)\text{div } \underline{\xi}^A \quad, \tag{163}$$

where I define:

$$s_1(r) = 2V_{00}r^4u^2 + (V_0 - \lambda)r^2 + 2V_{01}r^2u(3u + ru_r) + 3V_1$$
$$+ \left(2V_{01}r^2u + 2V_{11}(3u + ru_r)\right)(3u + ru_r) \tag{164}$$
$$= 3V_1 + 2V_{11}(3u + ru_r)^2 + 4V_{01}r^2u(3u + ru_r) + 2V_{00}r^4u^2 + (V_0 - \lambda)r^2$$

$$s_2(r) = rV_1 + \left(2V_{01}r^2u + 2V_{11}(3u + ru_r)\right)(ur + r^2u_r)$$
$$s_3(r) = V_1 + \left(2V_{01}r^2u + 2V_{11}(3u + ru_r)\right)u \tag{165}$$

When we substitute equation 163 back into equation 160, the final form of the Lagrange-Euler equation for g becomes:

$$h_1(r)g + h_2(r)g_r + rp_2(r)g_{rr} + V_1r^2\Delta g + h_3(r)\,\text{div}\,\underline{\xi}^A + rp_3(r)\partial_r(\text{div}\,\underline{\xi}^A) = 0 \quad , \tag{166}$$

where I define:

$$h_1(r) = p_1(r) + r(\partial_r p_1) - s_1(r)$$
$$h_2(r) = p_2(r) + rp_1(r) + r(\partial_r p_2) + \partial_r(V_1r^2) - s_2(r) \tag{167}$$
$$h_3(r) = p_3(r) - V_1 + r(\partial_r p_3) - s_3(r)$$

By substituting the definitions of the functions $p_i$ and $s_i$ back into equation 167, one might hope for a relatively tractable, more direct expression of $h_1$, $h_2$ and $h_3$ in terms of our original functions $u(r)$, $V_0(r)$, etc. However, that is not important for the goals of thise section. Furthermore, the computations involved in calculating $h_1$, $h_2$ and $h_3$ via the equations above, for specific numerical examples, would not be difficult. The reader should be warned that I have not tested these equations numerically, the way one should before performing extensive calculations. My goal here is to prove some general conclusions, which do not depend heavily on the numerical details.

### 4.1.3. Eigenfunction Decomposition

The purpose of section 4.1 is to show that $\frac{1}{2}\delta^2V^*$ has a complete eigenfunction decomposition based on eigenfunctions of the form given in equations 127.

To begin with, recall the well-known fact that the spherical harmonics, $Y_l^m(\theta,\phi)$, provide a complete basis system for functions defined on the surface of the sphere[19]. Thus any well-behaved localized 3-D vector field $\underline{\psi}(\underline{x})$ can be decomposed as:

$$\underline{\psi}(\underline{x}) = \sum_{\ell,m} \underline{a}_{\ell m}(r)Y_\ell^m(\theta,\phi) \tag{168}$$

However, this is not quite a suitable decomposition for our purposes here. For our purposes, it is important to decompose the vector $\underline{\psi}(\underline{x})$ at each point into its radial and antiradial components. To do this, we can use the alternative decomposition:

$$\underline{\psi}(\underline{x}) = \sum_{\ell,m} \left(G_{\ell m}(r)\underline{x}Y_\ell^m + E_{\ell m}(r)\nabla Y_\ell^m + H_{\ell m}(r)(\underline{x} \times \nabla Y_\ell^m)\right) = \sum_{\ell,m} \underline{\psi}^{[\ell m]}(\underline{x}) \tag{169}$$

Use of the vector cross product insures that we have all three orthogonal directions represented, so long as $|\nabla Y_l^m| \neq 0$. In fact, $Y_l^m$ has a nonzero gradient almost everywhere except for the case l=m=0, where $Y_l^m$ is just a constant. Equation 169 does include two useless (zero) basis functions in that case; however, they do not affect the conclusions which follow. (Formally, they reflect the fact that continuous fields of antiradial vectors -- vectors tangent to

the sphere -- over the sphere always possess fixed points; thus the only "constant term" possible here is a term for radial vector fields.)

When we insert equation 169 back into the eigenvector equation for $\frac{1}{2}\delta^2 V^*$ -- equations 154 and 166 -- we will find that the equations for each component $\underline{\psi}^{[lm]}$ are independent of the values of $\underline{\psi}^{[l'm']}$ for (l',m') (l,m)! This shows that the eigenvector equation can be solved only if it can be solved separately for each combination of (l,m). Since we know that $\frac{1}{2}\delta^2 V^*$ must have a complete set of eigenfunctions, this proves that it has a complete eigenfunction decomposition based on the combination of the eigenfunctions for each (l,m) combination. Each of those eigenfunctions has the form for $\underline{\psi}^{[lm]}$ given in equation 169. Thus it proves that $\delta^2 V^*$ has a complete set of eigenfunctions of that form.

To complete the proof of the initial claim of this section, I must now do two things: (1) actually perform the insertion described in the previous paragraph, to validate the first sentence of that paragraph; (2) for any combination (l,m), show that the eigenfunctions can be further decomposed into the two forms given in equation 127 (which are a special case of equation 169).

The first step involves a straightforward but tedious substitution of equation 169 into the two eigenvector equations, starting with equation 154. I will substitute only one $\underline{\psi}^{[lm]}$ component, and show (as claimed) that it does not generate any (l',m') terms for (l',m') (l,m). First, we must carry out a series of preliminary calculations. I will suppress the indices l and m in most cases, in order to simplify the appearance of the equations. From the definition of the radial decomposition, it follows immediately that:

$$g(\underline{x}) = G(r)Y(\theta,\phi) = G(r)Y(\underline{x})$$
$$\underline{\xi}^A = E(r)\nabla Y + H(r)(\underline{x} \times \nabla Y) \tag{170}$$

Equation 170 makes use of the well-known vector cross-product, which I recall for convenience:

$$\underline{x} \times \nabla Y = (x_2 \partial_3 Y - x_3 \partial_2 Y, x_3 \partial_1 Y - x_1 \partial_3 Y, x_1 \partial_2 Y - x_2 \partial_1 Y) \tag{171}$$

The first nontrivial term in equation 154 is:

$$\mathbf{div}\ \underline{\xi}^A = \sum_i \partial_i \left( E(r)(\partial_i Y) + H(r)(\underline{x} \times \nabla Y)_i \right)$$
$$= \sum_i \frac{x_i}{r} \left( E_r(r) \partial_i Y + H_r(r)(\underline{x} \times \nabla Y)_i \right) + E(r)\Delta Y + H(r) \sum_i \partial_i (\underline{x} \times \nabla Y)_i \tag{172}$$

In working this out, recall that the $\nabla Y$ and $\underline{x} \times \nabla Y$ are antiradial vectors, such that the sum on the left will go to zero. The rightmost term in equation 172 also equals zero, as can be verified by a brute force calculation from equation 171. Finally, the spherical harmonic $Y_l^m$ is well known to be an eigenfunction of the operator $\Delta$ [19]. Putting all this together, we obtain:

$$\mathbf{div}\ \underline{\xi}^A = -E(r)\ell(\ell+1)Y \tag{173}$$

The second nontrivial term in equation 154 is $\Delta \underline{\xi}^A$. It is well known that:

$$\Delta = (\partial_r^2 + \tfrac{2}{r}\partial_r) + \Delta_{\theta,\phi}\quad, \tag{174}$$

where $\Delta_{\theta,\phi}$ is the portion of $\Delta$ which operates on the sphere (e.e., on the $\theta$ and $\phi$ coordinates). Therefore:

$$\Delta\left(E(r)\nabla Y_\ell^m(\theta,\phi)\right) = (E_{rr} + \tfrac{2}{r}E_r)\nabla Y - E\ell(\ell+1)\nabla Y = (E_{rr} + \tfrac{2}{r}E_r - E\ell(\ell+1))\nabla Y$$
$$\Delta(H(r)(\underline{x}\times\nabla Y)) = (H_{rr} + \tfrac{2}{r}H_r)(\underline{x}\times\nabla Y) + H\Delta(\underline{x}\times\nabla Y) \quad (175)$$

Likewise, a brute force calculation based on equation 171 shows that:

$$\Delta(\underline{x}\times\nabla Y) = \sum_i \partial_i\partial_i(\underline{x}\times\nabla Y) = (\underline{x}\times\Delta\nabla Y) = -\ell(\ell+1)(\underline{x}\times\nabla Y) \quad (176)$$

Putting this together, we get:

$$\Delta\underline{\xi}^A = \left(E_{rr} + \tfrac{2}{r}E_r - E\ell(\ell+1)\right)\nabla Y + \left(H_{rr} + \tfrac{2}{r}H_r - H\ell(\ell+1)\right)(\underline{x}\times\nabla Y) \quad (177)$$

Using this information, equation 154 reduces to:

$$\nabla\left(f_1(r)GY + f_2(r)G_r Y - 2V_{11}u^2\ell(\ell+1)EY\right)$$
$$+V_1\left(\left(E_{rr} + \tfrac{2}{r}E_r - E\ell(\ell+1)\right)\nabla Y + \left(H_{rr} + \tfrac{2}{r}H_r - H\ell(\ell+1)\right)(\underline{x}\times\nabla Y)\right) \quad (178)$$
$$+(\partial_r V_1)(E_r\nabla Y + H_r(\underline{x}\times\nabla Y)) = (V_0 - \lambda)(E\nabla Y + H(\underline{x}\times\nabla Y)) - V_1\nabla(GY)$$

The large term on the left side of equation 178 reduces to:

$$\left(f_1(r)G + f_2(r)G_r - 2V_{11}u^2\ell(\ell+1)E\right)\nabla Y + Y\left(\frac{\underline{x}}{r}\partial_r\left(f_1 G + f_2 G_r - 2V_{11}u^2\ell(\ell+1)E\right)\right) \quad (179)$$

It is extremely important that all terms in equation 178 reduce to functions of r multiplied by one of our three basis functions for (l,m): $\underline{x}Y$, $\nabla Y$ and $\underline{x}\times\nabla Y$. This verifies our first main claim above, for the $\underline{\xi}^A$ equation. Furthermore, equation 178 is equivalent to the following set of three scalar equations, one for each of its basis components:

$$\tfrac{1}{r}\partial_r\left(f_1 G + f_2 G_r - 2V_{11}u^2\ell(\ell+1)E\right) = -V_1(\tfrac{1}{r}\partial_r G) \quad (180)$$

$$f_1 G + f_2 G_r - 2V_{11}u^2\ell(\ell+1)E + V_1(E_{rr} + \tfrac{2}{r}E_r - E\ell(\ell+1))$$
$$+ (\partial_r V_1)E_r = (V_0 - \lambda)E - V_1 G \quad (181)$$

$$V_1\left(H_{rr} + \tfrac{2}{r}H_r - H\ell(\ell+1)\right) + (\partial_r V_1)H_r = (V_0 - \lambda)H \quad (182)$$

Once again, by inspection, it is clear that the $\underline{x}\times\nabla Y$ component of the eigenfunction equation is independent of the $\underline{x}$ and $\nabla Y$ components. Thus the expected three eigenfunctions of this system, for any (l,m) combination (except l=m=0), would consist of one eigenfunction from equation 182, of the form $H(r)(\underline{x}\times\nabla Y)$, and two based on equations 180 and 181. All three would fit one of the two forms given in equation 127.

In order to complete the proof, it is necessary to repeat the same sort of substitutions into equation 166. Using the same preliminary calculations as before, we can substitute into equation 166 to get:

$$h_1 GY + h_2 G_r Y + rp_2 G_{rr} Y + V_1 r^2\left(G_{rr} + \tfrac{2}{r}G_r - G\ell(\ell+1)\right)Y$$
$$- h_3 E\ell(\ell+1)Y - rp_3 E_r\ell(\ell+1)Y = 0 \quad (183)$$

As before, all terms in the equation depend only on the eigenfunctions for (l,m).

This time, the equation reduces to only one scalar equation:

$$h_1 G + h_2 G_r + rp_2 G_{rr} + V_1 r^2 \left(G_{rr} + \tfrac{2}{r} G_r - G\ell(\ell+1)\right) - h_3 E\ell(\ell+1) - rp_3 E_r \ell(\ell+1) = 0 \quad (184)$$

This equation goes with equations 180 and 181; in other words, equations 180, 181 and 184 must be satisfied as a set by any eigenfunction of the first form given in equation 127. But eigenfunctions of the second form need only satisfy equation 182.

In summary, by substituting the proposed form of eigenfunctions into the eigenvector equation, we have verified the claims at the beginning of this section. Above all, we have verified that equations 127 provide a complete eigenfunction decomposition of $\delta^2 V^*$.

I would conjecture that this same eigenfunction decomposition would apply to virtually any relativistic field theory, for solutions following the "radial vector ansatz"[4]. A more general and more elegant proof of that is probably possible. Nevertheless, I hope that the specific concrete details above will have some additional value, in developing numerical examples for this specific family of cases, and in performing second-order stability analysis. Conversely, if numerical examples turn out to be impossible, I hope that the analysis above would be a useful starting point in proving so. (for example, one might begin by reconsidering the requirement that $\delta^2 V^*=0$ for translational modes, expressed in terms of these basis functions.)

## 4.2. Second-Order Stability Analysis

The calculations of the previous section are disturbingly complex. Fortunately, they result in a fairly straightforward conclusion: in order to perform a second-order stability analysis for any proposed V*-minimizing state of the form $\underline{\psi}=\underline{x}u(r)$, we need only consider the issue of stability against perturbations of the form given in equation 127. The state will be SSOS stable if and only if it is stable against perturbations of the form $G\underline{x}Y+E\nabla Y$ and it is stable against perturbations of the form $H(\underline{x}\times\nabla Y)$. Both of these simpler stability questions can be decided by a purely one-dimensional analysis, as discussed in section 2.

Section 2 has already spelled out exactly how to perform such a one-dimensional stability analysis, once the functional to be minimized has been specified. This section will derive the functionals to be minimized in one dimension, in order to evaluate the stability of $\delta^2 V^*$ against the two families of perturbations.

First, to evaluate stability against the first family of perturbations, we need only insert $\underline{\xi}=G\underline{x}Y+E\nabla Y$ into the integral which we are trying to minimize (where we now set $\lambda=0$):

$$\tfrac{1}{2}\delta^2 V^* = \int_0^\infty r^2 dr \iint V'(r,\theta,\phi)(\Omega d\theta d\phi) \quad, \qquad (185)$$

where $(r^2\Omega)drd\theta d\phi$ is the usual volume integral measure for spherical coordinates. We need to derive the functional which we wll try to minimize across one-dimensional space (r):

$$V'' = r^2 \iint V'(r,\theta,\phi)(\Omega d\theta d\phi) \qquad (186)$$

To do this, first insert $\underline{\xi}=G\underline{x}Y+E\nabla Y$ (i.e., $g=GY$ and $\underline{\xi}^A=E\nabla Y$) into the expression for V' given in equation 148. A direct substitution yields:

$$V' = 2V_{00}r^4 u^2 G^2 Y^2 + 4V_{01}r^2 uGYS + 2V_{11}S^2 + V_0(r^2 G^2 Y^2 + E^2|\nabla Y|^2) + V_1 \sum_{i,j}(\partial_i \xi_j)^2 \quad (187)$$

Before integrating this, we must also derive expansions for S and for the sum in the last term. Substituting our form for $\underline{\xi}$ into equation 145, and exploiting equation 173, we get:

$$S = (3u + ru_r)GY + (ur + r^2 u_r)G_r Y + u(-E\ell(\ell+1)Y) \tag{188}$$

Substituting our form for $\underline{\xi}$ into equation 147, we get:

$$\sum_{i,j}(\partial_i \xi_j)^2 = 3G^2 + 2rGG_r Y^2 + 2GY(-E\ell(\ell+1)Y) + r^2 \sum_i g_i^2 - 2\sum_i g_i \xi_i^A + \sum_{i,j}(\partial_i \xi_j^A)^2 \tag{189}$$

To finish working this out, we need some further calculations:

$$g_i = \partial_i g = \partial_i (GY) = (\partial_i G)Y + G\partial_i Y = (\frac{x_i}{r}G_r)Y + G(\partial_i Y) \tag{190}$$

$$\nabla g = \frac{G_r}{r}\underline{x}Y + G\nabla Y \tag{191}$$

Because $\nabla Y$ is antiradial (i.e. $(\underline{x},\nabla Y)=0$), we may deduce:

$$\sum_i g_i^2 = |\nabla g|^2 = \frac{G_r^2}{r^2} r^2 Y^2 + G^2 |\nabla Y|^2 = G_r^2 Y^2 + G^2 |\nabla Y|^2 \tag{192}$$

Likewise:

$$\sum_i g_i \xi_i^A = (\nabla g \cdot \underline{\xi}^A) = \left(\left(\frac{G_r}{r}\underline{x}Y + G\nabla Y\right) \cdot E\nabla Y\right) = GE|\nabla Y|^2 \tag{193}$$

Finally we work out the most difficult term in equation 189:

$$\sum_{i,j}(\partial_i \xi_j^A)^2 = \sum_{i,j}(\partial_i(E\partial_j Y))^2 = \sum_{i,j}((\partial_i E)(\partial_j Y) + E(\partial_i \partial_j Y))^2$$

$$= \sum_{i,j}((\frac{x_i}{r}E_r)(\partial_j Y) + E(\partial_i \partial_j Y))^2$$

$$= \sum_{i,j}((\frac{x_i^2}{r^2}E_r^2)(\partial_j Y)^2 + 2(\frac{x_i}{r}E_r)(\partial_j Y)E(\partial_i \partial_j Y) + E^2(\partial_i \partial_j Y)^2)$$

$$= E_r^2 |\nabla Y|^2 + \frac{2EE_r}{r}\sum_{i,j} x_i(\partial_j Y)(\partial_i \partial_j Y) + \sum_{i,j} E^2(\partial_i \partial_j Y)^2 \tag{194}$$

The middle term here is zero, because:

$$2\sum_{i,j} x_i(\partial_j Y)(\partial_i \partial_j Y) = \sum_{i,j} x_i \partial_i (\partial_j Y)^2 = \sum_i x_i \partial_i \sum_j (\partial_j Y)^2 = r\partial_r |\nabla Y|^2 \tag{195}$$

Combining all these results together, we may deduce that:

$$V' = I_0 Y^2 + I_1 |\nabla Y|^2 + V_1 E^2 \sum_{i,j}(\partial_i \partial_j Y)^2 \quad , \tag{196}$$

where:

$$I_0(r) = 2V_{00} r^4 u^2 G^2 + 4V_{01} r^2 u G S_0(r) + 2V_{11} S_0^2(r) + V_0 r^2 G^2$$
$$+ V_1\left(3G^2 + 2rGG_r - 2GE\ell(\ell+1) + r^2 G_r^2\right)$$

$$S_0(r) = (3u + ru_r)G + (ur + r^2 u_r)G_r + -uE\ell(\ell+1) \tag{197}$$
$$I_1(r) = V_0 E^2 + V_1(r^2 G^2 - GE + E_r^2)$$

Finally, in order to calculate V", we need to integrate V' over the sphere, for a constant r. For purposes of stability analysis, positive constant scaling factors do not affect the conclusions; thus we can safely assume that the spherical harmonics have been normalized to an integral of one over the sphere dΩ. The remaining integrals required here are familiar and elementary; we can work them out by integration by parts in schematic form:

$$\partial_i(Y\partial_i Y) = (\partial_i Y)^2 + Y(\partial_i^2 Y)$$
$$\int \partial_i(Y\partial_i Y) = 0 = \int (\partial_i Y)^2 + \int Y(\partial_i^2 Y)$$
$$\sum_i \int (\partial_i Y)^2 = -\sum_i \int Y(\partial_i^2 Y) = -\int Y\Delta Y = \ell(\ell+1)\int Y^2 = \ell(\ell+1) \tag{198}$$

$$\partial_i((\partial_j Y)\partial_i(\partial_j Y)) = (\partial_i \partial_j Y)^2 + (\partial_j Y)\partial_i^2(\partial_j Y)$$
$$\sum_i \int (\partial_i \partial_j Y)^2 = \sum_i \int (\partial_j Y)\partial_i^2(\partial_j Y) = -\int (\partial_j Y)\Delta(\partial_j Y) = -\int (\partial_j Y)\partial_j \Delta Y = \ell(\ell+1)\int (\partial_j Y)^2$$
$$\sum_{i,j} \int (\partial_i \partial_j Y)^2 = \ell(\ell+1)\sum_j (\partial_j Y)^2 = \ell^2(\ell+1)^2 \tag{199}$$

Substituting 198 and 199 into 196, we get:

$$V'' = \left(I_0(r) + \ell(\ell+1)I_1(r) + V_1 E^2 \ell^2(\ell+1)^2\right)r^2 \tag{200}$$

Our task is to decide whether the integral of V", over r ranging from 0 to ∞, is SSOS definite as a function of the two functions E(r) and G(r). We can proceed here exactly as we did in section 2, simply by working out the Lagrange-Euler equations for this minimization problem, and so on. The analysis must be repeated, however, for different values of l, as discussed in [4].

For completeness, we must also perform a similar stability analysis for perturbations of the form:
$$\underline{\xi} = H(r)(\underline{x} \times \nabla Y) \tag{201}$$

However, equation 173 tells us that div $\xi^A$ is zero for perturbations of this sort. Of course, g is also zero. Thus from equation 145, we see that S must be zero. Equation 148 (with λ=0) reduces to:

$$V' = V_0 H^2 |\underline{x} \times \nabla Y|^2 + V_1 \sum_{i,j} (\partial_i \xi_j)^2 \tag{202}$$

Likewise, equation 147 reduces to:

$$\sum_{i,j}(\partial_i \xi_j)^2 = \sum_{i,j}(\partial_i \xi_j^A)^2 = \sum_{i,j}\left((\frac{x_i}{r}\partial_r H)(\underline{x}\times\nabla Y)_j + H\partial_i(\underline{x}\times\nabla Y)_j\right)^2$$
$$= H_r^2|\underline{x}\times\nabla Y|^2 + H^2 \sum_{i,j}\left(\partial_i(\underline{x}\times\nabla Y)_j\right)^2 + \frac{2HH_r}{r}\sum_{i,j} x_i(\underline{x}\times\nabla Y)_j\left(\partial_i(\underline{x}\times\nabla Y)_j\right) \tag{203}$$

The last term drops out, because of a result parallel to equation 195; thus equation 202 expands to:

$$V' = (V_0 H^2 + V_1 H_r^2)|\underline{x} \times \nabla Y|^2 + H^2 \sum_{i,j}(\partial_i(\underline{x} \times \nabla Y)_j)^2 \tag{204}$$

Assuming that **r**×**p**=**L**, as in ordinary quantum mechanics[19], this yields (to within an irrelevant scaling factor):

$$V'' = r^2\left(\ell(\ell+1)(V_0 H^2 + V_1 H_r^2) + H^2 \ell^2 (\ell+1)^2\right) \quad, \tag{205}$$

an expression which clearly has some possibility of being positive definite when integrated over r. As with equation 200, the numerical methods of section 2 can be used to decide whether the system is stable. It is important to recall that these particular modes of perturbation are meaningful only for integers l≥1.

The Lagrange-Euler equations derived from equations 200 and 205 should match the eigenvector equations of section 4.1.3 (with λ=0). This provides a kind of double-check of calculations, which would be essential for anyone actually constructing this kind of numerical example.

### 4.3 Dipole Moments and Rybakov's Argument

In Appendix D of [1], Rybakov argues that V*-minimizing states cannot exist in ordinary 3-D vector fields, or in a wide variety of other "topologically trivial" field theories[4]. His argument makes a number of implicit assumptions, which put his final conclusion in doubt. Among those assumptions is the asumption (discussed in [4]) that the following integral can never equal zero across all combinations of i, j, k and l:

$$\mu_{jkl}^i = \int B_{jl}^i(\underline{x}) x_k \, d^3\underline{x} \quad, \tag{206}$$

where B is a special auxiliary quantity [1,4]. For the vector field theory $V(f_0, f_1)$ described above, his definition for B becomes:

$$2B_{jl}^i = \sum_k \left((\partial_j \Pi_k^i)(\partial_l \psi_k) - (\partial_l \Pi_k^i)(\partial_j \psi_k)\right) \quad, \tag{207}$$

where:

$$\Pi_k^i = \frac{\delta V}{\delta(\partial_i \psi_k)} = V_0 \frac{\delta f_0}{\delta(\partial_i \psi_k)} + V_1 \frac{\delta f_1}{\delta(\partial_i \psi_k)} = 2V_1(\partial_i \psi_k) \tag{208}$$

Substituting equation 208 into 207, we derive:

$$B_{jl}^i = \sum_k ((\partial_j V_1)(\partial_i \psi_k)(\partial_l \psi_k) + V_1(\partial_j \partial_i \psi_k)(\partial_l \psi_k) \\ - (\partial_l V_1)(\partial_i \psi_k)(\partial_j \psi_k) + V_1(\partial_l \partial_i \psi_k)(\partial_j \psi_k)) \tag{209}$$

Exploiting the assumption that $\underline{\psi}=\underline{x}u(r)$, for the examples considered in this section, we get:

$$B_{jl}^i = \frac{x_j}{r}(\partial_r V_1)\sum_k (\partial_i \psi_k)(\partial_l \psi_k) - \frac{x_l}{r}(\partial_r V_1)\sum_k (\partial_i \psi_k)(\partial_j \psi_k) \\ + V_1 \sum_k \left((\partial_i \partial_j \psi_k)(\partial_l \psi_k) - (\partial_i \partial_l \psi_k)(\partial_j \psi_k)\right) \tag{210}$$

Substituting in from equation 137, we get:

$$\sum_k (\partial_i \psi_k)(\partial_l \psi_k) = \sum_k (\delta_{ik} u + \frac{x_k x_i}{r} u_r)(\delta_{lk} u + \frac{x_k x_l}{r} u_r)$$
$$= \delta_{il} u^2 + \frac{x_i x_l}{r^2} r^2 u_r^2 + \sum_k \left( \delta_{ik} u \frac{x_l x_k}{r} u_r + \delta_{lk} u \frac{x_k x_i}{r} u_r \right) = \delta_{il} u^2 + x_i x_l u_r^2 + 2 \frac{u u_r}{r} x_l x_i \quad (211)$$

Substituting equation 211 (and the equivalent of equation 211 with l↔j) into the first two terms on the right-hand side of equation 210, we get:

$$\frac{\partial_r V_1}{r} \left( x_j (\delta_{il} u^2 + x_i x_l u_r^2 + 2 \frac{u u_r}{r} x_l x_i) - x_l (\delta_{ij} u^2 + x_i x_j u_r^2 + 2 \frac{u u_r}{r} x_j x_i) \right)$$
$$= \frac{\partial_r V_1}{r} (u^2 x_j \delta_{il} - u^2 x_l \delta_{ij}) \quad (212)$$

Differentiating equation 137, and substituting into the first part of the third term on the right hand side of equation 210, we get:

$$\sum_k (\partial_i \partial_j \psi_k)(\partial_l \psi_k) = \sum_k (\delta_{jk} \frac{x_i}{r} u_r + \frac{\delta_{ij} x_k}{r} u_r + \frac{x_j \delta_{ik}}{r} u_r + \frac{x_i x_j x_k}{r} \partial_r (\frac{u_r}{r}))(\delta_{lk} u + \frac{x_k x_l}{r} u_r)$$
$$= \sum_k (\delta_{jk} \delta_{lk} \frac{x_i}{r} u u_r + \delta_{lk} \delta_{ij} \frac{x_k}{r} u u_r + \delta_{lk} \delta_{ik} \frac{x_j}{r} u u_r + \delta_{lk} \frac{u x_i x_j x_k}{r} \partial_r (\frac{u_r}{r})$$
$$+ \delta_{jk} \frac{x_i x_l x_k}{r^2} u_r^2 + \delta_{ij} \frac{x_l x_k^2}{r^2} u_r^2 + \delta_{ik} \frac{x_j x_l x_k}{r^2} u_r^2 + \frac{x_i x_j x_k^2 x_l}{r^2} u_r \partial_r (\frac{u_r}{r}))$$
$$= \delta_{jl} \frac{x_i}{r} u u_r + \delta_{ij} \frac{x_l}{r} u u_r + \delta_{il} \frac{x_j}{r} u u_r + \frac{u x_i x_j x_l}{r} \partial_r (\frac{u_r}{r})$$
$$+ \frac{x_i x_l x_j}{r^2} u_r^2 + \delta_{ij} x_l u_r^2 + \frac{x_j x_l x_i}{r^2} u_r^2 + x_i x_j x_l u_r \partial_r (\frac{u_r}{r}) \quad (213)$$

By interchanging j and l in this expression, we can see that the last term in the last sum in equation 210 would equal:

$$\delta_{lj} \frac{x_i}{r} u u_r + \delta_{il} \frac{x_j}{r} u u_r + \delta_{ij} \frac{x_l}{r} u u_r + \frac{u x_i x_l x_j}{r} \partial_r (\frac{u_r}{r})$$
$$+ \frac{x_i x_j x_l}{r^2} u_r^2 + \delta_{il} x_j u_r^2 + \frac{x_l x_j x_i}{r^2} u_r^2 + x_i x_l x_j u_r \partial_r (\frac{u_r}{r}) \quad (214)$$

Subtracting equation 214 from 213, and noticeing all the terms which cancel out in this subtraction, we can work out the final term in equation 210:

$$V_1 (\delta_{ij} x_l u_r^2 - \delta_{il} x_j u_r^2) \quad (215)$$

In summary, the first two terms of the right-hand side of equation 210 were worked out in equation 212; the remaining large term reduces to equation 215. Combining these two results, we find that equation 210 reduces to:

$$B^i_{jl} = (u^2 \frac{\partial_r V_1}{r} - V_1 u_r^2)(x_j \delta_{il} - x_l \delta_{ij}) \quad (216)$$

Going back to equation 206, and then substituting in from 216, we may deduce:

$$\mu^i_{jkl} = \int_0^\infty r^2 dr \int_{sphere} x_k B^i_{jl}(r,\theta,\phi) d\Omega$$

$$= \int_0^\infty r^2 dr (u^2 \frac{\partial_r V_1}{r} - V_1 u_r^2)(\tfrac{1}{3}\delta_{jk}\delta_{il} - \tfrac{1}{3}\delta_{kl}\delta_{ij})$$

$$= (\tfrac{1}{3}\delta_{jk}\delta_{il} - \tfrac{1}{3}\delta_{kl}\delta_{ij})\int_0^\infty r^2 dr (u^2 \frac{\partial_r V_1}{r} - V_1 u_r^2) \tag{217}$$

This proves, as claimed in [4], that all of these moments can become zero if a single scalar quantity (the integral on the right-hand side) is tuned to equal zero. This observation (and Rybakov's asymptotic arguments in general) may be useful in finding numerical examples of V*-minimizing states, if they exist.

## 5. Lagrangians Containing Second Derivatives

Even if the negative claims of Rybakov should be absolutely correct, there is an interesting class of systems which these claims do not address:

$$\mathcal{L} = \mathcal{L}(\underline{\varphi}, \partial_\mu \underline{\varphi}, \partial_\mu \partial_\nu \underline{\varphi}) \quad , \tag{218}$$

where $\underline{\varphi}$ is some mathematical vector (not necessarily covariant!!). Because the energy density of such a system involves second derivatives of the field, Rybakov's claims about additive Liapunov functions [4] do not address the possibility of minimizing this energy density. The components of $\underline{\varphi}$ will be denoted as $\varphi_\alpha$, in order to emphasize the fact that these components do not necessarily form a covariant vector, like $\varphi_\mu$.

Lovelock and Rund[8] discuss such systems at length, and provide general equations for their energy density and Lagrange-Euler equations (equations 7.13 and 7.20). Unfortunately, most Lagrangians of this form require that third derivatives of $\underline{\varphi}$ be included as part of the state description, and violate our goal that energy be positive definite near the vacuum state [4]. To avoid these complications, I have looked more closely at Lagrangians of the type:

$$\mathcal{L} = g(\underline{\varphi}, \underline{\dot\varphi}, \nabla \underline{\varphi}) + \sum_\alpha (\ddot\varphi_\alpha - \Delta\varphi_\alpha) F_\alpha(\underline{\varphi}, \underline{\dot\varphi}, \nabla\underline{\varphi}) \quad , \tag{219}$$

where $\Delta$ is the usual Laplacian operator (sum of second derivatives with respect to space). Relativistic theories without third derivatives in their state description will always be of this form, but theories of this form also include nonrelativistic theories.

General relativity itself can be expressed in this form. Zinn-Justin [13, p.514] describes the Einstein Lagrangian as $g^{1/2}R$, while Lovelock and Rund [9, equation 3.49] use a slightly more complicated expression; in either case, however, the second derivatives of the field ($g^{\mu\nu}$) appear only in "R", which is multiplied by a function of the field itself (not its derivatives). Lovelock and Rund [9, equation 3.16] explicitly write out the equation for R as a function of the field $g^{\mu\nu}$ -- a function which clearly fits the form of equation 219.

Inserting equation 219 into Lovelock's equation 7.13, I derive:

$$\mathcal{H} = -\mathcal{L} + \sum_\alpha \left( (\frac{\delta g}{\delta \dot\varphi_\alpha})\dot\varphi_\alpha + \sum_\beta (\ddot\varphi_\beta - \Delta\varphi_\beta)\frac{\delta F_\beta}{\delta \dot\varphi_\alpha}\dot\varphi_\alpha - \dot\varphi_\alpha \partial_0 (F_\alpha) + \ddot\varphi_\alpha \right) \tag{220}$$

(I feel somewhat unsure about the sign of one of these terms, but that will not affect the reasoning here.) Even without writing out the expression for $\delta^2 H$, we can easily see that the term ($\partial_t^2 \xi_\alpha$) can only appear underline{linearly} in that expression. If the expressions which multiply $\partial_t^2 \xi_\alpha(\underline{x})$ at any point $\underline{x}$ are nonzero, for any choice of the functions $\underline{\xi}(\underline{x})$ and $\partial_0\underline{\xi}(\underline{x})$, then we can pick $\partial_t^2\xi_\alpha$ as large or as small as we like in a neighborhood as close as we like to $\underline{x}$, and thereby make $\delta^2 H$ positive or negative. This implies instability. But if these expressions are always zero, for all choices of $\underline{\xi}$ and $\partial_0\underline{\xi}$, then we do not fulfill the conditions for SSOS; the system is totally indefinite with respect to all perturbations of $\partial_0^2\underline{\varphi}$! Note that this does not necessarily imply instability -- only a failure to meet the strict form of stability stressed in this paper.

## 6. Quantum Foundations: 3 Formalisms for Quantizing Solitons

Sections 2 through 5 focused on the relatively narrow issue of the existence of nontopological solitons. This section will deal with the more general problem of how to quantize solitons -- either topological or nontopological solitons. It will discuss the basic principles of three alternative approaches to quantization -- the standard approach, a stochastic approach, and the heretical "neoclassical" approach described in section 6 of [4]. It will assume that we are using soliton models of some sort, in order to explain or characterize those elementary particles which have a rest state (such as electrons and hadrons). No consideration will be given here to the issue of soliton models for photons, neutrinos, etc.; it is not yet known whether such models would be helpful in any way, in the end.

The "quantization of solitons" refers to the procedures which we use in order to translate a classical soliton model, based on some classical field theory (CFT), into the corresponding quantum field theory (QFT).

Historically, the standard literature on the quantization of solitons has stressed the technical aspects of the procedures[2,1,3]. However, the success of these technical procedures has very deep implications for the foundations of quantum theory. After all, if we use these procedures to derive the actual mathematical formalisms used to confront experiments, then these procedures really define what the foundations of quantum theory actually are. Many classical discussions of the foundations of quantum theory implicitly assume a set of procedures very different from what is commonly used today.

This section will suggest that the standard methods for quantizing solitons, in the usual functional integral approach, implicitly assume that the universe is a set of classical fields defined over 4-dimensional space, governed by what I would call an "iMRF" -- imaginary Markhov Random Field -- model with parallels to fuzzy logic. This particular formalism may sound radical to some readers, but it is little more than a matter of noticing what has already been going on for years in field theory. (But then again, even special relativity could be seen as a matter of noticing the significance of invariance with respect to Lorentz transformations, a topic which Lorentz may have pursued before Einstein.) The outlines of this approach will be discussed below, but considerable future research would be needed in order to establish its axiomatic validity (as with field theory in general!).

Before explaining this idea, I will first review some basic facts about Markhov Random Fields, Markhov Processes, scattering and Bell's Theorem, in section 6.1. Then in section 6.2 I will describe the iMRF interpretation of standard QFT, for the case of time-symmetric bosonic fields. Section 6.3 will discuss a few aspects of fermionic fields, related to a more extensive discussion in section 6 of [4]; it will also discuss how all of these formalisms permit consideration of a wider range of physical theories. Section 6.4 will discuss how a slight modification of this approach (a true MRF model) might be used to construct a more classical stochastic field theory, with some possible relation to stochastic quantization[20,21] and stochastic electrodynamics; this formulation seems to remove some "show-stoppers" with that general approach, but it still leaves a lot of open questions. Section 6.5 will discuss how the zero-noise ($\sigma=0$) limit of the MRF model appears to be a viable alternative to the iMRF model (as discussed further in [4]). Adoption of the $\sigma=0$ model would be tantamount to a complete acceptance of the Einstein position in the classic Einstein-Bohr debate (except for Einstein's surprisingly careless statements about the prediction of EPR experiments).

These subsections will not be independent of each other. Each will depend heavily on information from previous subsections.

This section will say nothing about the very important issue of renormalization and regularization. This has already been discussed at some length in [4]. Therefore, the discussion below may be viewed as a discussion of the three formalisms either for the case of a finite field theory, or for the case of a particular finite value of the renormalization/regularization parameter Λ in a regularized/renormalized field theory.

### 6.1. MRFs Versus Markhov Processes and Bell's Theorem

### 6.1.1. Classical MRFs Across a Spatial Grid

The Markhov Random Field (MRF) model across space has been used for decades, under various names, as one of the standard models in engineering and in physics. For example, Laveen Kanal helped introduce the idea to image processing decades ago; the resulting literature is extremely rich, and the discussion of "cliques" and "elites" is very powerful and rigorous. In physics, the usual Ising lattice models [3,22] are an example of MRF models. ("Spin-glass" lattice models of this sort have also become popular in parts of the neural network field.) More recently, MRF lattice models across space and imaginary time have been used in the renormalization of nuclear field theories; they have become essential to numerical calculation of the predictions of quantum chromodynamics[3,22].

MRF models take a very simple form, when applied to spatial lattices in two dimensions. Let us assume a spatial grid made up of points (x,y), where x and y are integers. Crudely speaking, we may say that the stochastic law which "generates" the field values $\underline{\varphi}(x,y)$ is:

$$\mathbf{Pr}(\underline{\varphi}(x,y)) = f(\{\underline{\varphi}(x',y')\} \text{ such that } (x',y') \in N(x,y)) \tag{221}$$

In other words, the probability of $\underline{\varphi}$ taking on a particular value at the point (x,y) is a function of the field values at the neighboring points (x',y'), where N(x,y) refers to the set of "neighbors" of (x,y). The exact specification of the function N is part of the specification of the particular model; however, ordinary MRF models tend to be local and translationally invariant, so that (x',y') will not be a neighbor of (x,y) unless the length of (x'-x,y'-y) is less than some small bound, such as 2 or 3.

In formulating an MRF model, we essentially assume that equation 221 applies independently (in a statistical sense) at different points (x,y). In other words, we invoke the classical assumption from statistics that the random noise or disturbance at each point (x,y) is uncorrelated with the random noise at each other point. (See [23] for a discussion of noise independence across time, which is a fundamental assumption of time-series analysis; this assumption does not limit the stochastic processes which can be generated as a result of such random inputs.) Crudely speaking, then, the probability of a configuration of values of $\underline{\varphi}(x,y)$, across a region R, should equal the product of $\Pr(\underline{\varphi}(x,y))$ across all points (x,y) in that region. In addition, we may generalize equation 221 so as to permit a more implicit stochastic relation between $\underline{\varphi}(x,y)$ and its neighbors; thus instead of requiring that $\Pr(\underline{\varphi}(x,y))$ be expressed as a function of neighboring field values, we may simply require that the model must tell us the probability, p(x,y), of any combination of values for $\underline{\varphi}(x,y)$ and for its neighbors:

$$p(x,y) = F(\underline{\varphi}(x,y), \{\underline{\varphi}(x',y')\} \text{ such that } (x',y') \in N(x,y)) \tag{222}$$

Obviously equation 221 is a special case of equation 222.

These considerations lead directly to the well-known classical formula:

$$\mathbf{Pr}(\Phi) = \frac{\prod_{(x,y) \in R} p(x,y)}{Z} = Z^{-1} e^{-\sum_{(x,y) \in R} s(x,y)} = Z^{-1} e^{-S} \quad , \tag{223}$$

where I have defined:

$$s(x,y) = -\log p(x,y) \qquad (224)$$

$$S = \sum_{(x,y) \in R} s(x,y) \qquad (225)$$

$$Z = \sum_{\Phi} e^{-S(\Phi)} \qquad (226)$$

and where $\Phi$ represents the configuration, the set of values of $\varphi(x,y)$ across all points $(x,y)$ in R. Note that S, Z and $\Pr(\Phi)$ usually depend implicitly on the values of $\varphi(x,y)$ at points $(x,y)$ just outside the region R, because of an implicit dependence through equation 225. However, when proving theorems, we often adopt special boundary conditions such as periodic boundary conditions which eliminate such external dependencies.

### 6.1.2. Extensions to Continuous Space-Time

There is no reason why equations 222 through 226 should only be applied to coordinates which represent "space" as opposed to "time". Thus instead of applying these equations over points $(x,y)$, we may apply them over points $(\underline{x},t)$, forming a four-dimensional grid, representing space-time. For example, we may define the region $R(t_-:t_+)$ in space-time as the set of <u>all</u> grid points $\underline{x}$ at times t which obey:

$$t_- \leq t \leq t_+ \qquad (227)$$

If we assume that the region R is actually a subset of an infinite space-time grid, governed by a translationally invariant local MRF, then we may deduce (under reasonable assumptions):

$$\mathbf{Pr}(\Phi) = P_+(\Phi_+) P_-(\Phi_-) e^{-S(t_-:t_+)} / Z \quad , \qquad (228)$$

where $\Phi_+$ represents the set of values of $\varphi(\underline{x},t_+)$ across all grid points $\underline{x}$ at the time $t_+$, where $\Phi_-$ is the same idea for time $t_-$, where S is essentially the same as in equation 225, where $P_+$ represents the probability of $\Phi_+$ <u>as determined by</u> the effects of $P(\underline{x},\tau)$ across times $\tau > t_+$, and $P_-$ represents the effects of times before $t_-$. For technical reasons, equation 228 is valid for general choices of cliques and elites only if we take great care in assigning times t correctly to S, to $P_+$ and to $P_-$.

The continuous space-time version of the MRF is also very straightforward, conceptually. We may replace equations 223 and 225 by:

$$\mathbf{Pr}(\Phi) = e^{-S/\sigma} Z^{-1} \qquad (229)$$

$$S = \int_{t_-}^{t_+} \int s(\underline{x},t) d^3\underline{x}\, dt \quad , \qquad (230)$$

where $\sigma$ is a scalar parameter introduced for convenience. Equation 228 remains unchanged, except that "S" should be replaced by "S/$\sigma$."

This general approach of treating time like just another dimension was a key element of Einstein's vision. It appeared even in his PhD thesis, which explained the photoelectric effect as the result of behavior which is symmetric with respect to time-reversal.

### 6.1.3. MRFs Versus Markhov Processes

A local Markhov Process (MP), defined over discrete space-time, may be written as:

$$\Pr(\underline{\varphi}(\underline{x},t)) = f(\{\underline{\varphi}(\underline{x}',t-1),...,\underline{\varphi}(\underline{x}',t-k)\} \text{ such that } \underline{x}' \in N(\underline{x})) \quad , \tag{231}$$

where k is some positive integer and where N($\underline{x}$) refers to neighbors of the point $\underline{x}$. (Markhov processes in a strict sense require that k=1, but the representation of $\underline{\varphi}$ can always be altered to meet that strict requirement, for other finite integer choices of k.)

Clearly, a local MP is a special case of local MRF; in fact, equation 231 is just the special case of equation 221 for which we impose the restriction that ($\underline{x}$',t')$\notin$ N($\underline{x}$,t) whenever t'≥t. Equation 221, in turn, is a special case of equation 222.

There is a common tendency for many people to assume that all well-behaved dynamical models must be local MPs, or equivalent to MPs in the statistical sense [23]. This is absolutely not correct. The differences between local MRFs and local MPs are crucial to the foundations of quantum theory. They are also well-known in many other fields.

For example, in image processing, MP-based processing of images is still popular in some applications. A system may scan a range of pixels, in some arbitrary sequential order, and represent the statistics for any pixel as a function of k preceding pixels. It is now well-known that such a sequential representation (for finite k) cannot incorporate the <u>forwards and backwards</u> statistical relations implied by an MRF model of the image.

Perhaps the easiest example from another field is that of the numerical simulation of heat-flow equations, by finite element approximation, across space-time grids[24, ch.1]. Early, naive efforts to simulate such equations used a simple feedforward approximation, in which the temperature T(x,t+1) would be calculated as an explicit function of nearby temperatures at time t. This calculation was intended to be deterministic as a function of those earlier temperatures; however, roundoff errors of various types added a small degree of random noise, which converted this calculation to something very much like a local MP. Even this small amount of noise, in this MP process, accumulated very rapidly for reasonable choices of grid size, leading to gross instability and explosion of the process.

More modern finite element methods [24] often specify <u>relations</u> between T(x,t+1) and the temperature at other points at time t+1 (as well as earlier times), such that one simply cannot calculate T(x,t+1) in a direct, feedforward manner. Instead, one must solve the entire set of nonlinear equations representing these relations, across all space for time t+1, before one can know even that one number, T(x,t+1). The cost of calculations from time t to t+1 is far greater in this implicit approach, but the stability and improved accuracy are often well worth the cost. Because the value for T(x,t+1) depends on T at all other points in space at time t+1, the correlations here are highly nonlocal, in principle. This would be even more true if the random noise terms were of moderate size.

Neural network researchers may be interested to note that these results are closely related to the result that cellular feedforward networks cannot approximate the same, wide range of functions than can be approximated by cellular Simultaneous Recurrent Networks[25,26]. Even more general results should be possible with SRNs exploiting a more general relational notion of locality [11] rather than the usual grid-oriented notions.

For people trained in time-series analysis (where spatial localization is not an issue), the difference between MPs and MRFs may sound somewhat paradoxical at first. After all, the Markhov property merely states that Pr($\psi$(t+1)| {$\psi$($\tau$), $\tau$≤t})=Pr($\psi$(t+1)|$\psi$(t)). Wouldn't an MRF with a small neighborhood N satisfy this requirement? Wouldn't this imply that an MRF is an MP? The answer to these questions is: "Yes (unless there is a stability problem with the MP or MRF representation), but locality is not preserved." In essence, the example of finite element computation shows why this is. It is also possible to create simple linear examples which bring out this phenomenon, such as the scalar example:

$$\varphi(x,t+1) = b\varphi(x,t) + a\varphi(x+1,t+1) + a\varphi(x-1,t+1) + e(x,t+1), \tag{232}$$

where "e" is normally distributed white noise. It is easy to verify that this process results in a situation where $\varphi$(x,t+1) depends on e(y,t+1), even for y far away from x, resulting in correlations between $\varphi$(x,t+1) and $\varphi$(y,t+1), even conditional upon complete knowledge of time t. (To see this, one simply needs to solve symbolically for $\varphi$(x,t+1) as a function of all the values for $\varphi$(y,t) and e(y,t).)

For high-energy physics, the most important class of MRF models would be <u>time-symmetric</u> models. In equations 221 or 222, this would mean that we could interchange the arguments φ(y,t+k) with φ(y,t-k), across all arguments, and would always end up with the same value of the function f or F, regardless of the actual values of the arguments. (Symmetry with respect to a combination of such time-reversal T with some time-independent operator M yields similar results.) Systems of this sort are radically different in behavior from local MPs!!

For an MP system, $P_+$ in equation 228 is essentially a uniform distribution. Noise injected after time $t_+$ cannot have any causal effect or correlation with φ at time $t_+$ or earlier. (This is an example of the usual time-forwards "causality" assumption of statistics[23].) In effect, the $P_+$ term simply disappears from equation 228. On the other hand, for a time-symmetric MRF model, $P_+$ would be the exact same function as $P_-$!! (In a TM-symmetric MRF model, $P_+$ is the same as $P_-$, but with a similarity transform of the arguments based on M.)

### 6.1.4. Bell's Theorem, MRFs and Principles of Scattering

All three formalisms discussed in section 6 rely heavily on Classical Field Theory (CFT) to some extent. This leads to an obvious question: what ever happened to all those results in the foundations of physics, which were interpreted to mean that Einstein was wrong, and that CFT no longer has anything useful to contribute to basic physics?

By far the most important result along those lines was the classic Bell-Shimony Theorem[27], popularized in [28] and supported by a long strong of experiments (e.g.[29,30]).

The Bell-Shimony theorem, in the original language [27], rules out all "local, causal hidden variable theories." Each of these three terms ("local," "causal" and "hidden variable") is defined more precisely in [27], as required to permit proof of the theorem. The three formalisms of this section probably do all fit their definition of "hidden variable" (even though there are no hidden variables assumed!), insofar as they do describe reality as a collection of fields across four-dimensional space-time. Likewise, they are entirely local. However, the Bell-Shimony definition of "causality" is essentially just time-forwards causality, the kind of causality assumed in a Markhov Process! More precisely, when Shimony et al rule out the combination of "locality" and "causality," they are actually ruling out local MP models of reality. But time-symmetric local MRF models do not fit this assumption. Therefore, the Bell-Shimony Theorem does not rule them out as models of reality. (This is not so surprising, since the functional integral formalism is the main foundation of modern QFT!)

More concretely, it is important to see how this class of theory can fit experiments like the Bell's Theorem experiments.

In scattering experiments, we may think of the input state $\Phi_-$ as a combination of two classes of variables -- the measured, known variables $\underline{M}_-$ and the other variables $\underline{u}_-$. (For example, for a beam of point particles, we might choose to control/measure the momentum -- which then becomes $\underline{M}$ -- while leaving the position variable unmeasured, i.e. part of $\underline{u}$. Many other examples could be considered, but these would be a distraction at this point.) To describe the output state, especially, we may further subdivide $\underline{M}$ into two collections of variables: $\underline{M}^{[m]}$ (which specify which measurement is being performed or controlled), and $\underline{M}^{[r]}$ (which give the results of the measurement, or specify the value AT WHICH it is controlled). The state $\Phi_-$ is then specified by specifying all three pieces of information, $\underline{M}^{[m]}(t_-)$, $\underline{M}^{[r]}(t_-)$ and $\underline{u}(t_-)$. Likewise for $\Phi_+$.

In this notation, the Bell's Theorem experiments depend on our ability to specify $\underline{M}^{[m]}(t_+)$, the choice of polarization direction to be used in the measurement at the final time $t_+$. $\underline{M}^{[m]}(t_+)$ is manipulated, physically, by inserting polarized pathways, which the outgoing light must orient to. In effect, these polarizers simply provide part of the channel or boundary conditions which the light must satisfy, across space and time. $\underline{M}^{[r]}(t_+)$ then tells us which of the allowed polarizations is then "selected" by the photons in the experiment.

The Bell's Theorem experiments defy everyday notions of causality, because the probability distributions for the results of measurement -- $\underline{M}^{[r]}(t_+)$ -- along one output channel of the experiment depend on the choice of measurement -- $\underline{M}^{[m]}(t_+)$ -- along the other channel, even when the former takes place before the latter. The experimental results contradict any theory for which $\underline{M}^{[r]}(t_+)$ in one channel depends only on $\underline{M}^{[m]}(t_+)$ in that channel and on $\Phi_-$.

However, if we describe reality as an MRF across space-time, then the state of the system just before time $t_+$ will depend on <u>all</u> the specified boundary conditions -- $\mathbf{M}^{[m]}(t_-)$, $\mathbf{M}^{[r]}(t_-)$ and $\mathbf{M}^{[m]}(t_+)$, across all channels -- which influence the entire space-time region under study. In other words, the choice of $\mathbf{M}^{[m]}(t_+)$ <u>causes</u> a kind of change in the probability distribution for the state of the fields between times $t_-$ and $t_+$, just as the choices of $\mathbf{M}^{[m]}(t_-)$ and $\mathbf{M}^{[r]}(t_-)$ do. (In this case, the word "change" refers to a change relative to what would have been the case with a different value of the boundary condition. It does not refer to a change over time.) This follows directly from equation 228, when we recall that time-symmetry implies $P_+(\Phi')=P_-(\Phi')$ for any spatial state $\Phi'$. This symmetry phenomenon is central to what we actually observe in quantum experiments in general.

The essential role of time symmetry in Bell's Theorem experiments was noticed decades ago by DeBeauregard [31] and myself [32,33]. Many subsequent authors (such as Leggett) have elaborated on this point.

## 6.2. The Standard Formalism as an iMRF Model in the Bosonic Case

This section (6.2) will propose a simple interpretation of the standard formalism for QFT, based on functional integrals, used in modern high-energy physics [3,34]. It will not address the issue of renormalization for these formalisms, because that was already discussed in [4], as discussed in the beginning of this section (6). It will consider only the case of bosonic, time-symmetric theories. Section 6.3 will discuss the extensions to fermionic and time-asymmetric theories.

In general, the modern functional integral approach is actually closer to Schwinger's source theory [35] than to the early Feynman path-integral model of point particles. Schwinger's procedures for calculating things have been clarified and refined tremendously in recent years [3,34]; however, much of the original rationale <u>behind</u> his approach appears to be forgotten. This section will essentially try to articulate a piece of Schwinger's original vision, a vision of <u>deriving</u> QFT from a relatively simple axiomatic starting point. In the end, I myself do not find Schwinger's vision quite so compelling as the more extreme (as yet unproven) alternative in section 6.5; however, his vision certainly offers a major improvement over earlier formalisms which sometimes seemed to introduce new implicit assumptions (axioms) in every other equation over hundreds of pages. My interpretation of Schwinger's vision here is based more on observing him in the classroom, circa 1970, rather than his books on source theory[35].

Schwinger's way of reformulating QFT has a striking parallel to the system of "fuzzy logic" promulgated by Lotfi Zadeh[36]. In developing fuzzy logic, Zadeh has stressed again and again that he is not trying to erase classical control theory or reasoning systems, etc. Instead, he is trying to make them more powerful by "fuzzifying" them. By "fuzzifying," he means pinpointing important cases where reasoning is based on a sharp 0/1, true-or-false measure of truth, and then replacing that binary truth variable with a continuous variable representing the degree of truth, in the continuous range between zero and one.

In much the same way, Schwinger built his formalism on a foundation of classical field theory (basically just the MRF model!), but <u>replacing</u> the usual measure of probability ranging from 0 to 1 by probability amplitudes, complex numbers of length ≤1. In fact, Schwinger's logic could be linked more closely to Zadeh's, if we think of the square length of the probability amplitude as representing the probability dimension, and we think of the remaining continuous dimension (theta) as a kind of truth measure. Would such an approach to reasoning be more like circular logic than fuzzy logic? In actuality, Kosko has mentioned to me that there has been serious work in the reasoning community which claims to find some value in using angular measures of truth. But all of that is far beyond the scope of this paper.

When we apply this kind of Schwinger logic to modify ("fuzzify") equation 229 -- the fundamental equation of the MRF model -- we arrive at:

$$\boldsymbol{Pa}(\Phi) = e^{iS/\hbar}\boldsymbol{Z^{-1}}, \qquad (233)$$

where "Pa" refers to probability amplitude, and where "for convenience" I have changed the label of the arbitrary scalar from "σ" to "ℏ". This is an imaginary MRF (iMRF), in that we now multiply S by an imaginary scalar instead of a real one, and interpret the overall result as a probability amplitude. Our discussion so far helps to interpret and explain equation 233, but it is not intended as any kind of derivation; instead, the equation itself is the main assumption (or axiom) of the model as such.

Equation 233 is extremely well-known as the fundamental equation of the functional integral approach. (E.g., see equation 3.39 of [34].) Usually, this equation is presented with an additional term included, "J." However, later on, when the equation is actually used, it is used for the case J=0. The J term is used as a kind of scaffolding, so as to simplify a later step -- the calculation of the "G" functions -- which will be discussed below. (Zinn-Justin devotes the first part of the first chapter [3] to explain this kind of scaffolding as a trick used in integration.)

Unfortunately, equation 233 is not powerful enough by itself to provide the axiomatic basis for deriving the rest of field theory. Most of the discussions starting out from equation 233 proceed in a very intuitive fashion, adding additional implicit assumptions. (Zinn-Justin's account[34] is far more rigorous than most, but does not yet provide the more compact foundation I am asking for here.) To some extent, additional assumptions are inevitable. However, it is possible to minimize the use of additional assumptions by continuing the analogy to fuzzy logic (and hoping that practitioners of more crisp logic will fill in some of the details).

In fuzzy logic, a fundamental task in actually <u>using</u> the new logic in practical engineering [36] is to find some way to connect it up to real-world sensors or actuators, which we control by use of more conventional logic and numerical I/O. This task is called "fuzzification and defuzzification." Our problem here is simply to "defuzzify" -- to translate abstract probability amplitudes (whose dynamics are fully specified by equation 233) into predictions which we can test involving ordinary probabilities.

My claim is that we can minimize the number of additional assumptions required here, by using only one additional assumption beyond equation 233 (and the choice of the function s): a partial definition of what a probability amplitude actually "means." In other words, we need only specify what the relation between a probability amplitude and a probability is, for a simple class of possible states or configurations Φ. (In effect, this is just a minimal defuzzification rule.)

The defuzzification rule that I propose is as follows. Suppose that we are trying to estimate the probability that we will see particles ("solitons" or "instantons") at a set of space-time points $(q_1,t),...,(q_n,t)$ <u>or</u> at a set of points $(q_1,t_+),...,(q_n,t_+)$, $(q_1,t_-),...,(q_m,t_-)$. Suppose that we have chosen a Lagrangian function s which generates solitons. Consider the situation where the probability amplitude is zero for all states Φ(t) or {Φ($t_-$),Φ($t_+$)} which would have solitons at those points, except for one state (and its immediate neighbors). I propose the following simple rule for such a state: that the probability of observing particles at these locations equals the square of the probability amplitude of that state (with the proper scalar normalization). Certainly this axiom will need a more formal statement, before an axiomatic version of this idea could be possible; however, I hope that the axiom itself would not have to be extended too far. As additional assumptions go, this one would appear relatively weak. Again, we may even regard it as part of the <u>definition</u> of what equation 233 <u>means</u> (what "Pa" means).

Notice the key role of solitons in this formulation. This formulation assumes that we will try to explain the existence of massive particles (but not necessarily photons!) by modeling them as solitons. Most accounts of functional integral methods [3,34] focus on <u>correlations</u> across space-time, rather than solitons as such; however, <u>when</u> solitons are present, correlations measure joint probabilities of the location of such solitons. When solitons are not present, one may not deduce that particles exist at all in any location; at best, any association between field correlations and particle locations would involve an additional axiomatic assumption -- and a rather big one. (In the case of photons, however, we can reproduce existing theory simply by assuming that they are intermediate stages of processes where what we observe in the end are excited states of atoms or other massive particles.)

In summary, a soliton-based approach should make it possible not only to "explain" why particles exist and what their properties are, but also to minimize the number of axiomatic assumptions in the formalism itself.

In the standard descriptions of the functional integral approach, the next step after describing equation 233 is to discuss the functions $G(x_1,...,x_n)$, where the $x_i$ are points in 4-dimensional space-time. These functions are simply the n-th order statistical moments of the field $\underline{\varphi}$, weighted by probability amplitudes rather than probabilities. In other words, they are the Schwinger-logic equivalent of ordinary statistical moments. The functions G are defined mathematically for arbitrary collections of points, but the fundamental scattering theory only requires that we consider cases where the $x_i$ are all at the same time t (as t goes off to $\pm\infty$, away from the zone of interaction) or where they are a mixture of points at times $t_-$ and $t_+$. If the field condenses into solitons away from the interaction zone, and if interference between nearby states becomes less and less of a factor away from the interaction zone, then our defuzzification rule tells us that the joint probability of observing particles at $q_1,...,q_n$ at $t_+$ and at $q'_1,...,q'_m$ at $t_-$ will equal the square of the corresponding probability amplitude. That probability amplitude will essentially be the G function for those points, for systems which do condense into solitons. One can then derive the usual reduction formulas (which give the actual scattering probabilities) by first normalizing the G function (to account for the values of $\int\varphi^2$ in different soliton types) and then deriving the conditional probabilities from these joint probabilities.

The usual dynamics of the G functions over time follow directly from their mathematical definitions and from equation 233, as shown in the existing literature. Their physical significance, however, is deduced from their physical meaning, coming in and out of scattering experiments, based on equation 233 and the defuzzification rule.

In actuality, scattering calculations based on probability amplitudes really represent only one side of modern QFT. Theoreticians who rely solely on that part of the story cannot properly understand realistic scattering experiments. Empirical studies of condensed matter physics[37] and quantum computing[38] show that the inputs and outputs of many scattering experiments (as well as solid objects!) must be described as "mixed states." In other words, one uses a density matrix rather than a wave function or simple probability amplitudes to characterize such states. However, authors like Harrison[37], drawing on quantum statistical mechanics, have argued that the density matrix formalism, in turn, can be deduced as the large-scale statistical consequence of a universe governed in the large by the more fundamental principles discussed above.

Clearly, this formalism requires further axiomatic development. In the meantime, there are two more interesting connections here to the foundations of physics, one minor and one major.

The minor point: in this treatment of scattering, it is easiest to think of Bell's Theorem polarizers simply as part of the scattering process itself, as part of the interaction zone. The final quantized soliton observation takes place later, in the photon counters. This is the simplest way to think about those experiments, and everything presumably goes forwards as described in section 6.1.4.

The major point: this formalism has an interesting analogy to the many-worlds formalism originally proposed by Hugh Everett[39]. The major goal of Everett's thesis was to prove that the usual measurement formalisms of QFT could be derived directly (approximately) as a consequence of the universe obeying a Schrodinger equation over infinite dimensions. That idea has new advantages, now that Mandel [30] has performed experiments which definitively show that conscious intelligent "observers" are not part of the apparatus necessary for generating the usual measurement effects. In my view (section 6.5), Everett's theorem really does not prove his larger claim; the usual Schrodinger equation by itself does not contain sufficient information, in my view. However, the iMRF equation comes much closer; one can rederive the quantum measurement results by adding just a simple defuzzification axiom, which can be regarded as part of the definition of equation 233. It is already quite significant that this approach eliminates the need for the metaphysical observer.

## 6.3. From Bosons to Fermions, Time Asymmetry and Nonperturbative Approaches

Standard treatments of QFT based on functional integration usually begin by discussing bosonic theory (as above), and then move on to add fermions. Advanced treatments also go beyond the standard model (which is quasilinear and time-symmetric) to consider topics such as superweak interactions, gravity and cosmology. This section will discuss those extensions, but far more briefly, mainly by citing other sources.

### 6.3.1. From Bosons to Fermions

When the field $\underline{\varphi}(\underline{x},t)$ is a mathematical vector made up of some mixture of ordinary real and complex numbers, then the procedures described in section 6.2 always yield a bosonic field theory. In order to represent fermions, Schwinger proposed[35] that we follow the same procedures, but introduce new field components which are "classical anticommuting numbers." Zinn-Justin[3], for example, follows this same path, but provides far more rigorous detail on the principles of Grassmannian algebra and on how to make sure that we specify the usual procedures exactly. These procedures have led to impressive empirical results. However, there is a difference between rigorously specifying procedures and rigorously specifying a model of a process to be described by the procedures. There is reason to question what the procedures really mean, from a mathematical point of view.

Many physicists would immediately ask: who cares? If these procedures have provided impressive empirical results, why should we care whether they can be interpreted in terms of a meaningful mathematical picture of reality? Certainly we should not care so much that we stop using these methods, in the absence of a working alternative! But there are reasons for at least some of us to try to develop that alternative. There are many domains in physics (like much of nuclear physics) where we are still far away from being able to predict empirical reality with high precision. In those domains, it would be very useful to have some choices, to have a source of plausible, computable alternative models. In section 6 of [4], I described how the search for a purely bosonic foundation for QFT does point to interesting and testable possibilities, especially for nuclear physics. (I also described the role of the more modern literature on spin and statistics there.) To be honest -- even if the only eventual benefit of such work were to be a more compact and rigorous foundation leading to the same set of empirical predictions, I would consider that a worthwhile achievement, for many reasons.

The next obvious question here is: what are the mathematical questions you want to raise about "classical anticommuting quantities?" Let me emphasize that these are questions, not disproofs; however, I have not myself seen satisfactory answers to these questions in the literature.

Zinn-Justin points out that the objects in a Grassmann algebra must all be linear combinations (like vectors) of simple products of the fundamental generating elements $\theta_i$. If we begin by imagining a grid in space-time (for simplicity), we have two main choices for how to go about formulating an iMRF or MRF model here.

First, we could assume a finite list of generators $\theta_i$, which in turn leads to a finite list of elementary products. In that case, the state of the field can be represented simply as a mathematical vector over ordinary real numbers, as Zinn-Justin mentions. Then the entire field theory itself can be represented as an MRF or iMRF based on that vector, which leads to a bosonic field theory. I have not been able to consistently derive fermionic statistics for such a situation (or at least not in the continuous limit).

The second major option would be to assume an infinite set of generators, indexed to the grid points. The effect of this would be to generate connections between distant points in space-time. (For example, the "field" at one point would normally contain a nonzero component of the "generator" belonging to another point.) This suggests the fascinating possibility of constructing more classical "pointer fields", $\underline{\varphi}(x,y)$, where x and y are points in 4-D space-time. This would be similar to the usual Fock-space wave functions (or "Q potentials"), but defined over a clean 8-dimensional space instead of a weird infinite-dimensional space. I have not found a way to make this option both useful and workable, but it certainly seems like an intriguing alternative. This alternative would be even more interesting if it could somehow be connected to some of the more intuitive or indirect arguments about dual-valued functions and fermion statistics reviewed in [1].

For the present, the boson-based strategy given in [4] seems far more tangible and workable to me than either of these two alternatives. However, there is certainly a possibility that new work on these alternatives might change the story.

## 6.3.2. Benefits of Time Asymmetry and Nonperturbative Methods

Before the functional integral formalism was developed. QFT was defined as a combination of classical perturbation theory and operator fields [40]. That combination may be called "CQFT" (classical quantum field theory). The functional integral approach still allows the use of classical perturbation theory, in situations where it works, but it also permits the use of nonperturbative methods (or methods based on a more general concept of perturbation theory). This should also be true for the two new formalisms to be discussed in sections 6.4 and 6.5; however, it has taken many years to develop and test the nonperturbative methods for the standard formalism, and similar efforts have yet to be undertaken for the new formalisms. Nonperturbative methods have played a crucial role in the progress of physics in the past two decades, both by expanding the range of theories which can be considered and by making it feasible to actually calculate the predictions of theories already under consideration.

This section will discuss how the new formalisms allow consideration of interesting theories beyond the scope of CQFT.

Up to this point, I have mainly discussed the case of time-symmetric theories. Even CQFT allows consideration of theories which are not symmetric with respect to the operator T (time reversal). However, all three components of the "Standard Model of Physics" (quantum chromodynamics (QCD), electroweak theory (EWT) and general relativity) are symmetric with respect to T. It will be an important task simply to prove that the two new formalisms can replicate QCD and EWT in detail, at least for the situations where these theories have been tested empirically. This is why I stressed the role of T symmetry in section 6.1.3, and will stress it again in section 6.5.

CQFT does require that theories must be symmetric with respect to CPT, the operator formed by combining T with a reversal of charge and parity. CPT is simply a special case of "TM" symmetry, discussed in section 6.1.3. Theories which obey TM symmetry do not require any fundamental change in approach, either with functional integration or with the new formalisms; they only require a slight increase in complexity, as we must keep track of the similarity transformations (based on M) whenever we switch between forwards time and backwards time. There is no fundamental reason why a change of charge or of parity should change the direction of causal flows in time; again, this follows directly from a careful consideration of section 6.1.3.

Classical perturbation theory also tends to require that the higher-order derivatives in a theory enter in a purely quadratic term [34,ch.3]. Theories of that sort are a special case of a more general class of theories called "quasilinear" [4]. But general relativity is nonquasilinear (NQL). Likewise, most field theories capable of generating solitons (topological or nontopological) in 3+1 dimensions are nonquasilinear [4]. Equations 229 and 233 do not in any way require that "s" represent a quasilinear theory; however, when it does represent an NQL theory, we must use nonperturbative methods in order to work out the predictions of the theory. This is the reason why nonperturbative methods based on functional integrals have been the primary tool used to quantize soliton models [2,3]. Nevertheless, if we model particles as "point particles," as the limiting case of a soliton model as the core radius of the soliton goes to zero, then we may still use classical perturbation methods, as described in section 6.3. of [4]; this is probably the easiest way to recover the standard model in the rigorous soliton-based version of all three formalisms.

Finally, there is no reason why we must always restrict ourselves to theories which obey CPT symmetry or which possess a positive-definite energy H, in an MRF or iMRF formalism. Some of the theories which violate these conditions would be highly unstable and unacceptable, even in an MRF context; however, others might be of serious potential value in physics.

For example, in [12], I proposed a sketch of a theory to address a very fundamental question: why does causality appear to run forwards in time, in macroscopic life, even though the fundamental laws of physics appear to follow CPT symmetry?

This alternative theory grew out of a conversation with Schwinger's lead research assistant years ago. (I would want to credit him here, but hesitate to give his name in this kind of context without his permission.) He was complaining, over lunch, about the sticky assignment Schwinger had given him: to work out the physics of tachyons according to source theory. Classical tachyons were supposed to have imaginary "rest mass," but never to stay at rest; because they always travel at a speed >c, they would have positive mass-energy. However, he pointed out, one could always view them in an alternate rest frame, and generate modes with negative mass energy. This, in turn, would imply a certain rate of spontaneous vacuum decay. (In other words, there would be a spontaneous emission of balanced sets of positive-energy particles and negative energy particles, even in the vacuum.) This vacuum decay would be predicted both in source theory and in CQFT,

I asked myself: is this such an unacceptable prediction? If there did exist a very low level of vacuum decay, what would this vacuum decay do to us that we could observe? Since the amount of decay is essentially proportional to the volume of space that we look at, the biggest effect should simply be in the biggest volume of space -- the deep vacuum of intergalactic space. I asked: how do we know what is going on in intergalactic space? How do we know that there are no negative energy modes out there? How would we see them if they were there?

The answers are straightforward. Our direct, empirical knowledge of intergalactic space comes almost entirely from observing the light which passes through that space. If there were a low but significant density of negative energy modes, interacting at a low level with this light, then we would tend to expect a low level of energy drain from that light, essentially proportional to the distance traveled. This is not at all inconsistent with what we actually see!! In fact, it provides an alternative possible explanation for the well-known cosmological redshift phenomenon. It provides an alternative to the usual view that the universe originated in a Big Bang.

The Big Bang theory is currently one of the main pillars of astrophysics; however, major difficulties have begun to surface in recent years. These include the age-of-the-universe problem, issues associated with new data from the COBE observatory, and recent evidence of "accelerating explosion" of the universe (based on the usual Doppler interpretation of the redshifts). The alternative theory would erase these problems very quickly, and would also provide a possible mechanism to solve the missing neutrino problem. The classic blackbody distribution of background radiation (the 3 degrees K radiation) would be a natural prediction of the new theory, assuming a long-standing thermodynamic equilibrium of positive and negative energy modes in deep space. The prediction of steady (exponential) decay in energy with respect to distance in deepest space could be useful in comparing this theory with the conventional theories. Also, depending on the average assumed size of collision between light and negative energy modes, it is possible that a very small angular dispersion of light might result.

However: how could we explain the macroscopic arrow of time, without a Big Bang? In [41], Prigogine has proposed that the arrow of time actually results from a kind of spontaneous symmetry breaking in time, rather than a Big Bang. As an alternative, I suggested [12] that there may exist some interactions in field theory which are extremely weak (in any small region of space) but highly asymmetric with respect to time. These interactions would be related to the superweak interactions and to the proposed negative energy particles or radiation. In effect, they would provide a kind of small feedback loop which has a very large cumulative effect when integrated over the vastness of deep space for billions of years. I can even imagine Einstein asking: "If God wanted to make a universe, would he make it like a bomb or like a terrarium?"

Models of this sort could be pursued more seriously within the general framework provided by iMRF or MRF models. Whether they are valid or not is an empirical issue.

## 6.4. An Alternative Formalism: The MRF Model with $\sigma>0$

The great success of the iMRF formalism leads to an obvious question: why not try a true MRF model, based on equation 229 with $\sigma>0$? Is the fuzzy logic really necessary?

This section will argue that this alternative formalism is well worth pursuing; however, I will not provide answers to many of the questions which it raises.

The MRF model with $\sigma>0$ could be viewed as a classical field theory plus the addition of white noise. Models of that general sort have been explored for decades, with limited results in high-energy physics; however, the MRF formalism offers the possibility of overcoming the key difficulties.

In classical models, when white noise is injected in the traditional way, the resulting theory is a Markhov Process (see section 6.1). Theories of that sort can be rejected immediately, because of the Bell's Theorem experiments. But a time-symmetric MRF model would automatically be consistent with those experiments.

Perhaps the most successful effort in these directions has been "stochastic quantization"[20,21]. (Unfortunately, the term "stochastic quantization" has also been used for one or two very different ideas.) If one injects white noise, in a more time-symmetric way, into the interaction terms of a field theory, one generates an effect very similar to what QFT generates, but only in 6 dimensions. If we then postulate that the universe actually has 6 dimensions, we must then explain why we only see 3 or 4, and then we have to consider how the statistics would change as a result of that condensation mechanism. So far as I know, these fundamental problems have yet to be solved. Furthermore, section 6.5 will suggest that we do not need to introduce white noise anyway, in order to match the Standard Model.

The MRF formalism differs from the stochastic quantization formalism, because the white noise is injected into the Lagrangian itself, in effect, rather than the Lagrange-Euler equations. This has substantial implications.

Another fairly well-known approach is called "Stochastic Electrodynamics" (SED), pursued by researchers such as Puthoff. The SED community has claimed to explain many esoteric phenomena in electrodynamics which are almost ignored in conventional research. Phenomena such as unseen particles and Brownian motion are widely discussed in that community. I must confess a lack of knowledge about the details of that literature, based on my concern that mechanisms based on Brownian motion would be difficult to reconcile with special relativity. However, the SED community claim that they have recently found ways to overcome that problem. In any case, if their theories were modified so as to fit within a relativistic MRF framework, that problem would certainly go away. Likewise, the MRF framework would allow local temporary fluctuations in energy (exactly as with virtual particles in the iMRF framework) but would conserve energy in the large, in the same way that the iMRF formalism allows.

Nevertheless, an MRF model using the same Lagrangian as an iMRF model would of course yield different predictions! I personally have no idea how to construct the Lagrangians (if any) which could replicate such theories as EWT or QCD, for $\sigma>0$. On the other hand, I have not made a serious attempt in that direction.

In the case where $\sigma=0$, I do have an idea for how to replicate the standard results of QFT, based on classical perturbation theory. This will be discussed in section 6.5. A serious study of the $\sigma>0$ option should yield two important benefits: (1) an extension of the nonperturbative methods of the standard formalism, to be used for $\sigma=0$ or $\sigma>0$; (2) suggestions for how to measure $\sigma$ empirically, in the neighborhood of $\sigma=0$. If there is any validity to the claims of the SED community, this would also help us assimilate their insights into mainstream physics.

## 6.5. Reified Field Theory: The MRF Model With $\sigma=0$

There are several possible ways to compute the $\sigma=0$ limit of the MRF model. for example, in section 4 of [7], I described a procedure called "reification of moments." I showed that the reified statistical moments generated by a simple scalar field theory form a kind of wave function in Fock space, governed by the usual sort of Schrodinger equation from CQFT, complete with an Hermitian Hamiltonian expressible in terms of the usual sort of operator fields. According to Everett's thesis (discussed in section 6.2), this fact by itself should be enough to prove that qauntjm measurement should also work in that system; however, as I mentioned in section 6.2, it is necessary to prove that solitons/particles exist in a system before it is valid to associate field correlations across a set of points with probabilities of locating a particle at those points. This section will elaborate on these issues. One should note that the other sections of [7] are superseded by this paper and by [4].

Perhaps the $\sigma=0$ limit could be computed more effectively by using functional integral methods to compute (in the $\sigma\rightarrow0$ limit) such functions as:

$$G = \langle \varphi(\underline{x}_1,t_+)\varphi(\underline{x}_2,t_+)...\varphi(\underline{x}_n,t_+)\varphi(\underline{y}_1,t_-)...\varphi(\underline{y}_m,t_-) \rangle$$
$$= \int \varphi(\underline{x}_1,t_+)...\varphi(\underline{y}_m,t_-)\mathbf{Pr}(\Phi)d\Phi \tag{234}$$

However, in [7], I used more explicit, traditional statistical methods.

With traditional methods, correlation functions like G do not obey tractable dynamics over time. For example, the ordinary statistical moments of a classical field $\underline{\varphi}(\mathbf{x},t)$ at time t may be represented (roughly) as a vector in Fock space:

$$v(\Phi(t)) = \exp\left(\int \sum_\alpha \varphi_\alpha(\underline{x},t)a_\alpha^+(\underline{x})d^3\underline{x}\right)|0> \quad, \tag{235}$$

where $a_\alpha^+(\mathbf{x})$ is the usual creation operator of CQFT [40]. But the dynamics of v, deduced from the PDE governing $\underline{\varphi}(\mathbf{x},t)$, take the form:

$$\dot{v} = Av, \tag{236}$$

where "A" is not Hermitian. (Again, see [7] for details.) Thus I described a series of linear transformations on v, culminating in a projection to the eigenspace of H corresponding to the energy of the state Φ. I called these series of transformations "reification."

Let "z(Φ)" represent the reified version of v(Φ), normalized to unit length. For any ensemble of states Φ, I would define the statistical density function as:

$$\rho = \int |z(\varphi)\rangle\langle z(\varphi)|^H \mathbf{Pr}(\Phi)d\Phi \tag{237}$$

How do we compute the results of a scattering experiment, at time $t_+$, in this arrangement? If we know that the results of the experiment resolve themselves into solitons, then we can deduce that the possible outgoing states z(Φ) are orthogonal to each other unless they are close neighbors, representing essentially the same outgoing momentum. (As in section 6.2, I ask that polarizers and the like be treated as part of the interaction zone.) From equation 237, if we know which states $z_+$ represent solitons in the locations of interest, then we can simply "read out" the probabilities Pr(Φ) for any such states, by calculating Tr($\rho z_+ z_+^H$). In fact, this is almost -- but not quite -- the usual CQFT measurement procedure.

For a complete calculation, we must go back to equation 228. In order to implement equation 228, we may begin by calculating $\rho_-(t_-)$, the density matrix at time $t_-$ based on the probabilities Pr(Φ|**M**-). (This calculation corresponds exactly to the usual CQFT "encoding" of incoming states, the density matrix formulation discussed in section 6.2.) Then, because z obeys a Schrodinger equation, we can update this to the resulting $\rho_-(t_+)$ by using the ordinary S-matrix evolution operator procedures of CQFT!

But the resulting density matrix, $\rho_-(t_+)$, is again based a probability distribution for states Φ conditional only upon **M**-. We have to find a way to convolve these probabilities, in effect, to account for the term $\overline{P_+}$ in equation 228. We need to find the complete density matrix $\rho_+$ based on probabilities conditional upon the combination of **M**- and $\mathbf{M}_+^{[m]}$. Fortunately, the effect of $\mathbf{M}_+^{[m]}$ is relatively simple; it acts as a boundary condition, a requirement which all states must meet in order to have any probability at all. Thus we may compute the conditional probabilities (implicitly using Bayes Law) simply by throwing out all states which fail to meet the boundary condition at time $t_+$ and scaling up the probabilities of the rest so that they add up to one. In order to implement this calculation, we may simply project the matrix $\rho_-(t_+)$ into the allowed states, and renormalize the trace to one. Even more simply, we may construct a "quantum mechanical measurement operator M" as a weighted sum of the allowed states $z_+ z_+^H$, and follow the usual CQWFT procedures, which implement this kind of projection. Notice the close parallel here to section 6.2.

In summary, the calculations for scattering for a simple classical field theory end up matching CQFT exactly, if the field would condense into solitons.

Unfortunately, the reification procedures in [7] appear to work easily only for quasilinear theories. The σ=0 model itself would allow nonperturbative procedures, as discussed in previous sections, but not the reification procedures now known. Nevertheless, section 6 of [4] describes a strategy for building up a model of strong and electroweak interactions as the <u>limiting case</u> of a family of such quasilinear models. Considerable further research will be needed to develop the full potential of this approach, in a variety of directions.